\documentclass[reprint,amsmath,amssymb,aps,prb,superscriptaddress]{revtex4-2}
\usepackage{amsthm}
\usepackage{array}
\usepackage{bm}
\usepackage{xcolor}
\usepackage{float}
\usepackage{graphicx}
\usepackage{hyperref}
\hypersetup{colorlinks=true,allcolors=blue}
\usepackage{natbib}
\usepackage{newtxtext}
\usepackage{newtxmath}
\usepackage{mathrsfs}
\usepackage[caption=false]{subfig}

\begin{document}

\title{Mapping vortices to anyons in toric code phases of generalized Kitaev models}

\author{Li Ern Chern}
\affiliation{Max Planck Institute for the Physics of Complex Systems, 01187 Dresden, Germany}

\author{Roderich Moessner}
\affiliation{Max Planck Institute for the Physics of Complex Systems, 01187 Dresden, Germany}

\author{Claudio Castelnovo}
\affiliation{T.C.M.~Group, Cavendish Laboratory, University of Cambridge, Cambridge CB3 0HE, United Kingdom}

\begin{abstract}
We present a comprehensive theory of mapping flux excitations, or vortices, to electric and magnetic particles in the toric code phases of generalizations of the Kitaev honeycomb model in two spatial dimensions. Our method, which is formulated with the Majorana fermion representation, utilizes the fusion rule of the Abelian anyons and the physical constraint on the fermion parity, and applies to generic model parameters including any perturbative limit. Not only are we able to reproduce the known mapping scheme in the dimer limit, we also derive the conditions for the invariance of anyon species of individual vortices. We prove that the mapping of anyons is left unchanged by any continuous evolution of model parameters that does not close the fermion gap in both the vortex-free and two-vortex sectors, which enables precise demarcations between multiple regimes associated with different maps within a single phase characterized by a trivial Chern number. We illustrate our theory via extensive computations for a number of selected models, in particular those defined on the square-octagon lattice and the honeycomb lattice with a Kekul\'{e} structure. We also demonstrate that distinct mappings of anyons can nevertheless exhibit the same weak symmetry breaking, and further argue that they belong to the same symmetry-enriched topological order.
\end{abstract}


\maketitle


\section{\label{section:introduce}Introduction}

In the study of highly-entangled quantum states of matter, few models are as simple, yet at the same time as profound, as the toric code model \cite{KITAEV20032}. It admits an exact solution and hosts $\mathbb{Z}_2$ topological order \cite{RevModPhys.89.041004}, which is beyond the conventional Landau paradigm \cite{landautextbook}, with anyonic excitations \cite{PhysRevLett.49.957}, which are not exactly bosons or fermions. The importance of the toric code model is not confined to the intangible realm of abstract ideas. Since its fourfold ground-state degeneracy on a torus can encode two qubits that are robust against errors, it also finds potential applications in quantum computing \cite{pachostextbook}. Experimental realizations of the toric code model have recently been reported \cite{science.abi8378,science.abi8794}. While the original toric code model defined on the square lattice involves four-body interactions, the Kitaev honeycomb model \cite{KITAEV20062}, which consists only of two-body interactions, offers a pathway for its emergence as an effective theory in certain parameter regimes \footnote{Remarkably, both the toric code model and the Kitaev honeycomb model were invented by Alexei Kitaev, among several other exactly solvable models. To avoid ambiguity, we reserve the term ``Kitaev model'' for the Kitaev honeycomb model or its generalization in this article.}. The relation between these models is included in a larger framework known as Kitaev's sixteenfold way \cite{Kells_2011,PhysRevResearch.2.023334,PhysRevB.102.115130,PhysRevB.102.201111,SciPostPhys.14.5.087,PhysRevB.109.134412}, which states that the topological order of a $2+1\mathrm{D}$ system of free fermions coupled to $\mathbb{Z}_2$ gauge fields depends only on $\nu \, \mathrm{mod} \, 16$, where $\nu$ is the Chern number \cite{KITAEV20062}. In particular, the $\nu = 0$ phase of the Kitaev honeycomb model or any of its generalizations in two spatial dimensions is equivalent to a toric code model at low energies and long distances. They are described by the same topological quantum field theory \cite{simontextbook}, and they have the same set of anyonic excitations. \\

The toric code model has four superselection sectors \cite{KITAEV20062}, namely the vacuum $1$, the electric particle $e$, the magnetic particle $m$ \footnote{$e$ and $m$ are originally named the electric charge and the magnetic vortex, respectively. To avoid confusion with a generic flux excitation also known as a vortex, we refer the magnetic vortex to as the $m$ particle/anyon. The electric charge is similarly referred to as the $e$ particle/anyon.}, and their composite $\psi$. The $e$ and $m$ particles are bosonic with respect to their own species, but they have nontrivial mutual statistics, such that when an $e$ particle encircles an $m$ particle or vice versa, the wavefunction acquires a minus sign. As a result, $\psi$ is fermionic with respect to its own species, but it has nontrivial mutual statistics with $e$ and $m$. Therefore, $e$, $m$, and $\psi$ are anyons. When $\nu=0$, the different types of fractionalized excitations in the Kitaev honeycomb model and its generalizations are identified with these anyons, as follows. A complex fermion constructed from a pair of Majorana fermions belongs to the same superselection sector as $\psi$, while flux excitations or \textit{vortices} (i.e., elementary plaquettes where the $\mathbb{Z}_2$ gauge fluxes are not in accordance with the ground-state flux sector) correspond to $e$ and $m$ \cite{KITAEV20062,PhysRevB.90.134404}. For a fixed set of couplings, each elementary plaquette can only host an anyon of a particular species, i.e., either $e$ or $m$, but not both. Given that much of the topological information is encoded in the braiding properties of anyons, it is desirable to have a simple and robust mapping from the low-energy excitations of a microscopic model to the anyons of the topological order realized by the model. \\

In this work, we study the correspondence between the vortices and the $e$ and $m$ particles in the toric code phases of generalized Kitaev models \eqref{kitaevmodel} on planar graphs with odd coordination numbers. In the existing literature, such a correspondence is typically obtained for some anisotropic limit where the couplings on a subset of bonds are much stronger than the rest, which allows one to derive an effective Hamiltonian via a perturbative expansion \cite{KITAEV20062,PhysRevB.76.180404}. A simple and efficient scheme of mapping vortices to anyons is also provided by Ref.~\cite{Wootton_2015} in case the strong bonds form a dimer covering of the graph. However, the $\nu = 0$ phase possibly extends beyond the perturbative limit to a parameter regime that is less anisotropic or even near the isotropic point where all couplings are equal. In some models \cite{PhysRevB.76.180404,Kamfor_2010}, multiple dimer limits with apparently different mappings are included in a single phase characterized by $\nu = 0$, such that they can be smoothly connected to each other without a fermion-gap-closing transition in the ground state. This raises the questions of how the parameter space is partitioned to accommodate distinct mappings of anyons and what happens to the individual vortices when the boundary of such a partition is crossed, which have not been addressed thus far. Here, we develop a comprehensive theory of mapping anyons in toric code phases of generalized Kitaev models, building on a method that was first implemented by Ref.~\cite{PhysRevResearch.2.023334} to explore distinct topological orders in the isotropic Kitaev honeycomb model augmented by three- and four-body interactions. Formulated with the Majorana fermion representation, this method utilizes the fusion rules of anyons \cite{KITAEV20062,simontextbook} and the fermion parity of physical states \cite{PhysRevB.84.165414,PhysRevB.92.014403}, and it is able to treat generic parameters including any perturbative limit. To sketch the core idea, consider the creation of a pair of vortices by reversing the $\mathbb{Z}_2$ gauge fields along a string of bonds, which connects two elementary plaquettes sufficiently far away from each other such that their braiding statistics is well-defined. If this requires a (no) change in the parity of the complex fermions, then the vortices belong to different (the same) anyon species. With this method, we derive the criteria under which the anyon species of a given vortex must remain invariant as the model parameters are changed adiabatically, in addition to reproducing the known mapping scheme in the dimer limit. \\

The central results of our work are summarized by the following two statements. \vspace{4pt} \\
\noindent \textbf{Proposition 2.} Let $p_0$, $p$, and $p'$ be three elementary plaquettes such that (i) $p_0$ is well-separated from $p$ and $p'$ and (ii) the two-vortex sectors $(p_0 , p)$ and $(p_0 , p')$ have opposite fermion parities. If the fermion gaps $\Delta_\psi (p_0 , p)$ and $\Delta_\psi (p_0 , p')$ remain finite throughout an adiabatic transformation in the toric code phase, then the anyon species of each of the vortices at $p_0$, $p$, and $p'$ is invariant. \vspace{4pt} \\
\noindent \textbf{Corollary 2.} If the fermion gap of every two-vortex sector remains finite throughout an adiabatic transformation in the toric code phase, then the assignment of anyon species is invariant. \vspace{4pt} \\
\noindent The two-vortex sector $(p_0 , p)$ refers to the configuration of $\mathbb{Z}_2$ gauge fluxes with two flux excitations at the elementary plaquettes $p_0$ and $p$. Throughout this work, adiabatic transformations specifically refer to continuous changes of model parameters. We elaborate on the significance of these statements as follows. In the absence of a fermion-gap-closing transition in the vortex-free sector that defines the ground state, a change in the mapping of anyons is possible only if the fermion gaps of some two-vortex sectors vanish, as per Corollary 2. One may draw a loose analogy with the invariance of the ground-state Chern number $\nu$ as long as the fermion gap of the vortex-free sector remains finite. The enlarged set of critical parameters, at which the fermion gap closes in the vortex-free or two-vortex sectors, form the boundaries between distinct mappings of anyons. Each connected regime in the parameter space that results from the demarcation by these critical parameters now contains at most one dimer limit, where the mapping of anyons is most easily obtained. Proposition 2 further enables us to determine which vortices can be converted from one anyon species to the other and which must remain unchanged when the model parameters are tuned across some of these boundaries. We remark that Corollary 2 is an immediate consequence of Proposition 2, the proof of which requires the adiabatic theorem \cite{BF01343193,JPSJ.5.435,griffithstextbook}, and that it suffices to evaluate the fermion gaps for a small subset of all possible two-vortex sectors when translational symmetries are present. \\

\begin{figure}
\subfloat[]{\label{figure:octagonphase}
\includegraphics[scale=0.22]{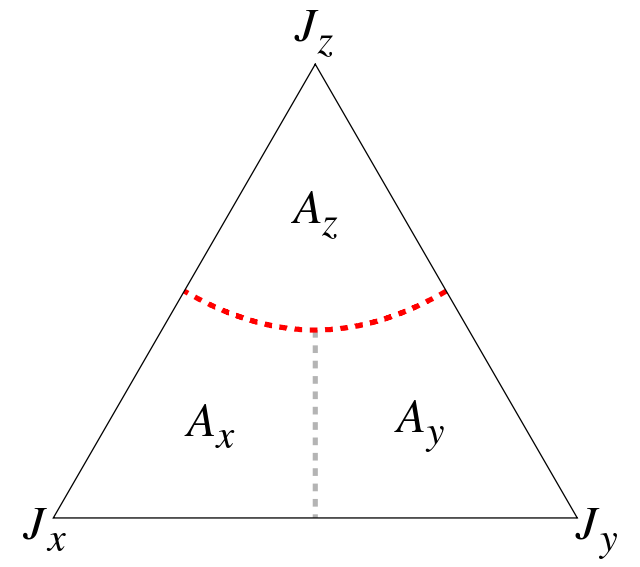}} \hspace{20pt}
\subfloat[]{\label{figure:kekulephase}
\includegraphics[scale=0.22]{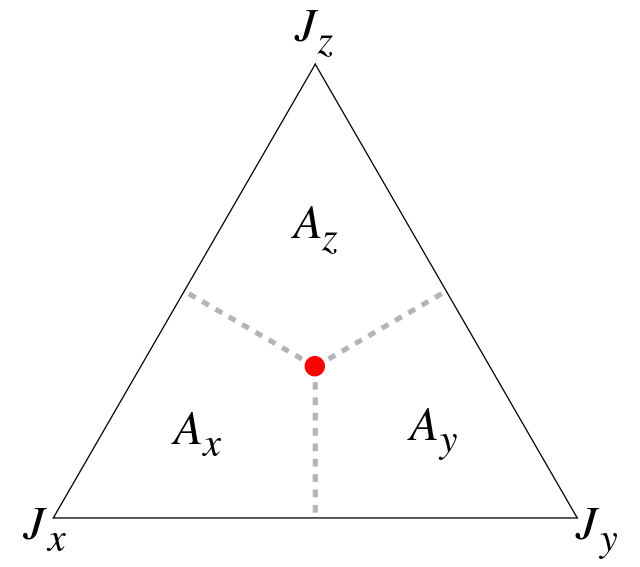}}
\caption{Toric code phases $A_x$, $A_y$, and $A_z$ that are characterized by different assignments of anyon species in the Kitaev (a) square-octagon and (b) Kekul\'{e} models. The red dashed curve and the red dot indicate the vanishing of the fermion gap in the vortex-free sector and all two-vortex sectors. The gray dahsed lines indicate the vanishing of the fermion gap in some two-vortex sectors, while the vortex-free sector has a finite fermion gap. The triangular parameter space is defined by $J_x + J_y + J_z = 1$, where the vertex labeled by $J_\lambda$ indicates the $J_\lambda \longrightarrow 1$ limit. All couplings are positive by the definition of the model \eqref{kitaevmodel}.}
\end{figure}

The generalized Kitaev models on the square-octagon lattice (Fig.~\ref{figure:octagonmodel}) and the honeycomb lattice with a Kekul\'{e} structure (Fig.~\ref{figure:kekulemodel}) offer excellent illustrations of our theory. In the former model, the fermion-gap-closing transition in the vortex-free sector only separates the strong $J_z$ limit from the strong $J_x$ and $J_y$ limits, but not the strong $J_x$ and $J_y$ limits from each other \cite{PhysRevB.76.180404}. In the latter model, the fermion gap of the vortex-free sector only closes at the isotropic point in the parameter space, such that the three distinct strong bond limits can be smoothly connected to each other \cite{Kamfor_2010,PhysRevB.91.134419}. In light of Corollary 2, we compute the fermion gaps of two-vortex sectors for each of these models. We find that the enlarged set of critical parameters indeed splits the parameter space into three parts, each of which contains a unique dimer limit, see Figs.~\ref{figure:octagonphase} and \ref{figure:kekulephase}. Applying Proposition 2 to the Kitaev Kekul\'{e} model further reveals that a change of anyon species only takes place at one-third of the unit hexagons when the model parameters are tuned across one of the three gray dashed lines in Fig.~\ref{figure:kekulephase}, such that the anyon species of every vortex would have been changed by varying the model parameters along a closed path that winds around the isotropic point. Our theory thus provides a simple understanding of the $e \longleftrightarrow m$ automorphism in the Kitaev Kekul\'{e} model \cite{PhysRevB.106.085122}, which is closely related to the honeycomb Floquet code \cite{q-2021-10-19-564}. Finally, we show that the necessity of a fermion-gap-closing transition in the vortex-free sector between different dimer limits can be understood from the corresponding patterns of weak symmetry breaking \cite{KITAEV20062,PhysRevResearch.3.023120}, where the $e$ and $m$ anyons are permuted by some symmetries of the model. We argue that two inequivalent mappings of anyons can be in the same symmetry-enriched topological order \cite{PhysRevB.100.115147,PhysRevX.14.021053} if there exist elementary plaquettes that are not related by symmetry. \\

From a broader perspective, our results suggest a finer classification of phases of matter on top of the existing hierarchy of frameworks. Given a microscopic model with a symmetry group \footnote{In this work, the term ``symmetry'' is used in the conventional sense, i.e., it refers to the global symmetry or the $0$-form symmetry, not the generalized symmetry or a higher-form symmetry.}, one may first determine whether the ground state breaks some symmetries and identify the relevant local order parameters as per the Landau-Ginzburg theory. If the ground state is symmetric and gapped, one may then determine whether it is short- or long-range entangled, i.e., whether it is topologically ordered. Topological order may be further enriched by symmetry, such that phases with the same topological order are distinguished by the symmetry actions on the anyonic excitations. Finally, phases with the same symmetry-enriched topological order may be distinguished by the mappings of anyons, as we have demonstrated for some generalized Kitaev models. In terms of applications, our results could be useful for building quantum memories as in the original toric code model, if we had a realization of a generalized Kitaev model on a torus. Let a non-contractible $e$ ($m$) string be an operator that creates a pair of $e$ ($m$) particles locally, moves one of them around a cycle of the torus, and finally recombines them into the vacuum. The four basis states in the ground-state subspace, which is the code space, can be chosen such that they are eigenstates of the parity of the number of non-contractible $e$ strings along each cycle of the torus. The non-contractible $e$ ($m$) string along a cycle thus acts as the logical $X$ ($Z$) operator on a qubit \cite{KITAEV20032,simontextbook}. The construction of these string operators would require the knowledge of which vortices correspond to $e$ particles and which correspond to $m$. Our theory reveals that such a correspondence is robust in the sense that it is invariant as long as the fermion gap remains finite in both the vortex-free and two-vortex sectors, which is helpful if the model parameters are only known approximately. \\

The rest of this paper is organized as follows. In Sec.~\ref{section:model}, we define generalizations of the Kitaev honeycomb model to planar graphs with odd coordination numbers via the $\Gamma$ matrices of the Clifford algebra. In Sec.~\ref{section:method}, we explain how the fusion rules of the Abelian anyons can be combined with the physical constraint on the fermion parity to identify the superselection sectors of the vortices. In Sec.~\ref{section:theory}, we present a comprehensive theory of mapping vortices to anyons, which covers not only the perturbative limit but also generic parameters in the toric code phase of any generalized Kitaev model. In Sec.~\ref{section:apply}, we apply our theory to a number of selected models. In particular, we compute and analyze the fermion gaps of two-vortex sectors for the Kitaev square-octagon and Kekul\'{e} models. We also examine these models, together with the Kitaev honeycomb and star models, from the viewpoint of symmetry-enriched topological orders. In Sec.~\ref{section:discuss}, we summarize our results, discuss their implications, and point out several open questions that may be of interest to future studies. We have also prepared a Supplemental Material \cite{supply} that contains various details of the background knowledge, e.g., the standard degenerate perturbation theory and the physical constraint that arises from the Majorana fermion representation of the $\Gamma$ matrices, the intermediate steps of some proofs, e.g., a few lemmas and the adiabatic theorem required for Proposition 2, and additional results, e.g., the demonstration of the correct braiding statistics of the $e$ and $m$ anyons in the dimer limit and the analyses of the full fermion spectra of the Kitaev square-octagon and Kekul\'{e} models.

\section{\label{section:model}Model}

\begin{figure}
\subfloat[]{\label{figure:honeycombmodel}
\includegraphics[scale=0.2]{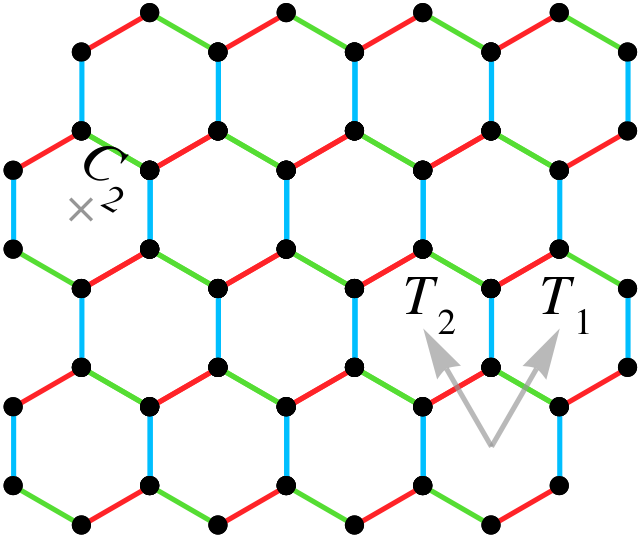}} \quad
\subfloat[]{\label{figure:octagonmodel}
\includegraphics[scale=0.2]{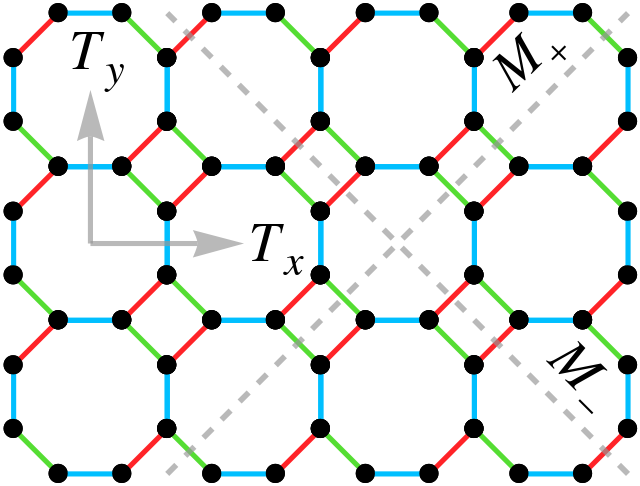}} \\
\subfloat[]{\label{figure:starmodel}
\includegraphics[scale=0.2]{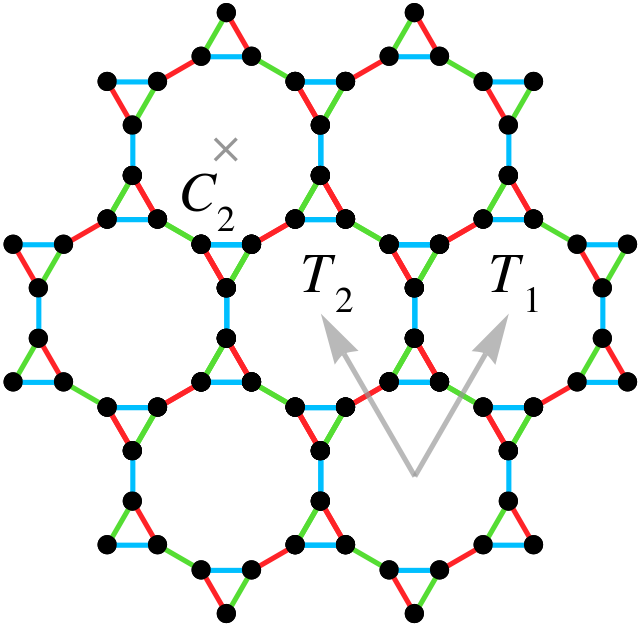}} \quad
\subfloat[]{\label{figure:kekulemodel}
\includegraphics[scale=0.2]{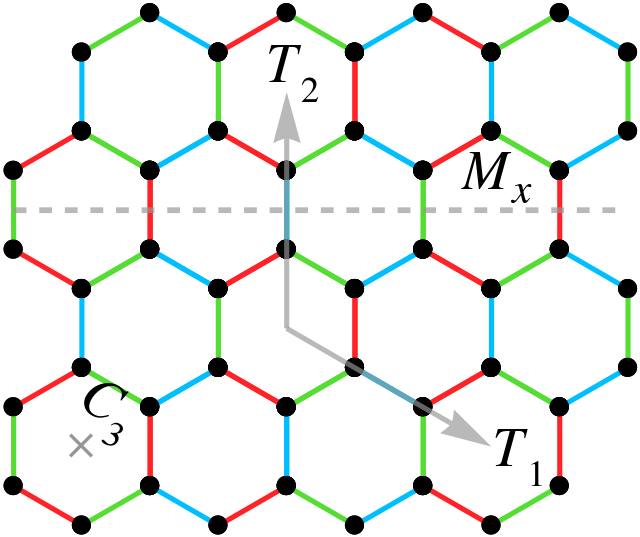}} \\
\subfloat[]{\label{figure:amorphousmodel}
\includegraphics[scale=0.2]{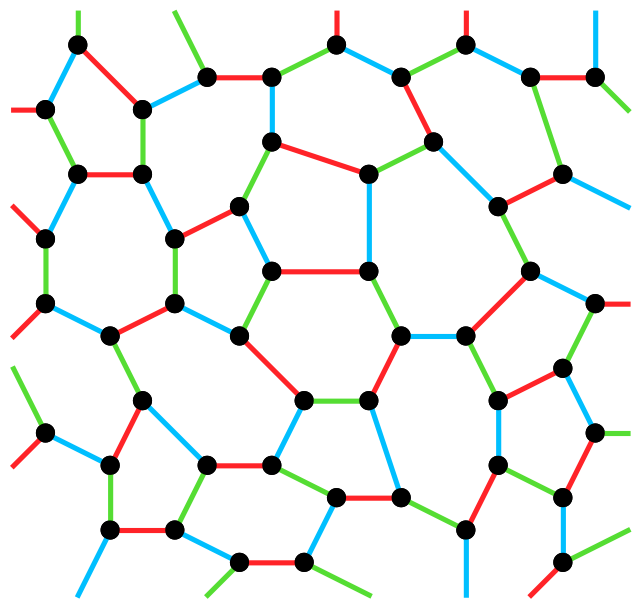}} \quad
\subfloat[]{\label{figure:shastrymodel}
\includegraphics[scale=0.2]{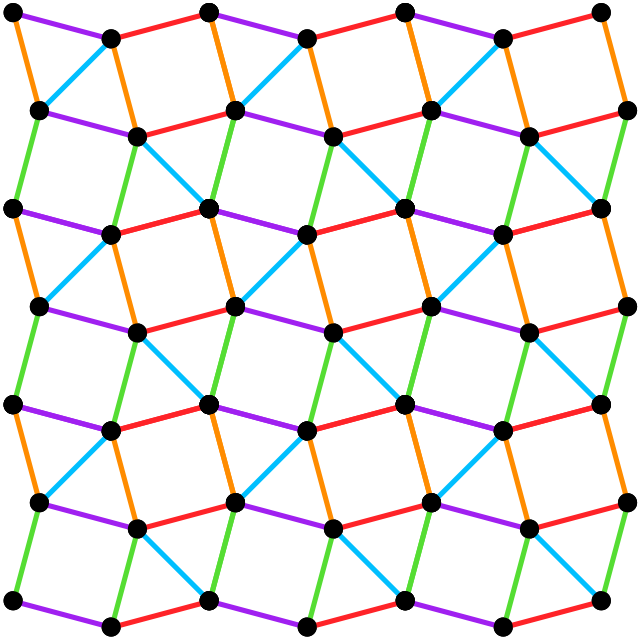}}
\caption{(a) The Kitaev honeycomb model and its generalizations on (b) the square-octagon lattice, (c) the star lattice, (d) the honeycomb lattice with a Kekul\'{e} structure, (e) a tri-coordinated amorphous solid, and (f) the Shastry-Sutherland lattice. The bonds are colored according to the types of interactions, see \eqref{kitaevmodel}. In (a)-(e), the red, green, and blue colors correspond to $x$, $y$, and $z$, respectively. In (a)-(d), the generators of the space groups of the corresponding models are indicated. $T$, $C$, and $M$ (with appropriate subscripts) represent translational, rotational, and reflection symmetries, respectively. Arrows, crosses, and dashed lines represent primitive vectors, centers of rotation, and mirror planes, respectively.}
\end{figure}

The Kitaev honeycomb model \cite{KITAEV20062} is defined by bond-dependent Ising interactions on the honeycomb lattice. It can be generalized to any graph with an odd coordination number $z=2n-1, n \geq 2$, which reads
\begin{equation} \label{kitaevmodel}
H = - \sum_\lambda \sum_{\langle ij \rangle \in \lambda} J_\lambda \Gamma_i^\lambda \Gamma_j^\lambda , \quad J_\lambda > 0 ,
\end{equation}
where $\lambda$ runs over a set of $2n-1$ distinct indices, and the $2^{n-1}$ dimensional $\Gamma$ matrices satisfy locality and the Clifford algebra $\lbrace \Gamma_i^\mu , \Gamma_j^\nu \rbrace = 2 \delta_{ij} \delta^{\mu \nu}$ \cite{PhysRevB.79.134427}. In the familiar case of $n=2$, the $\Gamma$ matrices are simply the Pauli matrices, which can be interpreted as the three components of a $S=1/2$ degree of freedom. Examples of \eqref{kitaevmodel} that we consider in this work are the Kitaev honeycomb model and its generalizations on the square-octagon lattice \cite{PhysRevB.76.180404}, the star lattice \cite{PhysRevLett.99.247203,PhysRevB.82.174412}, the honeycomb lattice with a Kekul\'{e} structure \cite{Kamfor_2010,PhysRevB.91.134419}, a tri-coordinated amorphous solid \cite{s41467-023-42105-9}, and the Shastry-Sutherland lattice \cite{PhysRevB.98.054432}, as shown in Figs.~\ref{figure:honeycombmodel}-\ref{figure:shastrymodel}. Although we focus here on planar graphs with $z=3$ and $5$ for simplicity, our results hold for networks with higher odd coordination numbers in two spatial dimensions. \\

The model \eqref{kitaevmodel} is exactly solvable via a representation of the local $\Gamma$ matrices by $2n$ flavors of Majorana fermions \cite{KITAEV20062,PhysRevB.79.134427,PhysRevB.79.075124,PhysRevLett.102.217202}, $\Gamma_i^\lambda = i b_i^\lambda c_i$. Since the bond operators $u_{ij}^\lambda \equiv i b_i^\lambda b_j^\lambda$ commutes with each other as well as with $H$, we can replace them by their eigenvalues $\pm 1$, which leads to the quadratic Hamiltonian
\begin{equation} \label{kitaevmodelquadratic}
H = \frac{i}{4} \sum_{ij} c_i A_{ij} c_j , \quad A_{ij} = \left \lbrace \begin{array}{ll} 2 J_\lambda u_{ij}^\lambda & \mathrm{if} \, \langle ij \rangle \in \lambda , \\ 0 & \mathrm{otherwise.} \end{array} \right.
\end{equation}
The fermion excitation spectrum is given by the non-negative eigenvalues $\varepsilon_k \geq 0$ of the matrix $i A$ \footnote{Since $u_{ji} = - u_{ij}$, $A$ is a real antisymmetric matrix. The complete eigenvalues of $iA$ come in positive-negative pairs.}. We refer to the minimum of the fermion excitation spectrum, $\min_k \varepsilon_k$, as the fermion gap $\Delta_\psi$. For clarity, we will keep the prefix ``fermion'' throughout this work, so that the fermion gap is not confused with the flux/vortex gap, which is a different quantity. \\

On each bond $\langle ij \rangle_\lambda$, $u_{ij}^\lambda = \pm 1$ is interpreted as a $\mathbb{Z}_2$ gauge field, which is not gauge-invariant. Instead, the $\mathbb{Z}_2$ gauge flux of an elementary plaquette $p$,
\begin{equation}
W_p \equiv \prod_{\langle ij \rangle_\lambda \in \partial p} \Gamma_i^\lambda \Gamma_j^\lambda = (- i)^{\lvert \partial p \rvert} \prod_{\langle ij \rangle_\lambda \in \partial p} u_{ij}^\lambda ,
\end{equation}
is a gauge-invariant and thus physical quantity, where $\partial p$ denotes the boundary of $p$, and $\lvert \partial p \rvert$ the length of $\partial p$ measured in units of bonds, e.g., $\lvert \partial p \rvert = 6$ for a unit hexagon on the honeycomb lattice. $W_p$ commutes with each other as well as with $H$, so the total Hilbert space can be divided into different flux sectors labeled by the eigenvalues $\pm (- i)^{\lvert \partial p \rvert}$ of all $W_p$. The flux sector that minimizes the ground-state energy can be determined analytically by Lieb's theorem \cite{PhysRevLett.73.2158,Macris1996} when the model fulfills certain geometric conditions or numerically by, e.g., quantum Monte Carlo simulations \cite{PhysRevLett.113.197205,PhysRevLett.115.087203,PhysRevB.96.125124}. Discussions of Lieb's theorem and the ground-state flux sectors of the Kitaev honeycomb, square-octagon, Kekul\'{e}, and star models can be found in Sec.~\ref{section:lieb} of the Supplemental Material \cite{supply}. \\

We study the toric code phases of generalized Kitaev models, which include the strongly anisotropic limit where one of the $2n-1$ couplings dominates over others, i.e., $J_\lambda \gg J_{\lambda'}$ for a given $\lambda$ and all $\lambda' \neq \lambda$. Since the $2n-1$ bonds confluencing at an arbitrary site carry all distinct indices, the strong bonds indexed by $\lambda$ form a dimer covering of the lattice of interest, such that each site is touched by one and only one dimer. These dimers are disconnected from each other, so the Majorana fermions are strongly localized on them, akin to the electrons bound to individual atoms in an atomic insulator \cite{RevModPhys.82.3045}. The ground-state Chern number $\nu$ is thus zero, which implies a toric code phase by Kitaev's sixteenfold way \cite{KITAEV20062}. We are interested in the problem of assigning the two anyon species $e$ and $m$ of the toric code model to the elementary plaquettes of generalized Kitaev models, not only in the dimer limit but also in the $\nu=0$ parameter regime where the anisotropy is weak or moderate. \\

To motivate our choice of methodology in the next section, we discuss how the dimer limit is usually treated in the existing literature. One may apply degenerate perturbation theory \cite{PhysRevB.69.064404,strongcoupleexpand,PhysRevB.99.104408} to derive an effective Hamiltonian at low energies and demonstrate its equivalence to a toric code model. This has been done, e.g., for the Kitaev honeycomb \cite{KITAEV20062} and square-octagon \cite{PhysRevB.76.180404} models in their respective strong $z$ bond limits. However, the standard degenerate perturbation theory, as well as its variants \cite{PhysRevLett.100.057208,PhysRevLett.100.177204,PhysRevB.78.245121,PhysRevB.90.134404}, has a number of limitations, the most prominent of which is that the effective Hamiltonian thus derived can never contain an operator that acts nontrivially only on an odd-length elementary plaquette \cite{supply}. This leads to difficulties in interpreting the model in question as a toric code and determining the anyon species of some vortices, when the graph contains odd-length elementary plaquettes, such as the unit triangles on the star lattice \cite{PhysRevB.78.125102,PhysRevB.81.104429} and the unit polygons on the regular hyperbolic tiling $\lbrace k , 3 \rbrace$ with odd $k$ \cite{PhysRevLett.134.256604,mx1t-74dm}. \\

If one is primarily interested in the correspondence between the vortices and the $e$ and $m$ anyons, one can circumvent degenerate perturbation theory together with its limitations by employing a simple mapping scheme, which appeared either explicitly or implicitly in several earlier works \cite{PhysRevB.90.134404,Wootton_2015,PhysRevB.108.195134}. A heuristic argument is given as follows. Let us take the $J_z \gg J_x , J_y$ parameter regime of the Kitaev honeycomb model as an example. The energy scale $\sim J_z$ of an excited dimer, where the two spins (measured in the $\sigma^z$ basis) on a $z$ bond are anti-aligned, is the same as a fermion excitation that would be obtained with the Majorana fermion representation of spins. One may thus identify an excited dimer with the presence of a high-energy fermion \cite{PhysRevB.90.134404}. Let $i$ be an arbitrary site. One finds that applying $\sigma_i^\lambda$ to the ground state creates two vortices at neighboring unit hexagons that share the $\lambda$ bond with one end at $i$ for every $\lambda \in \lbrace x , y , z \rbrace$ but excites a dimer only if $\lambda = x$ or $y$. Since a local operator can only create or annihilate fermions in pairs, one concludes that the vortices on two unit hexagons sharing a $z$ ($x$ or $y$) bond belong to the same (opposite) anyon species, such that they fuse into the vacuum (a fermion), cf.~Fig.~\ref{figure:honeycombanyonz}. This argument can be generalized to any dimer limit of any planar graph with an odd coordination number. \\

Nonetheless, the above scheme of mapping vortices to anyons is unable to treat model parameters away from the dimer limit yet still in the toric code phase. In particular, it does not determine the extent of a given map, the boundary of which does not necessarily coincide with a fermion-gap-closing transition in the ground state, in the parameter space. Interestingly, multiple dimer limits, which are associated with different maps according to the scheme, may be included in a connected parameter regime characterized by $\nu = 0$. To study the correspondence between vortices and anyons beyond the dimer limit, we employ a method formulated with the Majorana fermion representation of $\Gamma$ matrices, which is explained in the next section.

\section{\label{section:method}Methodology}

The Majorana fermion representation unfortunately introduces unphysical states, which double the dimension of the local Hilbert space. A projection back to the physical subspace is thus necessary. The projector is given by \cite{KITAEV20062}
\begin{equation} \label{projectphysical}
\mathcal{P} = \prod_{i=1}^N \frac{1+D_i}{2}, \quad D_i \equiv - (- i)^n \left( \prod_\lambda b_i^\lambda \right) c_i ,
\end{equation}
where $N$ is the total number of sites and we have chosen the $2n-1$ $\Gamma$ matrices such that $\prod_\lambda \Gamma^\lambda = i^{n-1}$. For $n=2$, e.g., $\sigma^x \sigma^y \sigma^z = i$, so $D_i = b_i^x b_i^y b_i^z c_i$. $D_i$ anticommutes with the gauge-field operator $u_{jk}^\lambda$ if and only if $j=i$ or $k=i$. In other words, it changes the gauge field on each bond with one end at $i$ by a minus sign, while preserving all the gauge fluxes. For this reason, each $D_i$, as well as each product of multiple $D_j$ for different $j$, is known as a gauge transformation \cite{KITAEV20062}. \\

According to the analysis in Ref.~\cite{PhysRevB.84.165414}, which is reviewed in Sec.~\ref{section:constraint} of the Supplemental Material \cite{supply}, for an eigenstate $\lvert \phi \rangle$ of the quadratic Hamiltonian \eqref{kitaevmodelquadratic}, $\mathcal{P} \lvert \phi \rangle$ is nonzero if and only if
\begin{equation} \label{physicalconstraint}
\prod_{i=1}^N D_i \lvert \phi \rangle = \lvert \phi \rangle .
\end{equation}
After some algebra \cite{supply}, one arrives at
\begin{equation} \label{projectproduct}
\begin{aligned}[b]
\prod_{i=1}^N D_i &= (-1)^\xi \left( \prod_{\langle ij \rangle_\lambda} u_{ij}^\lambda \right) \, \det Q \prod_{k=1}^{N/2} \left( 2 \psi_k^\dagger \psi_k - 1 \right) \\
&\equiv (-1)^\xi \, \mathcal{P}_b \det Q \, \mathcal{P}_\psi ,
\end{aligned}
\end{equation}
where $\xi \in \mathbb{Z}$ depends only on the geometry of the system, e.g., the connectivity of the graph and the total number of sites, the $-1$ factors arise from anticommuting the Majorana fermions past each other, $Q \in O (N)$ is the canonical transformation that brings $A$ into the following block diagonal form
\begin{equation} \label{canonicaltransform}
Q^\mathrm{T} A Q = \bigoplus_{k=1}^{N/2} \begin{pmatrix} 0 & \varepsilon_k \\ - \varepsilon_k & 0 \end{pmatrix} , \quad \varepsilon_k \geq 0 ,
\end{equation}
and $\psi_k = (\gamma_{2k - 1} + i \gamma_{2 k}) / 2$ are complex fermions constructed from the Majorana fermions $\gamma_i = \sum_j c_j Q_{ji}$. Each of $\mathcal{P}_b$, $\det Q$, and $\mathcal{P}_\psi$ is either $+1$ or $-1$, so they behave like parity operators. The implication of \eqref{projectproduct} is as follows. $\xi$ is determined by the geometry of the system, while the bond parity $\mathcal{P}_b$ and, for a fixed set of couplings, $\det Q$ are determined by the gauge-field configuration $\lbrace u_{ij}^\lambda \rbrace$. Therefore, once the geometry, the couplings, and the gauge fields are all specified, the fermion parity $\mathcal{P}_\psi$ of a physical state will be fixed by the constraint $\prod_i D_i = 1$, and any state with the opposite fermion parity will be unphysical \cite{PhysRevB.84.165414,PhysRevB.92.014403}. \\

\begin{figure}
\includegraphics[scale=0.36]{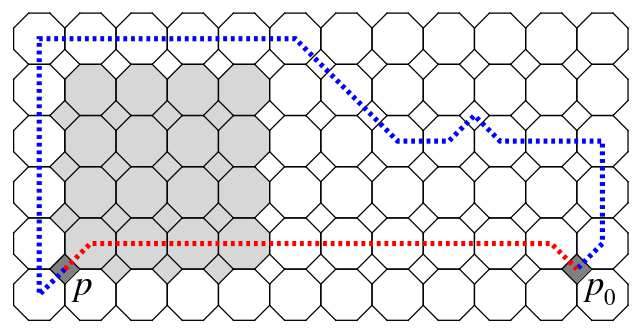}
\caption{\label{figure:twovortex}The creation of two vortices at the dark gray unit squares, $p_0$ and $p$, from the ground state of the Kitaev square-octagon model by flipping a string of bonds connecting them, crossed by the red dashed line. We fix the vortex at $p_0$ while moving the other over the light gray region, to map out the anyon species of both the unit squares and octagons in that region. The blue dashed line represents another string of bonds connecting the plaquettes $p_0$ and $p$.}
\end{figure}

On the other hand, the fusion rules of anyons in the toric code model reads \cite{KITAEV20062,simontextbook}
\begin{equation} \label{fusionrule}
\begin{gathered}[b]
1 \times 1 = 1 , \quad 1 \times e = e , \quad 1 \times m = m , \quad 1 \times \psi = \psi , \\
e \times e = 1 , \quad m \times m = 1 , \quad \psi \times \psi = 1 , \\
e \times m = \psi , \quad m \times \psi = e , \quad \psi \times e = m ,
\end{gathered}
\end{equation}
where the $\times$ operation is commutative. Importantly, the combination of $e$ ($m$) and $\psi$ belongs to the same superselection sector as $m$ ($e$). Based on the fusion rules \eqref{fusionrule} and the physical constraint \eqref{physicalconstraint}, Ref.~\cite{PhysRevResearch.2.023334} has developed a method of mapping the vortices in an extended Kitaev honeycomb model to the $e$ and $m$ particles, which we elaborate as follows. To simplify the discussions, we first introduce the terminology ``bond flip'' to represent the act of reversing the sign of the gauge field $u_{ij}^\lambda$ on a bond $\langle ij \rangle_\lambda$, i.e., $u_{ij}^\lambda \longrightarrow - u_{ij}^\lambda$. Our system is defined on a torus with a finite number of sites. Starting from the ground-state flux sector, we create two vortices at the elementary plaquettes $p_0$ and $p$ sufficiently far away from each other by flipping a string of bonds between them, as illustrated in Fig.~\ref{figure:twovortex} for the square-octagon lattice. We compare the physical fermion parities of this two-vortex sector and another one obtained by fixing the vortex at $p_0$ while moving the other to a different plaquette $p'$. If these fermion parities are same (opposite), then the vortices at $p$ and $p'$ belong to same (different) species. In more details, the physical constraint $\prod_i D_i = 1$ requires
\begin{equation} \label{tripodequal}
\mathcal{P}_b \det Q \, \mathcal{P}_\psi = \mathcal{P}_b' \det Q' \, \mathcal{P}_\psi' ,
\end{equation}
where, for a fixed set of couplings, $\mathcal{P}_b$ ($\mathcal{P}_b'$) and $\det Q$ ($\det Q'$) are determined by the set of gauge fields $\lbrace u_{ij}^\lambda \rbrace$ ($\lbrace u_{ij}^\lambda \rbrace'$) of the two-vortex sector involving the elementary plaquettes $p_0$ and $p$ ($p'$). We can thus calculate the relative fermion parity
\begin{equation} \label{relativefermionparity}
\frac{\mathcal{P}_\psi}{\mathcal{P}_\psi'} = \frac{\mathcal{P}_b' \det Q'}{\mathcal{P}_b \det Q} = \pm 1 ,
\end{equation}
so that $+1$ ($-1$) indicates that the vortices at $p$ and $p'$ belong to the same (different) species. Using this method, we can map out the anyon species $e$ and $m$ of the elementary plaquettes over a region sufficiently far from $p_0$. Different mapped regions can then be patched together to cover the entire graph. \\

Several comments are in order. First, we define a vortex to be a flux excitation, which is not necessarily a $\pi$-flux. The ground-state flux sector is always the vortex-free sector. Second, the separation of $p$ or $p'$ from $p_0$ should be, ideally, much greater than the correlation length \footnote{However, $p'$ does not have to be far away from $p$.} to ensure well-defined anyonic braiding statistics \cite{KITAEV20062}. The corresponding pair of vortices is then said to be well-separated. Third, two sets of gauge fields that differ by a gauge transformation lead to the same physical requirement on the fermion parity \cite{supply}. It is thus sufficient to determine the physical fermion parity of a given flux sector from a representative set of gauge fields among the sets generated by all possible gauge transformations. Fourth, the toric code phase has a global anyon permutation symmetry, under which the labels $e$ and $m$ can be freely exchanged without affecting the modular $S$ and $T$ matrices that define the topological quantum field theory \cite{simontextbook}. This gives an initial freedom of identifying, e.g., the vortex at $p$ either as $e$ or $m$, as long as we consistently distinguish the two different species throughout the system. However, once an assignment of anyon species has been specified for a set of couplings, the assignments for other sets of couplings is no longer independent. We will see some examples in Sec.~\ref{section:apply}. Finally, the method described in the previous paragraph is not limited to the toric code phase characterized by a trivial Chern number $\nu = 0$. It can also be applied to any phase with an even Chern number that falls under Kitaev's sixteenfold way \cite{PhysRevResearch.2.023334}. When $\nu \, \mathrm{mod} \, 2 = 0$, the vortices are Abelian anyons of two distinct species, which differ from each other by a fermion \cite{KITAEV20062}.

\section{\label{section:theory}Theory}

Our goal is to determine the anyon species of each vortex in the toric code phase, including parameter regimes beyond the perturbative limit, of any generalized Kitaev model. In light of the limited scope of applicability of degenerate perturbation theory and the simple mapping scheme discussed at the end of Sec.~\ref{section:model}, the method outlined in Sec.~\ref{section:method}, which is formulated with the Majorana fermion representation, offers a useful and general framework. We first demonstrate that this method is able to reproduce the known mapping scheme in the dimer limit (Proposition 1 and Corollary 1), and then derive the criteria under which the assignment of anyon species must be invariant when the model parameters are changed (Proposition 2 and Corollary 2). \\

Before proceeding, we mention a potential complication due to the topology of the torus. The implementation of periodic boundary conditions may cause an $e$ ($m$) particle to mutate into an $m$ ($e$) particle, which is known as anyon transmutation \cite{PhysRevB.90.134404}. When this indeed happens, one cannot consistently assign the $e$ and $m$ labels throughout the torus on which our system is defined. An example is given in Sec.~\ref{section:transmute} of the Supplemental Material \cite{supply}. Nevertheless, a well-defined mapping is always possible within a region that does not wrap around either cycle of the torus. \\

In the following, Lemma 1, Proposition 1, and Corollary 1 specifically refer to the anisotropy limit where strong bonds form a dimer covering of the planar graph, such that each site is touched by exactly one dimer. A weak bond is a bond that is not a strong bond. Lemma 2, Proposition 2, and Corollary 2 do not assume a dimer limit. For brevity, Lemmas 1 and 2 are stated hereafter, while their proofs are presented in the Supplemental Material \cite{supply}. \\

\noindent \textbf{Definition 1.} Let $l$ be a string of bonds that connects a pair of elementary plaquettes, such that if all the bonds are flipped, then two vortices are created at these elementary plaquettes. We define $n_w (l)$ to be the total number of weak bonds contained in $l$, and the parity of $n_w (l)$ to be $n_w (l) \, \mathrm{mod} \, 2$. \\

\noindent \textbf{Lemma 1.} Let $l_1$ and $l_2$ be two strings of bonds that connect the same pair of elementary plaquettes. Suppose that $l_1$ and $l_2$ form a contractible loop. Then, $n_w (l_1)$ and $n_w (l_2)$ have the same parity. \\

\noindent \textbf{Proof.} See Sec.~\ref{section:lemma1} of the Supplemental Material \cite{supply}. \\

In other words, $n_w (l_1)$ is odd (even) if and only if $n_w (l_2)$ is odd (even). Examples of $l_1$ and $l_2$ are the strings of bonds crossed by the red and blue dashed lines, respectively, in Fig.~\ref{figure:twovortex}. \\

\noindent \textbf{Lemma 2.} The determinant of the canonical transformation is invariant under an adiabatic transformation that does not close the fermion gap. \\

\noindent \textbf{Proof.} See Sec.~\ref{section:canonical} of the Supplemental Material \cite{supply}. \\

In this work, an adiabatic transformation specifically refers to a continuous change in the coupling strengths, i.e., the model parameters. Since the gauge fields only take discrete values $\pm 1$, a bond flip $u_{ij}^\lambda \longrightarrow - u_{ij}^\lambda$ is not an adiabatic transformation. \\

\noindent \textbf{Proposition 1.} Let $l$ ($l'$) be a string of bonds that connect the elementary plaquettes $p_0$ and $p$ ($p'$). If $n_w (l)$ and $n_w (l')$ have the same parity, then the vortices at $p$ and $p'$ belong to the same species. In contrast, if $n_w (l)$ and $n_w (l')$ have opposite parities, then the vortices at $p$ and $p'$ belong to different species. \\

\textit{Remarks.} For a well-defined mapping of anyons over a region that is topologically equivalent to a disk, one should require $l'$ be obtained from $l$ by deformations only within that region, to avoid the potential complication due to periodic boundary conditions as discussed earlier \footnote{Alternatively, but to the same effect, one can allow the deformations to happen within any predefined disk that contains the region of mapping.}. By Lemma 1, the parity of $n_w (l')$ is the same for all such choices of $l'$ that connect the same pair of elementary plaquettes $p_0$ and $p'$. \\

\noindent \textbf{Proof.} Let $\lbrace \overline{u}_{ij} \rbrace$ be a set of gauge fields that produces the ground-state flux sector. For simplicity, we have suppressed the superscript on the gauge field that indicates the bond type. We index the $N$ sites of the system such that for every $1 \leq k \leq N/2$, the pair of sites $2k-1$ and $2k$ are connected by a strong bond, and the gauge fields $\overline{u}_{ij}$ satisfy $\overline{u}_{2k-1,2k} = +1$. The indexing of sites corresponds to a choice of basis $( c_1 , \ldots , c_N )$ \footnote{Once a choice has been made, it is crucial that we use the same basis throughout the mapping.}. We also index the strong bonds, naturally, by $1 \leq k \leq N/2$. On the other hand, the bond parity involves a product of gauge fields over all independent bonds $\langle ij \rangle$, each of which specifies an orientation $i \longrightarrow j$. We fix the orientation such that $\overline{u}_{ij} = +1$ for every independent bond $\langle ij \rangle$ \footnote{With such choices for the independent bond orientations, the bond parity in the vortex-free sector is $\mathcal{P}_b=+1$.}. \\

Let $\lbrace u_{ij} \rbrace$ ($\lbrace u_{ij}' \rbrace$) be the set of gauge fields obtained by flipping the string of bonds $l$ ($l'$) in $\lbrace \overline{u}_{ij} \rbrace$. Let $A$ ($A'$) be the $N \times N$ matrix in \eqref{kitaevmodelquadratic} constructed from $\lbrace u_{ij} \rbrace$ ($\lbrace u_{ij}' \rbrace$). Let $Q$ ($Q'$) be the canonical transformation of $A$ ($A'$), see \eqref{canonicaltransform}. As discussed in Sec.~\ref{section:method}, $\lbrace u_{ij} \rbrace$ ($\lbrace u_{ij}' \rbrace$) completely specifies the bond parity $\mathcal{P}_b$ ($\mathcal{P}_b'$) and, for a given set of couplings, the determinant $\det Q$ ($\det Q'$). We define the function $f_l: \lbrace 1 , \ldots , N/2 \rbrace \longrightarrow \lbrace -1 , 1 \rbrace$ such that $f_l (k) = - 1$ if the $k$th strong bond is contained in $l$ and $f_l (k) = 1$ otherwise. $f_{l'} (k)$ is similarly defined with $l'$. Let $\lambda$ be the type of the strong bonds. In the limit $0 < \epsilon \equiv \max_{\lambda' \neq \lambda} J_{\lambda'} / J_\lambda \ll 1$, we have
\begin{equation} \label{antisymmetryanisotropy}
\frac{A}{2 J_\lambda} = \bigoplus_{k=1}^{N/2} \begin{pmatrix} 0 & f_l (k) \\ - f_l (k) & 0 \end{pmatrix} + O (\epsilon) .
\end{equation}
The fermion gap $\Delta_\psi$ is finite and approaches $2 J_\lambda$ with increasing anisotropy. Since $\epsilon$ can be continuously decreased to $0$ without closing the fermion gap, $\det Q$ remains unchanged by Lemma 2. We can thus evaluate $\det Q$ at $\epsilon = 0$ in practice, which is equal to $\det Q$ for $0 < \epsilon \ll 1$, though in principle all the couplings should be finite. At $\epsilon=0$, $Q$ assumes the following block diagonal form \footnote{In the following expression of $Q$, $\vmathbb{1}$ can be more generally replaced by $\vmathbb{1} \cos \theta + i \sigma^y \sin \theta \in SO (2)$, and $\sigma^x$ by $\sigma^x \sin \vartheta + \sigma^z \cos \vartheta \in O (2) \setminus SO (2)$, for $\theta , \vartheta \in [0, 2 \pi)$. Such replacements do not alter $\det Q$.}
\begin{equation}
Q = \bigoplus_{k=1}^{N/2} \frac{[1 + f_l (k)] \vmathbb{1} + [1 - f_l (k)] \sigma^x}{2} .
\end{equation}
Its determinant evaluates to
\begin{equation} \label{determinantbasis}
\begin{aligned}[b]
\det Q &= \prod_{k=1}^{N/2} \det \frac{[1 + f_l (k)] \vmathbb{1} + [1 - f_l (k)] \sigma^x}{2} \\
&= (\det \vmathbb{1})^{N/2 - n_s (l)} (\det \sigma^x)^{n_s (l)} = (-1)^{n_s (l)} ,
\end{aligned}
\end{equation}
where $n_s (l)$ is the number of strong bonds contained in $l$. On the other hand, the bond parity evaluates to 
\begin{equation} \label{bondparitybasis}
\mathcal{P}_b = (-1)^{\lvert l \rvert} = (-1)^{n_w (l) + n_s (l)} .
\end{equation}
The two $(-1)^{n_s (l)}$ factors from \eqref{determinantbasis} and \eqref{bondparitybasis} thus cancel each other in the triple product $\mathcal{P}_b \det Q \, \mathcal{P}_\psi = ( -1 )^{n_w (l)} \mathcal{P}_\psi$. Similarly, we have $\mathcal{P}_b' \det Q' \, \mathcal{P}_\psi' = (-1)^{n_w (l')} P_\psi'$. By \eqref{tripodequal}, we are led to $(-1)^{n_w (l)} \mathcal{P}_\psi = (-1)^{n_w (l')} \mathcal{P}_\psi'$, or
\begin{equation} \label{weakbondparity}
\frac{\mathcal{P}_\psi}{\mathcal{P}_\psi'} = (-1)^{n_w (l) + n_w (l')} .
\end{equation}
If $n_w (l)$ and $n_w (l')$ have the same parity, then $n_w (l) + n_w (l')$ is even, and the relative fermion parity \eqref{weakbondparity} is $+1$, which indicates that the vortices at $p$ and $p'$ are of the same species. In contrast, if $n_w (l)$ and $n_w (l')$ have opposite parities, then $n_w (l) + n_w (l')$ is odd, and the relative fermion parity \eqref{weakbondparity} is $-1$, which indicates that the vortices at $p$ and $p'$ are of different species. \qed \\

\noindent \textbf{Corollary 1.} If two neighboring elementary plaquettes are separated by a weak (strong) bond, then the vortices hosted by them belong to different (the same) species. \\

\noindent \textbf{Proof.} Let us use the same notation as in Proposition 1. If $p$ and $p'$ are separated by a weak bond, then $l'$ can be constructed from $l$ by adding the weak bond to the latter. We have $[n_w (l) + n_w (l')] \, \mathrm{mod} \, 2 = [2 n_w (l) + 1] \, \mathrm{mod} \, 2 = 1$, so the vortices at $p$ and $p'$ are of different species by Proposition 1. In contrast, if $p$ and $p'$ are separated by a strong bond, then $l'$ can be constructed from $l$ by adding the strong bond to the latter. We have $[n_w (l) + n_w (l')] \, \mathrm{mod} \, 2 = [ 2 n_w (l) ] \, \mathrm{mod} \, 2 = 0$, so the vortices at $p$ and $p'$ are of the same species by Proposition 1. \qed \\

\textit{Remarks.} To simplify the presentation, we have constructed the proofs of Proposition 1 and, by extension, Corollary 1 assuming that the strong bonds are of the same type $\lambda$. This assumption can in fact be relaxed. Proposition 1 and Corollary 1 remain valid even if the strong bonds are of different types, as long as they form a dimer covering of the graph, in a generalized Kitaev model that allows for different coupling strengths on bonds of the same type. The extension is plain and thus omitted for brevity. \\

The proof of the following proposition requires a version of adiabatic theorem \cite{BF01343193,JPSJ.5.435,griffithstextbook} that treats degenerate states carrying distinct good quantum numbers, namely the gauge fields. A detailed derivation of the adiabatic theorem is presented in Sec.~\ref{section:adiabatic} of the Supplemental Material \cite{supply}. \\

\noindent \textbf{Proposition 2.} Let $p_0$, $p$, and $p'$ be three elementary plaquettes such that (i) $p_0$ is well-separated from $p$ and $p'$ and (ii) the two-vortex sectors $(p_0 , p)$ and $(p_0 , p')$ have opposite fermion parities. If the fermion gaps $\Delta_\psi (p_0 , p)$ and $\Delta_\psi (p_0 , p')$ remain finite throughout an adiabatic transformation in the toric code phase, then the anyon species of each of the vortices at $p_0$, $p$, and $p'$ is invariant. \\

\textit{Remarks.} For clarity, the notation $(p_0 , p)$ refers to the two-vortex sector with vortices at the elementary plaquettes $p_0$ and $p$. $\Delta_\psi (p_0 , p)$ is the fermion gap of the two-vortex sector $(p_0 , p)$. We remind the readers that the term ``well-separated'' means that the distance between the vortices is much larger than the correlation length, such that their braiding statistics is well-defined. Also, the term ``toric code phase'' generically refers to an extended parameter regime with a trivial Chern number, which may contain different assignments of anyon species, as in the Kitaev square-octagon and Kekul\'{e} models. \\

\noindent \textbf{Proof.} Let us use the same notation as in Proposition 1. Let $\mathcal{P}$ be the projector to the physical subspace, see \eqref{projectphysical}. Moreover, let $\mathscr{T}$ be an adiabatic transformation that does not close the fermion gap of the vortex-free sector (such that the system remains in the toric code phase) as well as the fermion gaps of the two-vortex sectors $(p_0 , p)$ and $(p_0 , p')$. Without loss of generality, we assume that the vortices at $p_0$ and $p$ are $e$ particles, while the vortex at $p'$ is an $m$ particle, initially. At any instant of $\mathscr{T}$, the eigenstates of $H (\lbrace u_{ij} \rbrace)$ or $H (\lbrace u_{ij}' \rbrace)$ \footnote{For brevity, we have suppressed the dependence of $H (\lbrace u_{ij} \rbrace)$ or $H (\lbrace u_{ij}' \rbrace)$ on the couplings $J_\lambda$.} are Fock states defined by the occupancies $n_k \in \lbrace 0 , 1 \rbrace$ of $N/2$ independent complex fermion modes $\psi_k$, which we also refer to as matter fermions to distinguish them from the composite object $e \times m$ where necessary \footnote{Additionally, the terms ``fermion gap'' and ``fermion parity'' specifically refer to the excitation gap and the parity of the matter fermions, not those of the composite object $e \times m$.}. \\

We first show that the physical fermion parities $\mathcal{P}_\psi$ and $\mathcal{P}_\psi'$ of the two-vortex sectors $(p_0 , p)$ and $(p_0 , p')$, respectively, are invariant under $\mathscr{T}$. The bond parity $\mathcal{P}_b$ is completely specified by the set of gauge fields $\lbrace u_{ij} \rbrace$. As $\Delta_\psi (p_0 , p) \neq 0$, the determinant $\det Q$ is invariant under $\mathscr{T}$ by Lemma 2. Since both $\mathcal{P}_b$ and $\det Q$ are left unchanged, the invariance of $\mathcal{P}_\psi$ under $\mathscr{T}$ follows from \eqref{physicalconstraint} and \eqref{projectproduct}. Similarly, $\mathcal{P}_\psi'$ is invariant under $\mathscr{T}$. Note that we have $\mathcal{P}_\psi / \mathcal{P}_\psi' = -1$ by the condition (ii). \\

Suppose that the physical subspace of the vortex-free sector is characterized by even fermion parity. Since the creation of two $e$ particles does not change the fermion parity, any physical wavefunction of the two-vortex sector $(p_0 , p)$ must have an even number of matter fermions. In particular, the lowest-energy state of $(p_0 , p)$ is given by $\mathcal{P} \lvert \phi \rangle$ (with a proper normalization factor), where $\lvert \phi \rangle$ is the instantaneous eigenstate of $H (\lbrace u_{ij} \rbrace)$ with $n_k = 0$ for all $k$, i.e., without any matter fermion. When acting on $\lvert \phi \rangle$, $\mathcal{P}$ symmetrizes over different gauge-field configurations that produce the same flux sector without changing the physical fermion parity $\mathcal{P}_\psi$, which is determined from the representative gauge-field configuration $\lbrace u_{ij} \rbrace$, and the occupancy $n_k$ of each fermion mode $\psi_k$. Suppose that the system is initially in $\mathcal{P} \lvert \phi \rangle$. By the adiabatic theorem \cite{supply}, the system remains in $\mathcal{P} \lvert \phi \rangle$, which is physical due to the invariance of $\mathcal{P}_\psi$, under $\mathscr{T}$. At any instant, there is no excitation (matter fermion or vortex) other than the two vortices at $p_0$ and $p$. Since each vortex does not absorb or emit any fermion (in the form of $\psi_k$ or $e \times m$) throughout $\mathscr{T}$, its anyon species cannot change. This, together with the invariance of $\mathcal{P}_\psi / \mathcal{P}_\psi' = -1$ under $\mathscr{T}$, further implies that the anyon species of the vortex at $p'$ cannot change. \\

Suppose that the physical subspace of the vortex-free sector is characterized by odd fermion parity. Since the creation of an $e$ particle and an $m$ particle changes the fermion parity, any physical wavefunction of the two-vortex sector $(p_0 , p')$ must have an even number of matter fermions. In particular, the lowest-energy state of $(p_0 , p')$ is given by $\mathcal{P} \lvert \phi' \rangle$, where $\lvert \phi' \rangle$ is the instantaneous eigenstate of $H (\lbrace u_{ij}' \rbrace)$ with $n_k = 0$ for all $k$, i.e., without any matter fermion. By the same line of reasoning as in the previous paragraph, we deduce that the anyon species of each of the vortices at $p_0$, $p'$, and $p$ is invariant under $\mathscr{T}$. \qed \\

The following corollary is an immediate consequence of Proposition 2. Its proof is omitted for brevity. \\

\noindent \textbf{Corollary 2.} If the fermion gap of every two-vortex sector remains finite throughout an adiabatic transformation in the toric code phase, then the assignment of anyon species is invariant. \\

Proposition 1 or Corollary 1 enables us to efficiently assign to each elementary plaquette one of the two anyon species in the dimer limit. Corollary 2 guarantees that this assignment is invariant as we continuously move toward a less anisotropic point in the parameter space, as long as we do not encounter a fermion-gap-closing transition in both the vortex-free and two-vortex sectors. If the fermion gaps of some, but not all, of the two-vortex sectors vanish, then we can track the changes in the assignment of anyon species using Proposition 2. For tri-coordinated graphs, Proposition 1 or Corollary 1 gives precisely the same rule of mapping anyons in the dimer limit as Ref.~\cite{Wootton_2015}, though we have used a quite different formalism to obtain it. Proposition 2 and Corollary 2, on the other hand, are novel results of this work.

\section{\label{section:apply}Application}

This section is devoted to the application of our theory to various models introduced in Sec.~\ref{section:model}. First, we use Corollary 1 to map vortices to $e$ and $m$ particles in selected dimer limits. Next, we compute and analyze the fermion gaps of distinct two-vortex sectors for the Kitaev square-octagon and Kekul\'{e} models. We demonstrate that two dimer limits with different assignments of anyon species are not necessarily separated by a fermion-gap closing in the vortex-free sector. Instead, they are separated by fermion-gap closings in some of the two-vortex sectors as per Corollary 2. Using Proposition 2, we also provide a simple picture for the implementation of an $e \longleftrightarrow m$ automorphism in the Kitaev Kekul\'{e} model upon varying the model parameters along certain closed paths. Finally, we show that the presence or absence of criticality in the vortex-free sector, which defines the ground state of the model in question, can be understood in the framework of symmetry-enriched topological (SET) orders.

\subsection{\label{section:strong}Dimer Limits}

\begin{figure*}
\subfloat[]{\label{figure:honeycombanyonz}
\includegraphics[scale=0.2]{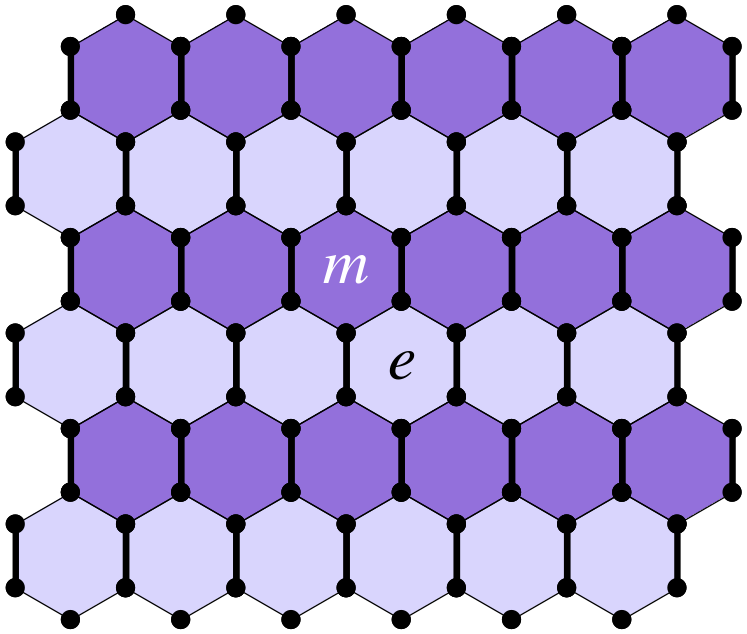}} \hspace{16pt}
\subfloat[]{\label{figure:octagonanyonz}
\includegraphics[scale=0.2]{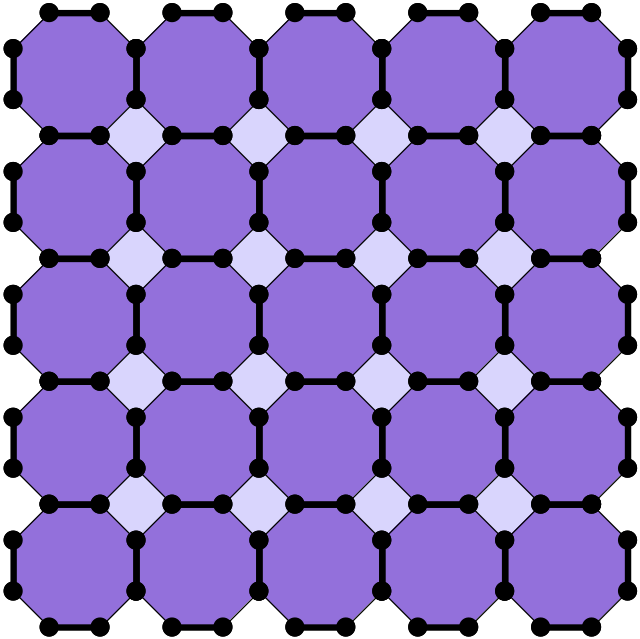}} \hspace{16pt}
\subfloat[]{\label{figure:octagonanyonx}
\includegraphics[scale=0.2]{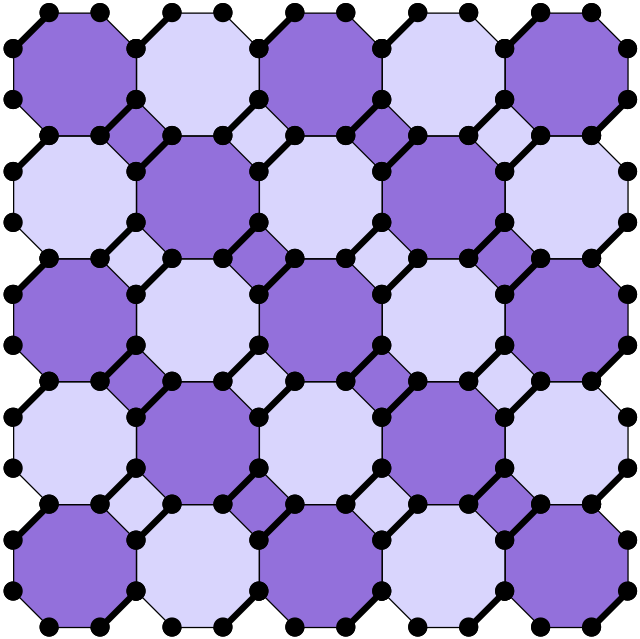}} \hspace{16pt}
\subfloat[]{\label{figure:octagonanyony}
\includegraphics[scale=0.2]{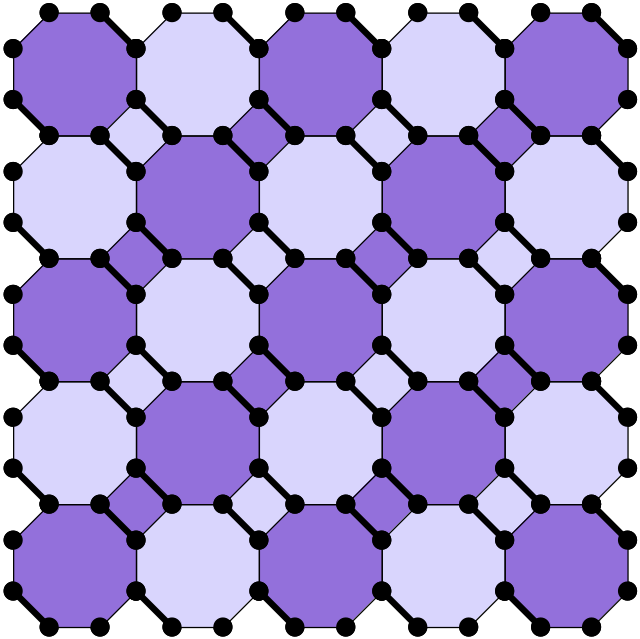}} \\
\subfloat[]{\label{figure:staranyonz}
\includegraphics[scale=0.2]{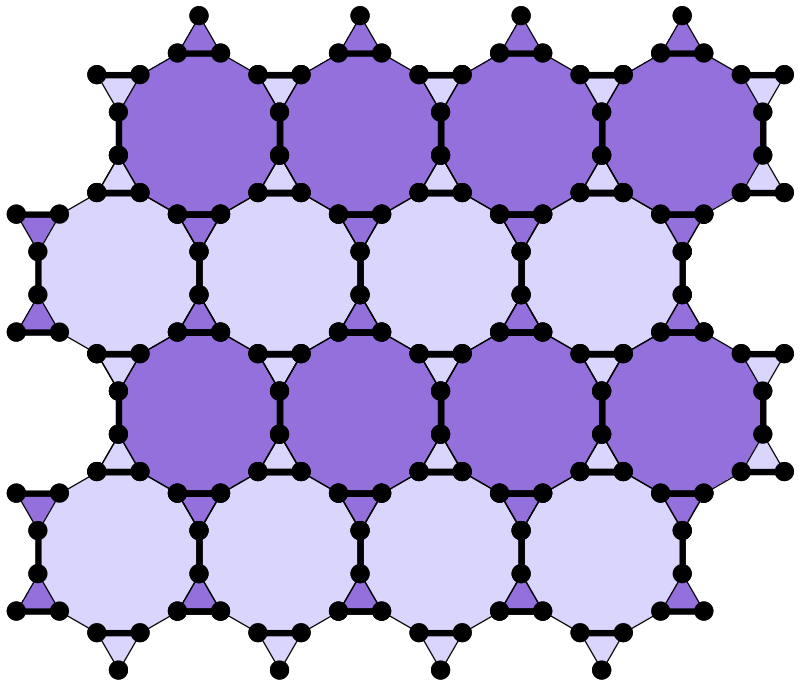}} \hspace{8pt}
\subfloat[]{\label{figure:kekuleanyonz}
\includegraphics[scale=0.2]{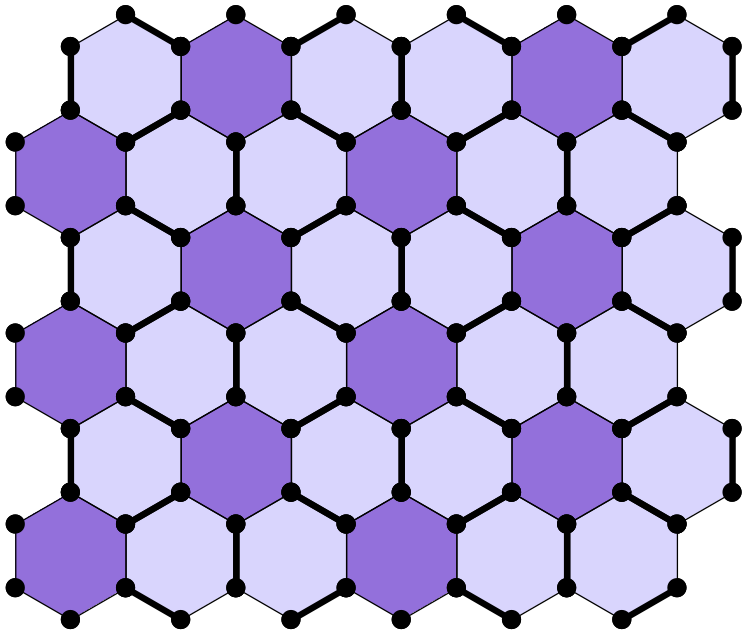}} \hspace{8pt}
\subfloat[]{\label{figure:amorphousanyonz}
\includegraphics[scale=0.2]{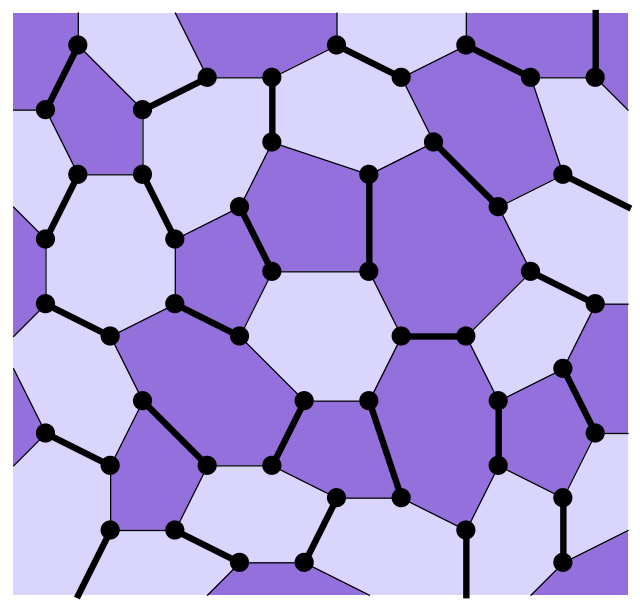}} \hspace{8pt}
\subfloat[]{\label{figure:shastryanyonzz}
\includegraphics[scale=0.2]{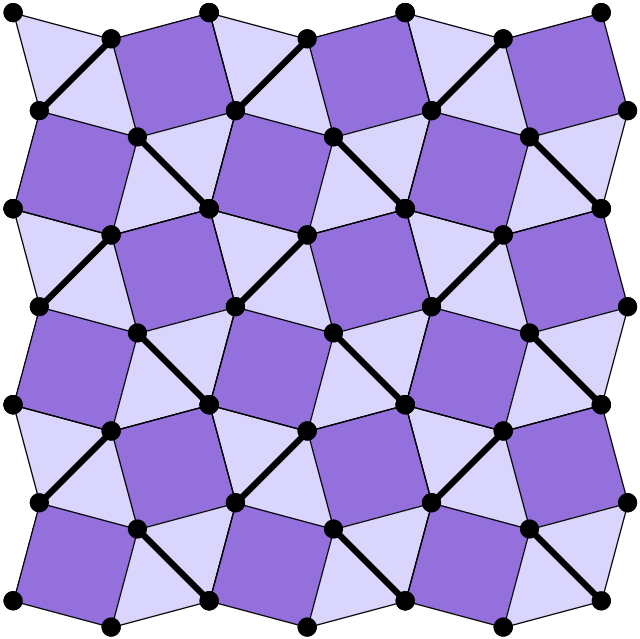}}
\caption{The assignments of anyon species for (a) the Kitaev honeycomb model in the strong $J_z$ limit, (b,c,d) the Kitaev square-octagon model in the strong $J_z$, $J_x$ and, $J_y$ limits, (e) the Kitaev star model in the strong $J_z$ limit, (f) the Kitaev Kekul\'{e} model in the strong $J_z$ limit, (g) the amorphous Kitaev model in the strong $J_z$ limit, and (h) the Kitaev Shastry-Sutherland model where the blue bonds in Fig.~\ref{figure:shastrymodel} are dominant. Thick lines indicate strong bonds of type $\lambda$ such that the coupling $J_\lambda \gg J_{\lambda'}$ for every $\lambda' \neq \lambda$. Elementary plaquettes filled with light and dark purple colors indicate that the respective vortices belong to distinct anyon species. Taking (a) as example, if the vortices on the light-colored hexagons are identified as $e$ particles, then the vortices on the dark-colored hexagons must be $m$ particles (and vice versa).}
\end{figure*}

Armed with Corollary 1, we map out the configurations of anyon species for the Kitaev honeycomb model (Fig.~\ref{figure:honeycombmodel}) in the strong $J_z$ limit, the Kitaev square-octagon model (Fig.~\ref{figure:octagonmodel}) in the strong $J_x$, $J_y$, and $J_z$ limits, the Kitaev star model (Fig.~\ref{figure:starmodel}) in the strong $J_z$ limit, the Kitaev Kekul\'{e} model (Fig.~\ref{figure:kekulemodel}) in the strong $J_z$ limit, the amorphous Kitaev model (Fig.~\ref{figure:amorphousmodel}) in the strong $J_z$ limit, and the Kitaev Shastry-Sutherland model (Fig.~\ref{figure:shastrymodel}) where the class of bonds between triangles dominates. The results are shown in Figs.~\ref{figure:honeycombanyonz}-\ref{figure:shastryanyonzz}. \\

Several comments are in order. First, the assignments of anyon species in the strong $J_z$ limits of the Kitaev honeycomb and square-octagon models indeed agree with the existing results derived from degenerate perturbation theory \cite{KITAEV20062,PhysRevB.76.180404}. Second, Ref.~\cite{Kells_2011} uses degenerate perturbation theory to investigate the $( J_x^2 + J_y^2 )^{1/2} \gg J_z$ limit of the Kitaev square-octagon model, where each unit square, instead of a dimer, is identified with a low-energy doublet. We compare their approach and result to ours in Sec.~\ref{section:square} of the Supplemental Material \cite{supply}. Third, it is unclear how degenerate perturbation theory can determine the anyon species of the unit triangles on the star and Shastry-Sutherland lattices, while its application is simply infeasible for amorphous solids, see related discussions in Sec.~\ref{section:introduce} and Sec.~\ref{section:perturb} of the Supplemental Material \cite{supply}. These limitations are easily overcome by Corollary 1. Finally, in Sec.~\ref{section:statistics} of the Supplemental Material \cite{supply}, we demonstrate the correct braiding statistics of the $e$ and $m$ particles assigned according to Corollary 1 in any dimer limit \cite{PACHOS20071254}.

\subsection{\label{section:setup}Computational Setup}

In the next two subsections and Sec.~\ref{section:addresult} of the Supplemental Material \cite{supply}, we will compute the fermion gaps of the vortex-free and two-vortex sectors for the Kitaev square-octagon, Kekul\'{e}, honeycomb, and star models. Here, we describe the common setup for our computations and introduce a few notations, to set the stage for later discussions. \\

For each model, the vortex-free sector admits a translationally invariant gauge choice, which allows one to study the thermodynamic limit via a Fourier transform \cite{supply}. Unless stated otherwise, the fermion gap of the vortex-free sector is computed in the thermodynamic limit. On the other hand, the fermion gaps of the two-vortex sectors are calculated on a finite system with periodic boundary conditions, i.e., a torus. We will use the Kitaev square-octagon model as an illustration, where the setup is schematically shown in Fig.~\ref{figure:twovortex} and elaborated in the following. Each unit cell contains two elementary plaquettes, namely a unit square and a unit octagon. Our system consists of $L_x$ ($L_y$) unit cells in the $x$ ($y$) directions. We create two vortices at the elementary plaquettes $p_0$ and $p_1 \equiv p$, which are separated from each other by $L_x/2 + L_x'/2$ unit cells in the $x$ direction, by flipping a string of bonds between them. We fix the vortex at $p_0$ while moving the vortex initially at $p_1$ over a region $\mathcal{R}$ of $L_x' \times L_y'$ unit cells, which does not wrap around either cycle of the torus, such that $L_x' \times L_y' \times 2$ different two-vortex sectors $(p_0 , p_i)$ are generated. Let $p_0$ be an elementary plaquette of type $\alpha \in \lbrace s , o \rbrace$, where $s$ ($o$) stands for square (octagon). We further distinguish these sectors according to the type $\beta \in \lbrace s , o \rbrace$ of the elementary plaquette $p_i$. We define the averaged fermion gap
\begin{equation}
\Delta_\psi^{\alpha \beta} = \frac{1}{L_x' L_y'} \sum_{p_i \in \beta} \Delta_\psi (p_0 , p_i) \end{equation}
and take the standard deviation
\begin{equation}
\delta^{\alpha \beta} = \sqrt{\frac{1}{L_x' L_y'} \sum_{p_i \in \beta} \left[ \Delta_\psi (p_0 , p_i) - \Delta_\psi^{\alpha \beta} \right]^2}
\end{equation}
as the error. Note that the summation runs only over the elementary plaquettes of type $\beta$ within $\mathcal{R}$ in each expression above. We will also refer to $\mathcal{R}$ as the region of mapping. For other models, we have $L_1 \times L_2$ and $L_1' \times L_2'$ for the sizes of the system and the region of mapping instead of $L_x \times L_y$ and $L_x' \times L_y'$, and the vortices are separated along the $\mathbf{a}_1$ direction instead of the $x$ direction. The Kitaev honeycomb model has only one elementary plaquette, while each of the Kitaev Kekul\'{e} and star models has three elementary plaquettes, per unit cell. \\

For each model, we plot various fermion gaps over the compactified parameter space $J_x + J_y + J_z = 1$, which is a triangle such that a vertex (an edge) corresponds to the $J_\lambda \longrightarrow 1$ ($J_\lambda \longrightarrow 0$) limit for some $\lambda$. We also provide a detailed convergence analysis of $\Delta_\psi^{\alpha \beta}$ with respect to the sizes of the system and the region of mapping, along selected paths in the triangular parameter space. The two endpoints of each path are chosen from $U$, $L$, and $R$, which correspond to the sets of couplings $\mathbf{J}^{(U)} = (0.1 , 0.1 , 0.8)$, $\mathbf{J}^{(L)} = (0.8 , 0.1 , 0.1)$, and $\mathbf{J}^{(R)} = (0.1 , 0.8 , 0.1)$, respectively. The path $LR$, for example, is parametrized as $\mathbf{J} (t) = (1-t) \mathbf{J}^{(L)} + t \mathbf{J}^{(R)}$ by a single variable $t \in [0 , 1]$. We typically choose $65$ equally spaced $t$ points for the computations of $\Delta_\psi^{\alpha \beta}$. Moreover, we compute the fermion parities of the two-vortex sectors $(p_0 , p_{i>1})$ relative to that of $(p_0 , p_1)$ at each $t$ point and map out the anyon species via the method outlined in Sec.~\ref{section:method}, as a numerical verification of our theory developed in Sec.~\ref{section:theory}. \\

We use $0\mathrm{v}$ and $2\mathrm{v}$ as shorthands of ``vortex-free'' and ``two-vortex'', respectively. $\Delta_\psi$ refers to a fermion gap in general without specifying the flux sector. $\Delta_\psi^{(0\mathrm{v})}$ specifically refers to the fermion gap of the vortex-free sector. $\Delta_\psi^{(2\mathrm{v})}$ refers to the fermion gap of a two-vortex sector, or the averaged fermion gap of multiple two-vortex sectors, in general without specifying the types of two elementary plaquettes that host the vortices. As defined in the last paragraph, $\Delta_\psi^{\alpha \beta}$ represents the averaged fermion gap of two-vortex sectors with one vortex at an elementary plaquette of type $\alpha$ and the other at an elementary plaquette of type $\beta$. When $\alpha \neq \beta$, $\Delta_\psi^{\alpha \beta}$ and $\Delta_\psi^{\beta \alpha}$ are the same, so we present only the former data. We have introduced the notation $\Delta_\psi (p_0 , p)$ for the fermion gap of the two vortex sector $(p_0 , p)$ with vortices at the elementary plaquettes $p_0$ and $p$. When the set of couplings is parametrized by $t \in \mathbb{R}$, we will use notations such as $\Delta_\psi^{\alpha \beta} (t)$ to indicate the dependence of the fermion gaps on $t$. Whether the argument in the bracket refers to a pair of elementary plaquettes or a real parameter should be clear from the context. Finally, by abuse of language, we will often refer to the anyon species of an elementary plaquette, which actually means the anyon species of the vortex at the elementary plaquette.

\subsection{\label{section:octagon}Kitaev Square-Octagon Model}

\begin{figure*}
\includegraphics[scale=0.22]{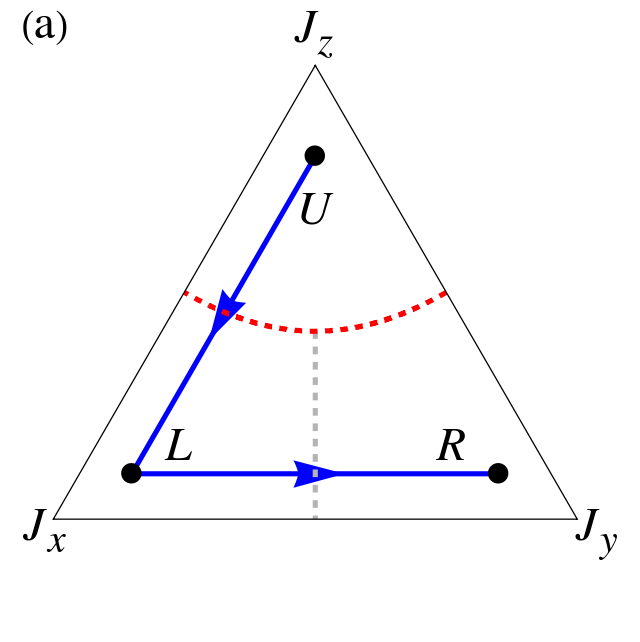} \hspace{20pt}
\includegraphics[scale=0.22]{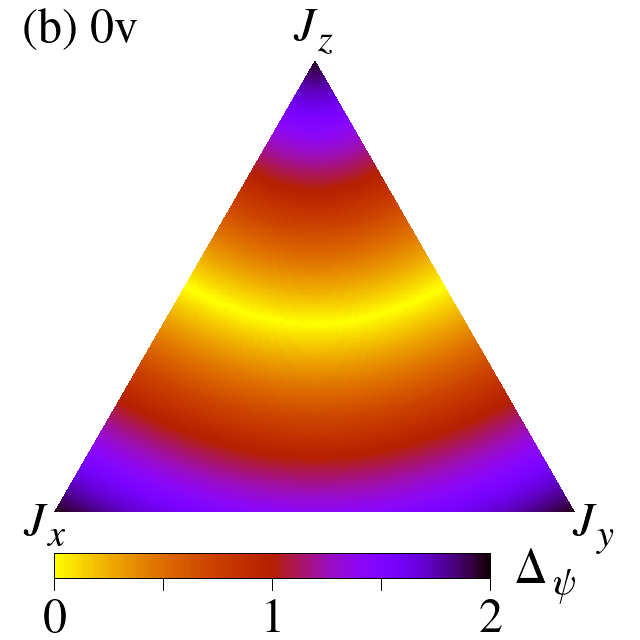} \hspace{20pt}
\includegraphics[scale=0.22]{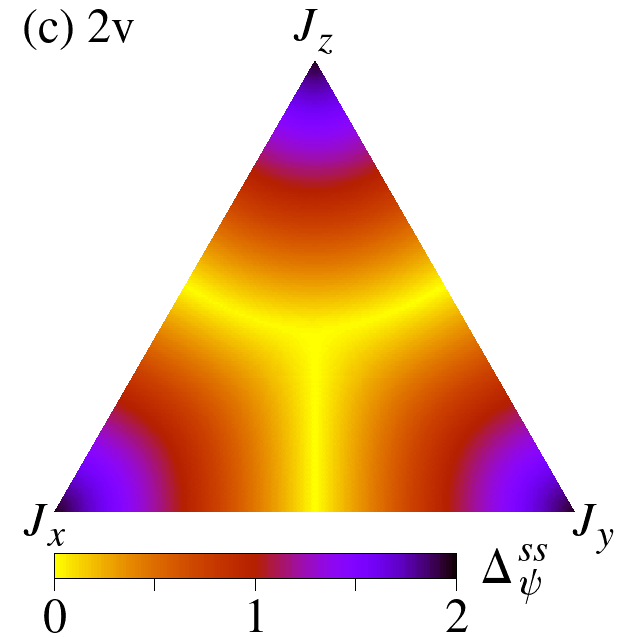} \hspace{20pt}
\includegraphics[scale=0.22]{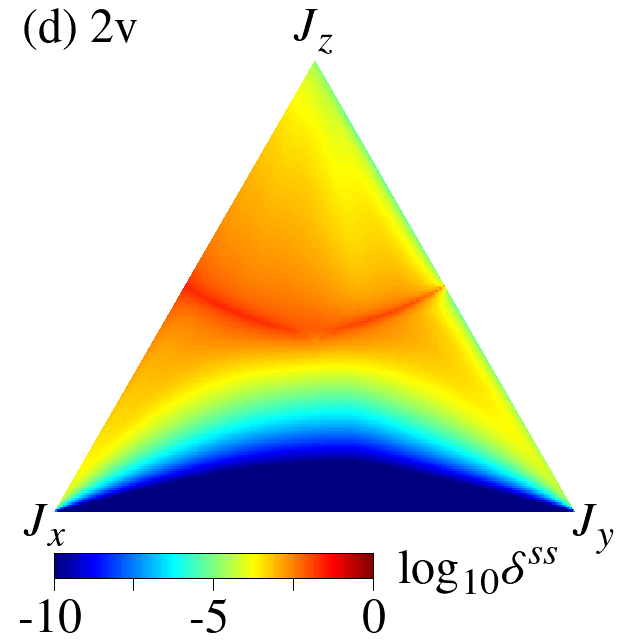} \\ \vspace{5pt}
\includegraphics[scale=0.22]{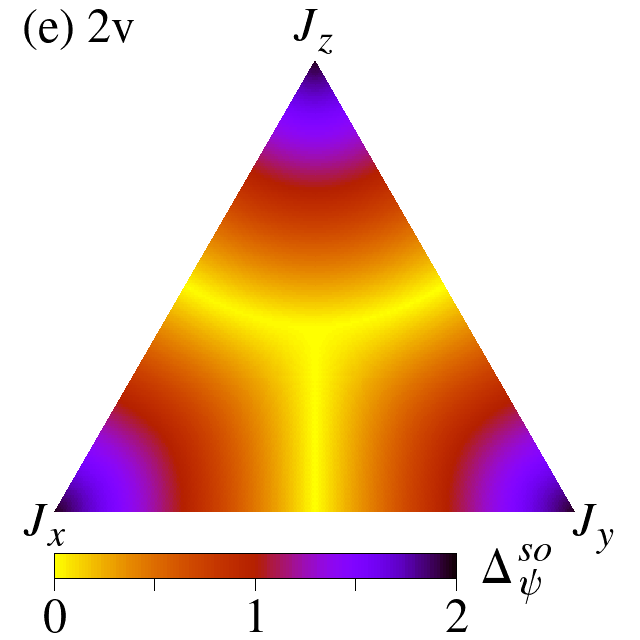} \hspace{20pt}
\includegraphics[scale=0.22]{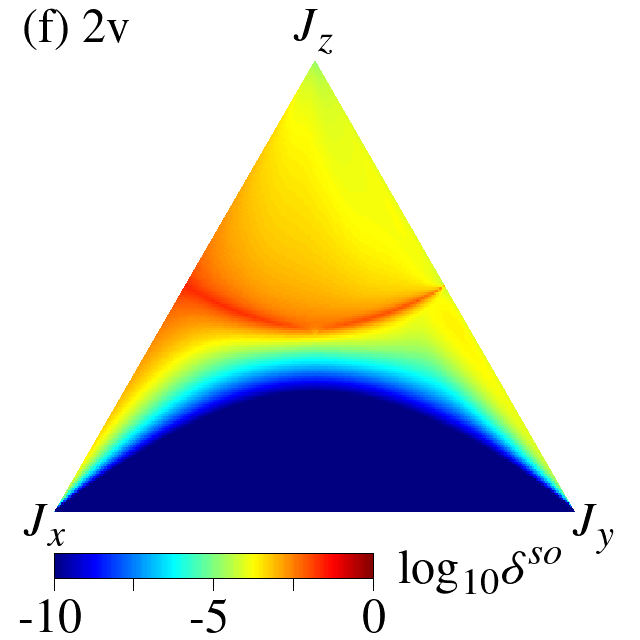} \hspace{20pt}
\includegraphics[scale=0.22]{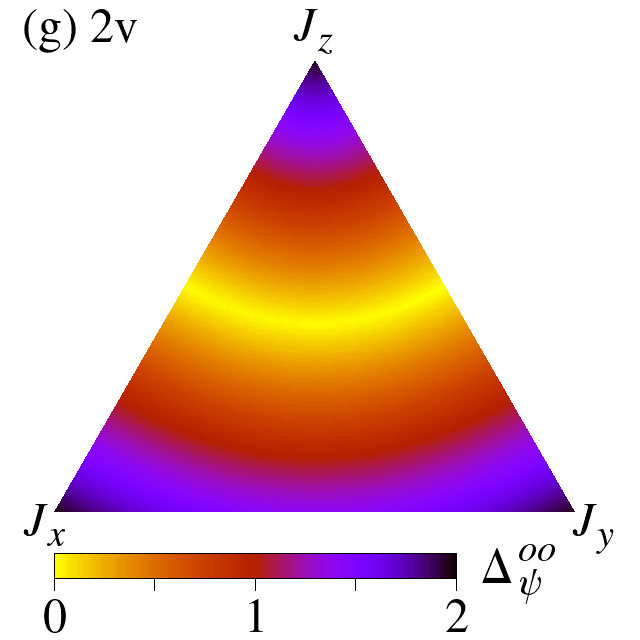} \hspace{20pt}
\includegraphics[scale=0.22]{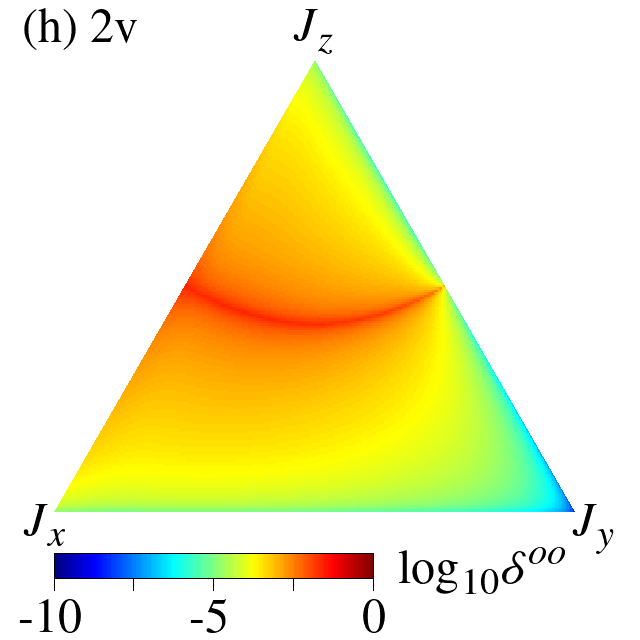} \\ \vspace{5pt}
\includegraphics[scale=0.25]{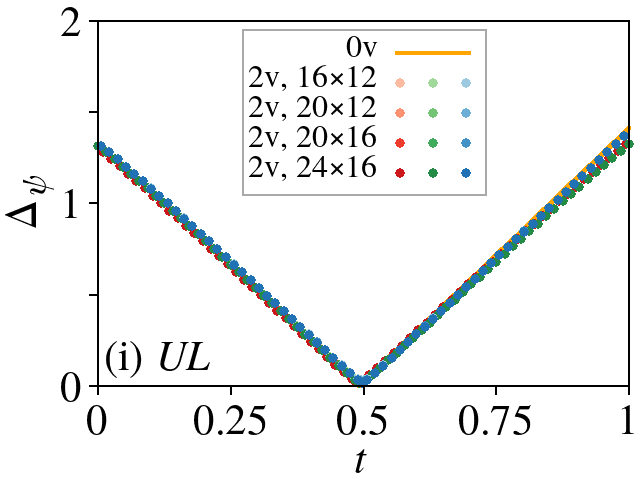} \;
\includegraphics[scale=0.25]{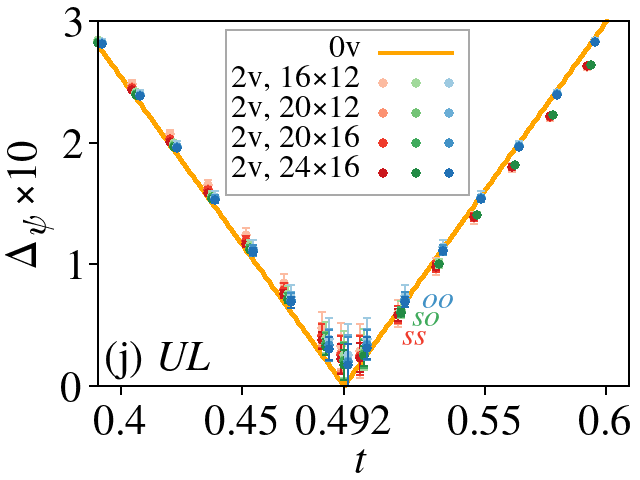} \;
\includegraphics[scale=0.25]{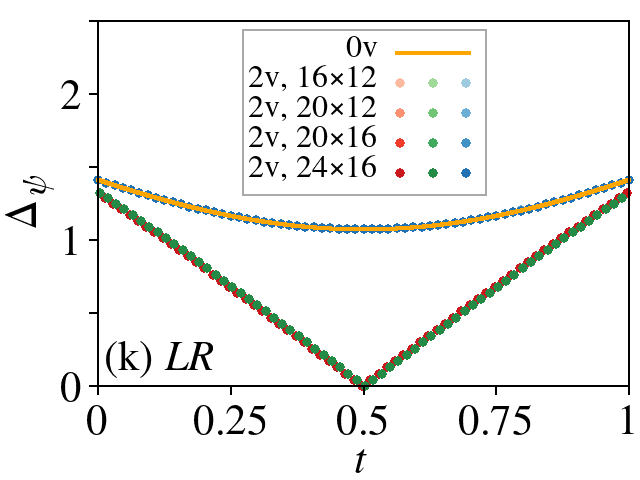} \;
\includegraphics[scale=0.25]{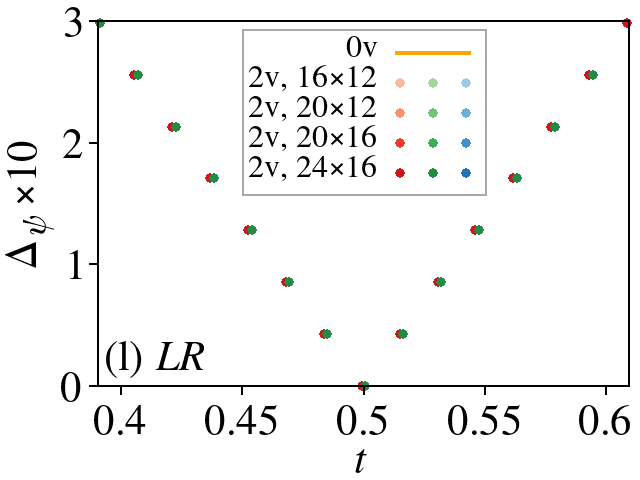}
\caption{\label{figure:octagondata}Various results from the study of the Kitaev square-octagon model. The triangular parameter space in (a)-(h) is defined by $J_x + J_y + J_z = 1$, where the vertex labeled by $J_\lambda$ corresponds to the $J_\lambda \longrightarrow 1$ limit. $0\mathrm{v}$ and $2\mathrm{v}$ stand for ``vortex-free'' and ``two-vortex'', respectively. (a) The red dashed curve indicates the closing of the fermion gap in the vortex-free sector and all two-vortex sectors. The gray dashed line indicates the closing of the fermion gap in some two-vortex sectors, while the vortex-free sector has a finite fermion gap. The blue solid lines with arrows represent selected paths along which we perform finite-size scaling analyses of the fermion gaps and determine the assignments of anyon species numerically. (b) The fermion gap $\Delta_\psi^{(0\mathrm{v})}$ of the vortex-free sector. (c,e,g) The averaged fermion gaps $\Delta_\psi^{ss}$, $\Delta_\psi^{so}$, and $\Delta_\psi^{oo}$ of two-vortex sectors with both vortices at unit squares, one vortex at a unit square and the other at a unit octagon, and both vortices at unit octagons, respectively. The corresponding standard deviations $\delta^{ss}$, $\delta^{so}$, and $\delta^{oo}$ are shown in (d,f,h) on a logarithmic scale. (i,k) The averaged fermion gaps $\Delta_\psi^{ss}$ (red data), $\Delta_\psi^{so}$ (green data), and $\Delta_\psi^{oo}$ (blue data) of two vortex sectors along the paths $UL$ and $LR$, each of which is linearly parametrized by $t \in [0,1]$. The error bars represent the standard deviations. The fermion gap $\Delta_\psi^{(0\mathrm{v})}$ of the vortex-free sector (orange data), which lies underneath other data in (i), is also plotted for comparison. (j,l) Zoom-ins of (i,k) near zero energy. In (i)-(l), the product of two integers associated with each set of 2v data indicates the size of the system ($L_x \times L_y$ unit cells) on which the computations are performed. The red/blue series in (i,j) and the red/green series in (k,l) are shifted slightly leftward/rightward for visibility.}
\end{figure*}

In the Kitaev square-octagon model (Fig.~\ref{figure:octagonmodel}), the assignments of anyon species are different in the strong $x$, $y$, and $z$ bond limits, see Figs.~\ref{figure:octagonanyonz}-\ref{figure:octagonanyony}. An earlier work \cite{PhysRevB.76.180404} found a fermion-gap-closing transition in the vortex-free sector, which separates the strong $z$ bond limit from the strong $x$ and $y$ bond limits, but not the latter two from each other, see Fig.~\ref{figure:octagondata}b. The condition for $\Delta_\psi^{(0\mathrm{v})}=0$ is $J_x^2 + J_y^2 = J_z^2$. The Chern number of each gapped parameter regime evaluates to $0$, which is expected as it is smoothly connected to a dimer limit. By Corollary 2, the strong $x$ and $y$ bond limits must be separated by fermion-gap-closing transitions in some two-vortex sectors. \\

With $L_x \times L_y = 16 \times 16$ and $L_x' \times L_y' = 4 \times 12$, we compute $\Delta_\psi^{\alpha \beta}$ and $\delta^{\alpha \beta}$ for $\alpha \beta = ss , so , oo$ and plot them over the triangular parameter space, as shown in Figs.~\ref{figure:octagondata}c-\ref{figure:octagondata}h. We find that all these $\Delta^{\alpha \beta}_\psi$ approach zero on the curve $J_x^2 + J_y^2 = J_z^2$. In addition, the fermion gap of a two-vortex sector vanishes on the line $J_x = J_y \geq J_z / \sqrt{2}$, when at least one of the vortices is hosted by a unit square, see Figs.~\ref{figure:octagondata}c and \ref{figure:octagondata}e. The combined set of critical parameters at which $\Delta_\psi^{(0\mathrm{v})}=0$ or $\Delta_\psi^{\alpha \beta}=0$ for some $\alpha \beta$ thus divides the triangular parameter space into an upper, lower-left, and lower-right parts, which contain the strong $J_z$, $J_x$, and $J_y$ limits, respectively. This is consistent, by Corollary 2, with the three different assignments of anyon species, as we noted in the beginning of this subsection. We also find that the error is largest in the vicinity of $J_x^2 + J_y^2 = J_z^2$, where it is of the order of $10^{-2}$. \\

The strong $J_x$ and $J_y$ limits can be continuously connected by a path (e.g., $LR$ as indicated in Fig.~\ref{figure:octagondata}a) along which $\Delta_\psi^{oo} \neq 0$ and $\delta^{oo} \ll \Delta_\psi^{oo}$ is essentially zero, indicating that any two-vortex sector involving a pair of well-separated unit octagons has a finite fermion gap. In either limit, half of the unit octagons are labeled by $e$ and the other half are labeled by $m$. For each unit octagon $p$, one can always find unit octagons $p'$ and $p_0$ such that the trio $p_0$, $p$ and $p'$ satisfy the conditions (i) and (ii) of Proposition 2. As the model parameters are varied along the path, $\Delta_\psi (p_0 , p)$ and $\Delta_\psi (p_0 , p')$ remain finite, so the anyon species of each of the trio cannot change, as per Proposition 2. In other words, the anyon species of every unit octagon is invariant. On the other hand, the anyon species of each unit square is allowed to change. Indeed, it must change upon crossing the line $J_x = J_y \geq J_z / \sqrt{2}$, so that the assignment of anyon species is in accordance with the appropriate dimer limit. \\

We further provide a detailed convergence analysis of $\Delta_\psi^{\alpha \beta}$ with respect to the sizes of the system and the region of mapping, along the paths $UL$ and $LR$ as indicated in Fig.~\ref{figure:octagondata}a. We also calculate the (relative) fermion parities of the two-vortex sectors to determine the anyon species of each elementary plaquette within the region of mapping. Our computations are performed with (i) $L_x \times L_y = 16 \times 12$, $L_x' \times L_y' = 4 \times 8$, (ii) $L_x \times L_y = 20 \times 12$, $L_x' \times L_y' = 6 \times 8$, (iii) $L_x \times L_y = 20 \times 16$, $L_x' \times L_y' = 6 \times 12$, and (iv) $L_x \times L_y = 24 \times 16$, $L_x' \times L_y' = 8 \times 12$. \\

Fig.~\ref{figure:octagondata}i shows $\Delta_\psi^{\alpha \beta} (t)$ for $\alpha \beta = ss , so , oo$ along the path $UL$, which is linearly parametrized by $t \in [0,1]$ such that $t=0$ ($1$) corresponds to $U$ ($L$). $\Delta_\psi^{(0\mathrm{v})} (t)$ is also plotted for comparison. Up to three significant digits, $\mathbf{J} (t)$ satisfies $J_x^2 + J_y^2 = J_z^2$ at $t=0.492$, which we add to the list of discrete $t$ points discussed earlier in Sec.~\ref{section:setup}. Starting from $t=0$ with a value of $\sim 1$, $\Delta_\psi^{\alpha \beta} (t)$ decreases monotonically as $t$ increases toward $0.492$, at which $\Delta_\psi^{(0\mathrm{v})} (t)$ goes to zero. For each $L_x \times L_y$ and $\alpha \beta$, we find that $\Delta_\psi^{\alpha \beta} (t)$ is smallest at or near $t=0.492$, where it is of the order of $10^{-2}$, which suggests that the fermion gap of every two-vortex sector also vanishes at $t=0.492$. As $t$ further increases from $0.492$ toward $1$, $\Delta_\psi^{\alpha \beta} (t)$ increases monotonically toward $\sim 1$. The discrepancies between the data of different system sizes, as well as the error bars, are largest in the vicinity of the critical point $t=0.492$, see Fig.~\ref{figure:octagondata}j. Far from the critical point, the data of different system sizes overlap rather perfectly, and the error bars are small. Overall, $\Delta_\psi^{\alpha \beta} (t)$ behaves similarly as $\Delta^{(0\mathrm{v})} (t)$. We also confirm that, for $t \in [0, 0.492)$, the assignment of anyon species is constant and given by Fig.~\ref{figure:octagonanyonz}, which is consistent with Corollaries 1 and 2. On the other hand, for $t \in (0.492, 1]$, the assignment of anyon species is constant and given by Fig.~\ref{figure:octagonanyonx}, which is again consistent with Corollaries 1 and 2. \\

Fig.~\ref{figure:octagondata}k shows $\Delta_\psi^{\alpha \beta} (t)$ for $\alpha \beta = ss , so , oo$ along the path $LR$, which is linearly parametrized by $t \in [0,1]$ such that $t=0$ ($1$) corresponds to $L$ ($R$). $\Delta_\psi^{(0\mathrm{v})} (t)$ is also plotted for comparison. Starting from $t=0$ with a value of $\sim 1$, $\Delta_\psi^{ss} (t)$ and $\Delta_\psi^{so} (t)$ decrease monotonically as $t$ increases toward $1/2$, at which they become zero. Numerically, we find that $\Delta_\psi^{ss} (1/2)$ and $\Delta_\psi^{so} (1/2)$ are of the order of $10^{-12}$ or lower. As $t$ further increases from $1/2$ toward $1$, $\Delta_\psi^{ss} (t)$ and $\Delta_\psi^{so} (t)$ increase monotonically toward $\sim 1$. On the other hand, $\Delta_\psi^{oo} (t)$ behaves similarly as $\Delta^{(0\mathrm{v})} (t)$ and remains finite at $\sim 1$ throughout $LR$. For each $\alpha \beta$, the data $\Delta_\psi^{\alpha \beta}$ of various system sizes overlap rather perfectly throughout $LR$. Besides, $\delta^{\alpha \beta}$ is extremely small such that the error bars are not visible even for $t$ near the critical point, see Fig.~\ref{figure:octagondata}l. We also confirm that, for $t \in [0, 1/2)$, the assignment of anyon species is constant and given by Fig.~\ref{figure:octagonanyonx}, which is consistent with Corollaries 1 and 2. On the other hand, for $t \in (1/2, 1]$, the assignment of anyon species is constant and given by Fig.~\ref{figure:octagonanyony}, which is again consistent with Corollaries 1 and 2.

\subsection{\label{section:kekule}Kitaev Kekul\'{e} model}

\begin{figure}
\includegraphics[scale=0.22]{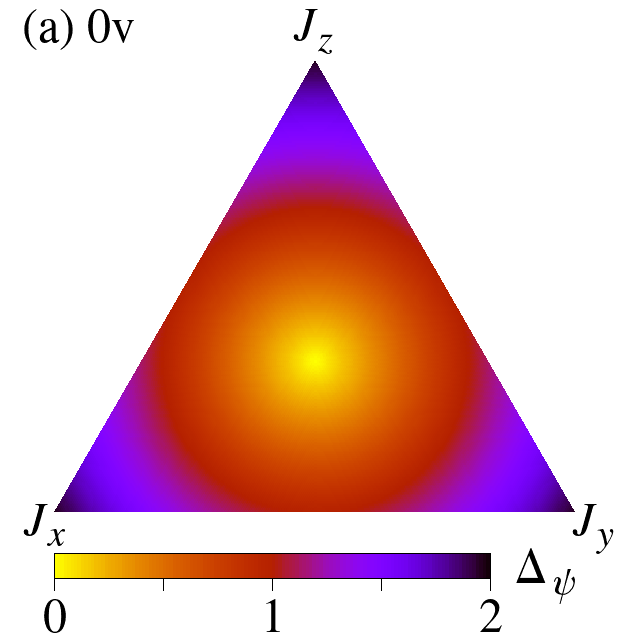} \hspace{20pt}
\includegraphics[scale=0.22]{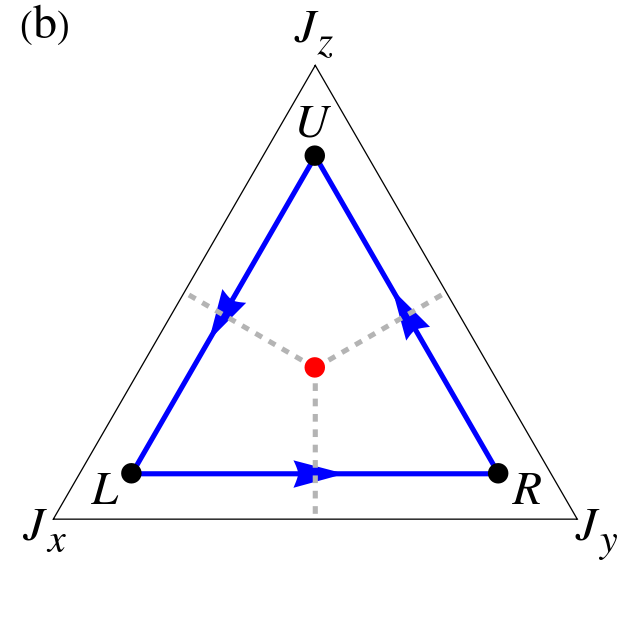}
\caption{\label{figure:kekulegap0v}The fermion gap $\Delta_\psi^{(0\mathrm{v})}$ of the vortex-free sector on the triangular parameter space $J_x + J_y + J_z = 1$, where the vertex labeled by $J_\lambda$ corresponds to the $J_\lambda \longrightarrow 1$ limit, for the Kitaev Kekul\'{e} model. (b) The red dot indicates the closing of the fermion gap in the vortex-free sector and all two-vortex sectors. The gray dashed lines indicate the closing of the fermion gap in some two-vortex sectors, while the vortex-free sector has a finite fermion gap. The blue solid lines with arrows represent selected paths along which we perform finite-size scaling analyses of the fermion gaps and determine the assignments of anyon species numerically.}
\end{figure}

In the Kitaev Kekul\'{e} model (Fig.~\ref{figure:kekulemodel}), the edges of each unit hexagon are formed by only two types of bonds. We thus classify a unit hexagon as $x$, $y$, or $z$ according to the type of bonds that is absent in its edges. For instance, a $z$ hexagon is bounded by alternating $x$ and $y$ bonds, and connected to other $z$ hexagons via $z$ bonds. The assignment of anyon species in the strong $J_\lambda$ limit is such that the $\lambda$ hexagons correspond to one anyon species (say $m$), while the remaining hexagons correspond to the other anyon species (say $e$), see Fig.~\ref{figure:kekuleanyonz}. The assignments of anyon species in the strong $J_x$, $J_y$, and $J_z$ limits are thus different from each other. An earlier work \cite{Kamfor_2010} found that the fermion gap of the vortex-free sector vanishes only at an isolated point, which is the isotropic point $J_x = J_y = J_z$, see Fig.~\ref{figure:kekulegap0v}a. The Chern number evaluates to $0$ in the rest of the triangular parameter space, which is expected as each point therein can be smoothly connected to a dimer limit. By Corollary 2, the three strong bond limits must be separated from each other by fermion-gap-closing transitions in some two-vortex sectors. \\

\begin{figure*}
\includegraphics[scale=0.22]{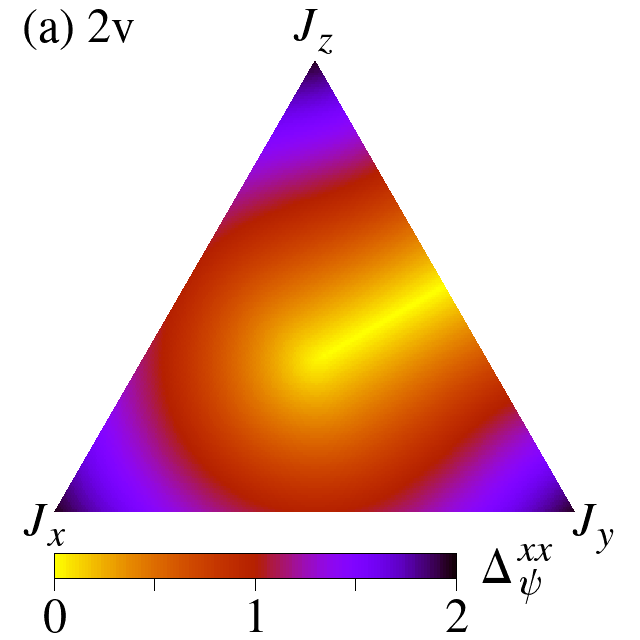} \hspace{20pt}
\includegraphics[scale=0.22]{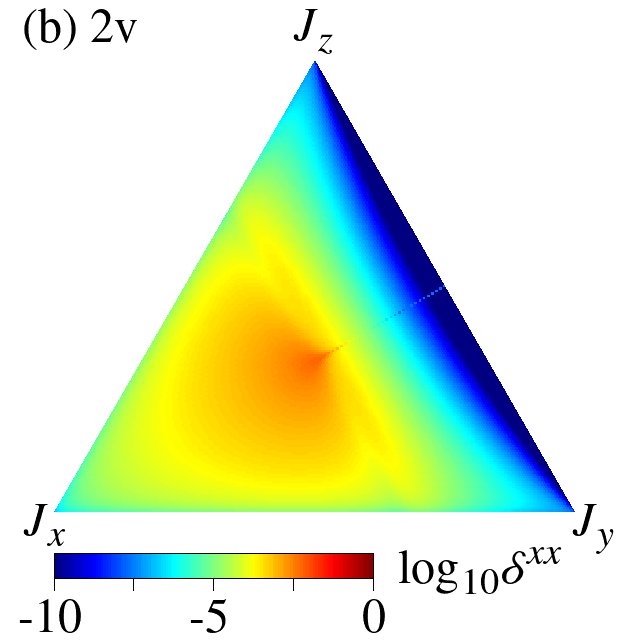} \hspace{20pt}
\includegraphics[scale=0.22]{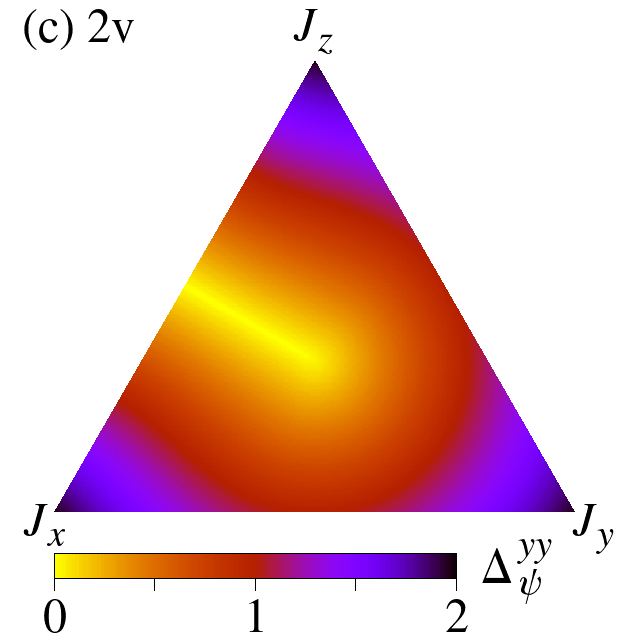} \hspace{20pt}
\includegraphics[scale=0.22]{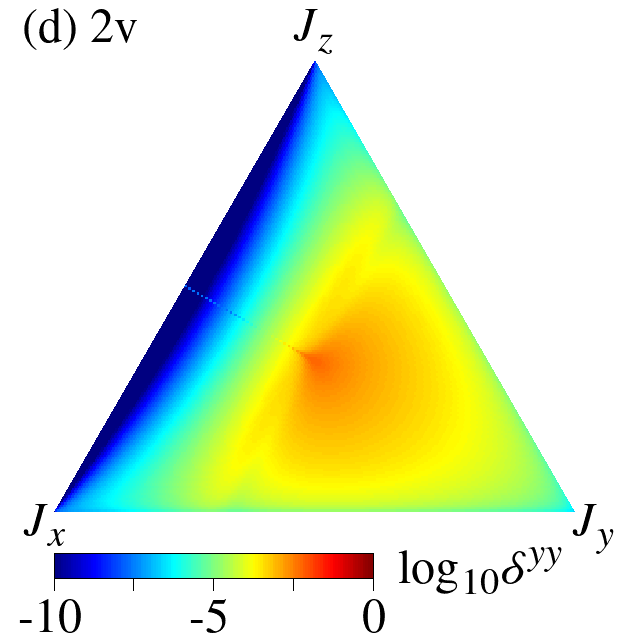} \\ \vspace{5pt}
\includegraphics[scale=0.22]{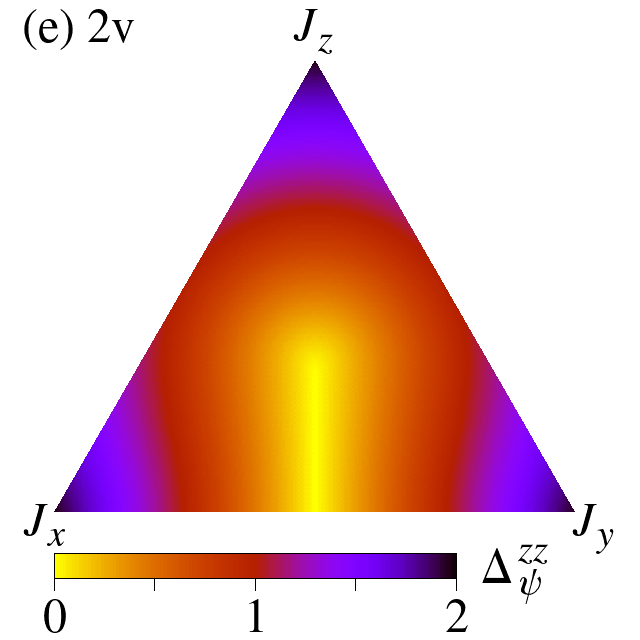} \hspace{20pt}
\includegraphics[scale=0.22]{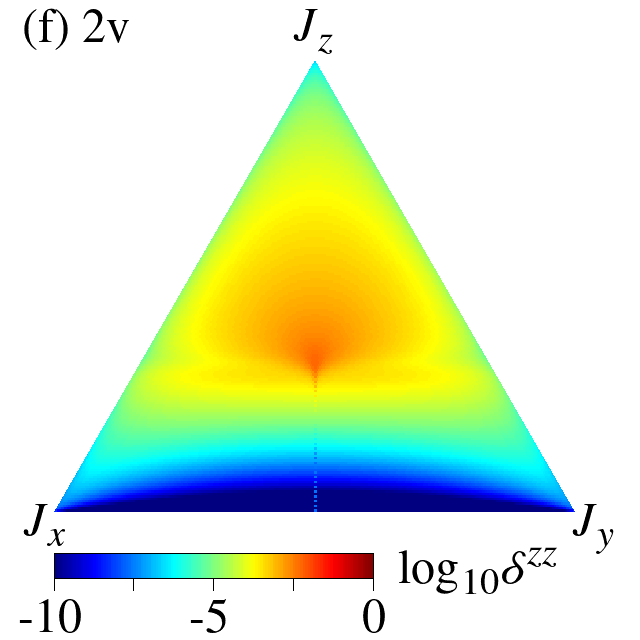} \hspace{20pt}
\includegraphics[scale=0.22]{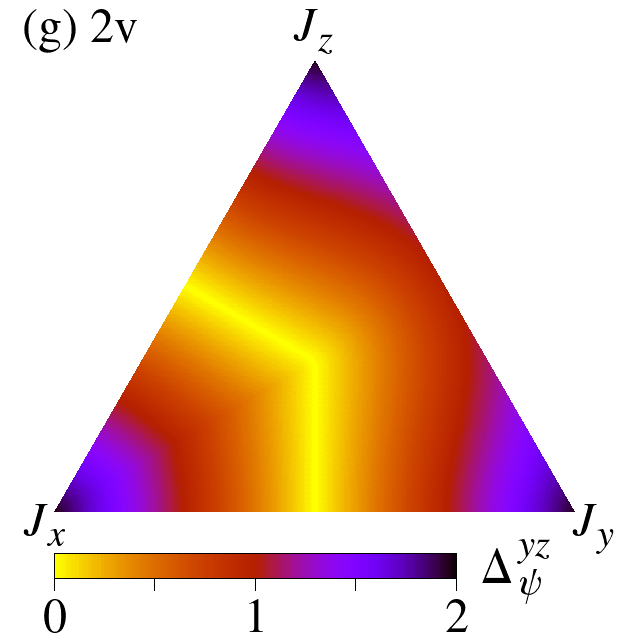} \hspace{20pt}
\includegraphics[scale=0.22]{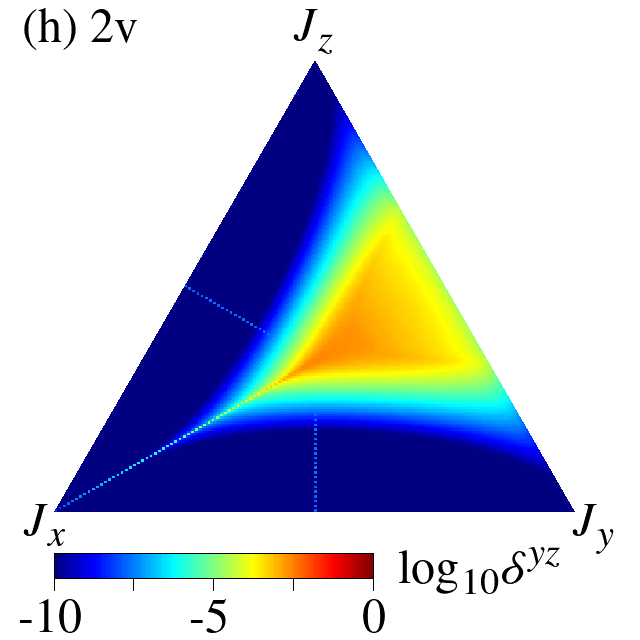} \\ \vspace{5pt}
\includegraphics[scale=0.22]{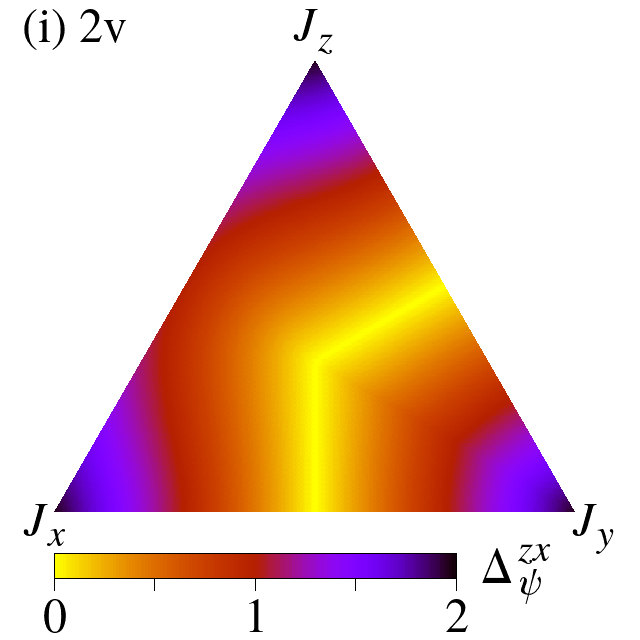} \hspace{20pt}
\includegraphics[scale=0.22]{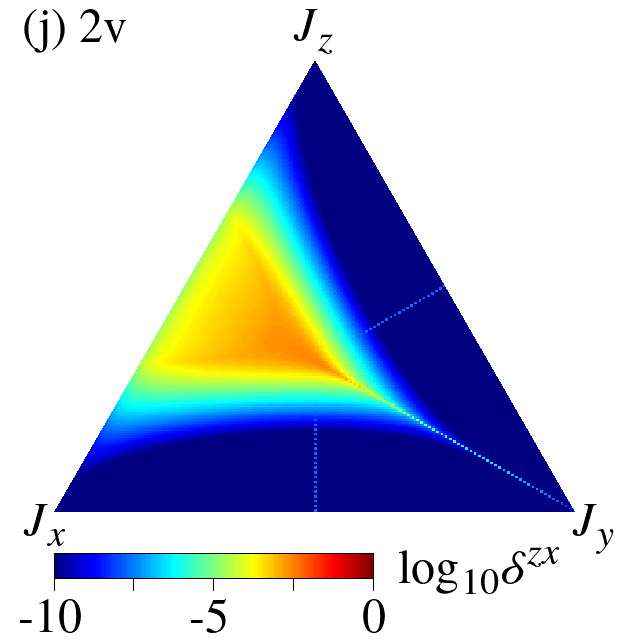} \hspace{20pt}
\includegraphics[scale=0.22]{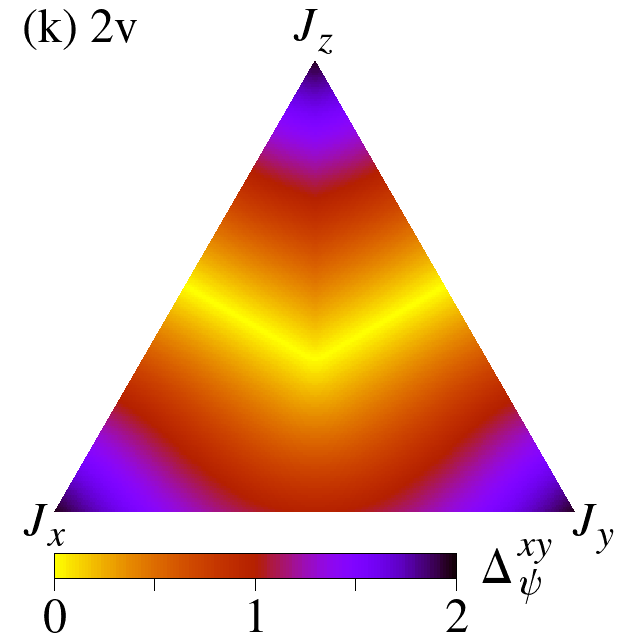} \hspace{20pt}
\includegraphics[scale=0.22]{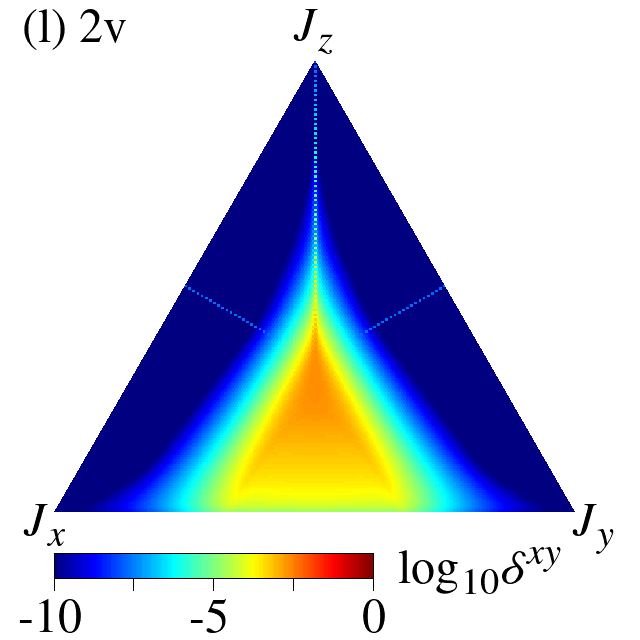}
\caption{\label{figure:kekulegap2v}The averaged fermion gaps $\Delta_\psi^{xx}$, $\Delta_\psi^{yy}$, $\Delta_\psi^{zz}$, $\Delta_\psi^{yz}$, $\Delta_\psi^{zx}$, and $\Delta_\psi^{xy}$ of two-vortex sectors with (a) both vortices at $x$ hexagons, (c) both vortices at $y$ hexagons, (e) both vortices at $z$ hexagons, (g) one vortex at a $y$ hexagon and the other at a $z$ hexagon, (i) one vortex at a $z$ hexagon and the other at an $x$ hexagon, and (k) one vortex at an $x$ hexagon and the other at a $y$ hexagon, respectively, on the triangular parameter space $J_x + J_y + J_z = 1$, where the vertex labeled by $J_\lambda$ indicates the $J_\lambda \longrightarrow 1$ limit, for the Kitaev Kekul\'{e} model. The corresponding standard deviations $\delta^{xx}$, $\delta^{yy}$, $\delta^{zz}$, $\delta^{yz}$, $\delta^{zx}$, and $\delta^{xy}$ are shown in (b), (d), (f), (h), (j), and (l) on a logarithmic scale.}
\end{figure*}

With $L_1 \times L_2 = 16 \times 16$ and $L_1' \times L_2' = 2 \times 12$, we compute $\Delta_\psi^{\alpha \beta}$ and $\delta^{\alpha \beta}$ for $\alpha \beta = xx, yy, zz, yz, zx, xy$ and plot them over the triangular parameter space, as shown in Figs.~\ref{figure:kekulegap2v}a-\ref{figure:kekulegap2v}l. We find that it is only possible for the fermion gaps of some two-vortex sectors to vanish on the three lines $l_{yz} \equiv \lbrace (J_x , J_y , J_z) \vert J_y = J_z \geq J_x \rbrace$, $l_{zx} \equiv \lbrace (J_x , J_y , J_z) \vert J_z = J_x \geq J_y \rbrace$, and $l_{xy} \equiv \lbrace (J_x , J_y , J_z) \vert J_x = J_y \geq J_z \rbrace$. More specifically, the fermion gap of a two-vortex sector vanishes on $l_{\mu \nu}$ if and only if at least one vortex is hosted by a $\lambda$ hexagon, where $(\lambda , \mu , \nu)$ is a cyclic permutation of $(x,y,z)$. The combined set of critical parameters $l_{yz} \cup l_{zx} \cup l_{xy}$ thus divides the triangular parameter space into three parts, which contain the strong $J_x$, $J_y$, and $J_z$ limits, respectively. This is consistent, by Corollary 2, with the three different assignments of anyon species, as we noted in the beginning of this subsection. Note that the intersection $l_{yz} \cap l_{zx} \cap l_{xy}$ is the isotropic point $J_x = J_y = J_z$, at which $\Delta_\psi^{(0\mathrm{v})} = 0$. The smallness of the error, which attains a maximum of the order of $10^{-3}$ near the isotropic point, indicates that the fermion gaps of the two-vortex sectors being averaged over are approximately equal. \\

\begin{figure*}
\includegraphics[scale=0.25]{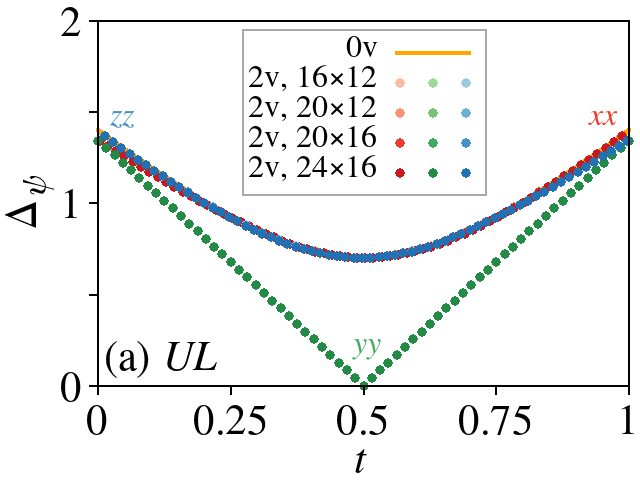} \;
\includegraphics[scale=0.25]{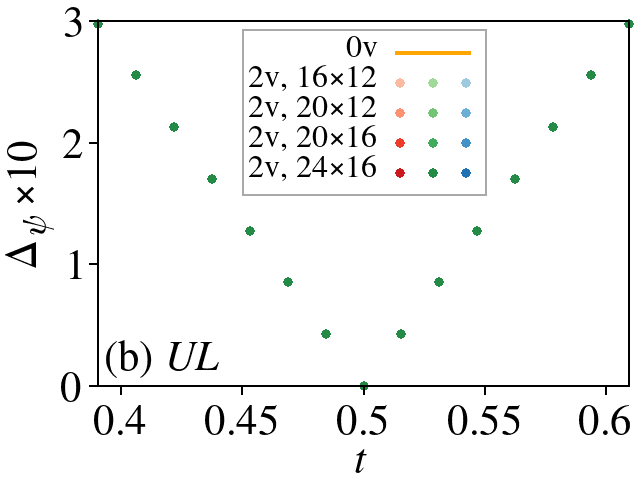} \;
\includegraphics[scale=0.25]{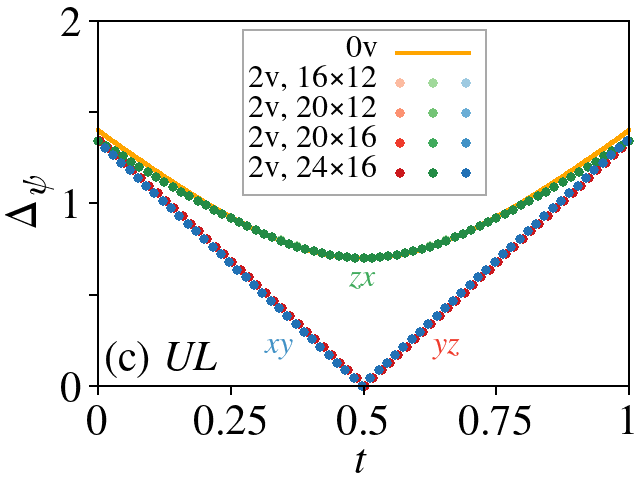} \;
\includegraphics[scale=0.25]{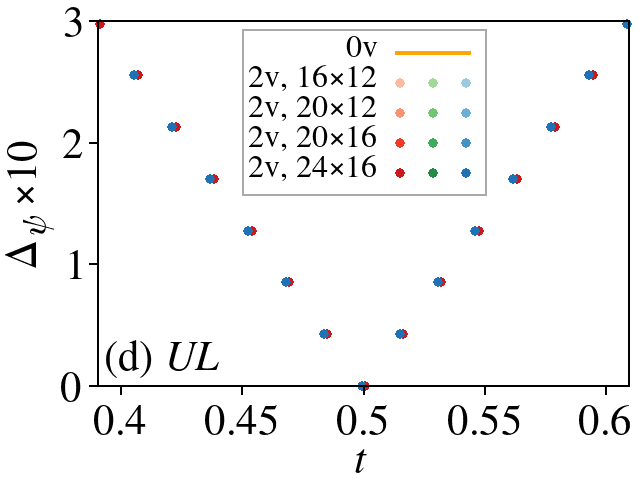} \\ \vspace{5pt}
\includegraphics[scale=0.25]{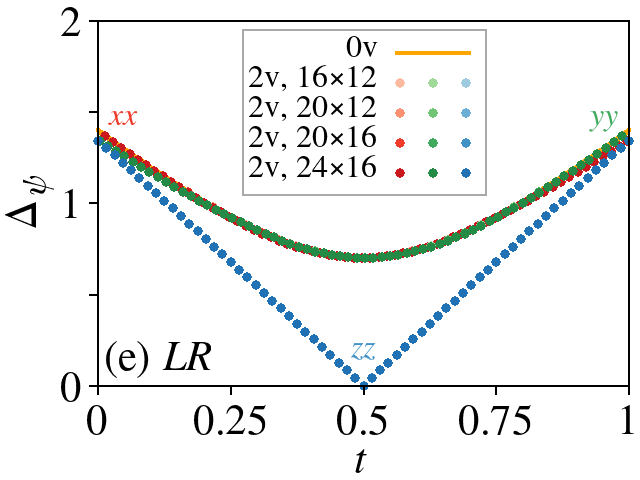} \;
\includegraphics[scale=0.25]{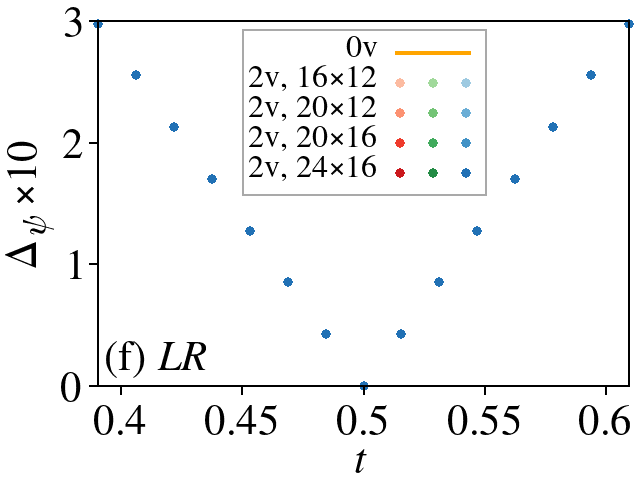} \;
\includegraphics[scale=0.25]{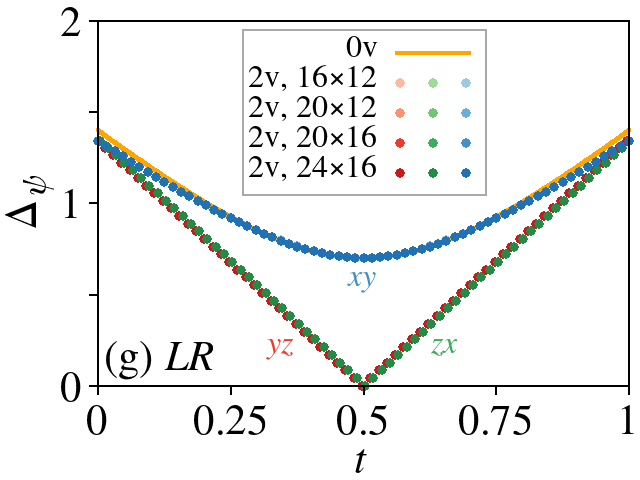} \;
\includegraphics[scale=0.25]{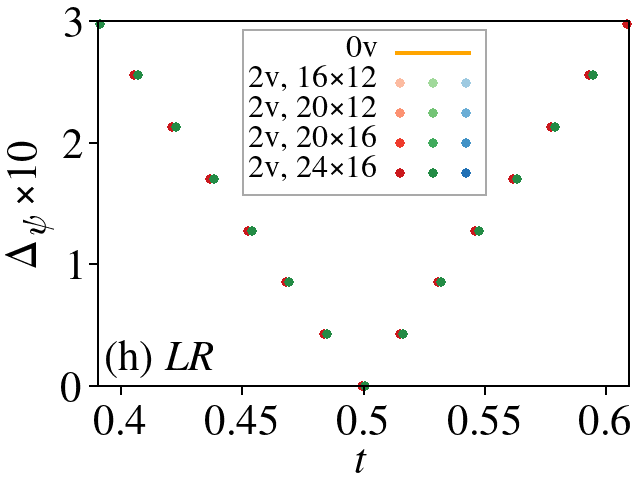} \\ \vspace{5pt}
\includegraphics[scale=0.25]{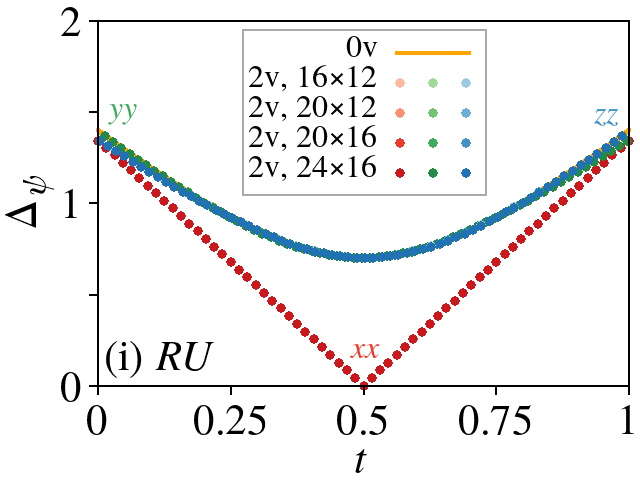} \;
\includegraphics[scale=0.25]{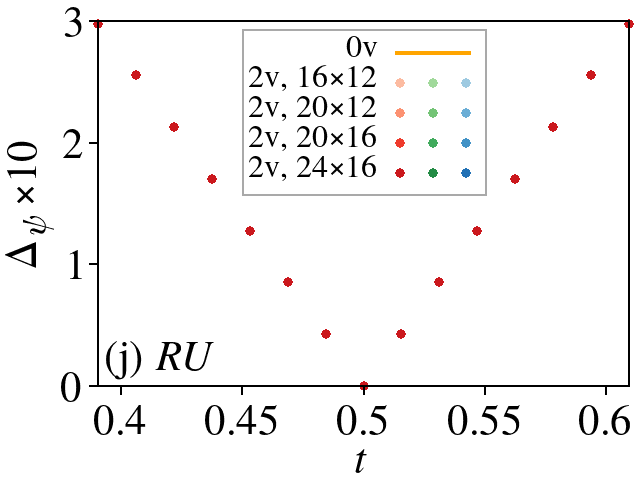} \;
\includegraphics[scale=0.25]{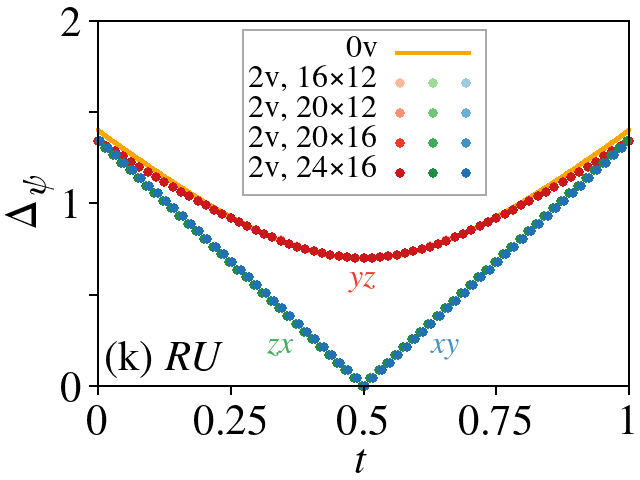} \;
\includegraphics[scale=0.25]{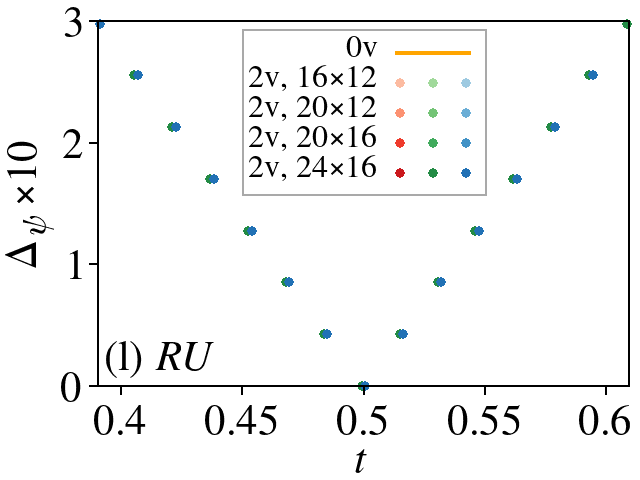}
\caption{\label{figure:kekulecut}(a,e,i) The averaged fermion gaps $\Delta_\psi^{xx}$ (red data), $\Delta_\psi^{yy}$ (green data), and $\Delta_\psi^{zz}$ (blue data) of two-vortex sectors with both vortices at $x$ hexagons, both vortices at $y$ hexagons, and both vortices at $z$ hexagons, respectively, along the paths $UL$, $LR$, and $RU$, each of which is linearly parametrized by $t \in [0,1]$, for the Kitaev Kekul\'{e} model. The error bars (not visible) represent the standard deviations. The fermion gap $\Delta_\psi^{(0\mathrm{v})}$ of the vortex-free sector (orange data) is also plotted for comparison. (b,f,j) Zoom-ins of (a,e,i) near zero energy. (c,g,k) The averaged fermion gaps $\Delta_\psi^{yz}$ (red data), $\Delta_\psi^{zx}$ (green data), and $\Delta_\psi^{xy}$ (blue data) of two-vortex sectors with one vortex at a $y$ hexagon and the other at a $z$ hexagon, one vortex at a $z$ hexagon and the other at an $x$ hexagon, and one vortex at an $x$ hexagon and the other at a $y$ hexagon, respectively, along the paths $UL$, $LR$, and $RU$, each of which is linearly parametrized by $t \in [0,1]$, for the Kitaev Kekul\'{e} model. The error bars (not visible) represent the standard deviations. The fermion gap $\Delta_\psi^{(0\mathrm{v})}$ of the vortex-free sector (orange data) is also plotted for comparison. (d,h,l) Zoom-ins of (c,g,k) near zero energy. In (a)-(l), the product of two integers associated with each set of 2v data indicates the size of the system ($L_1 \times L_2$ unit cells) on which the computations are performed. The blue/red series in (a)-(d), the red/green series in (e)-(h), and the green/blue series in (i)-(l) are shifted slightly leftward/rightward for visibility.}
\end{figure*}

We further provide a detailed convergence analysis of $\Delta_\psi^{\alpha \beta}$ with respect to the sizes of the system and the region of mapping, along the paths $UL$, $LR$, and $RU$ as indicated in Fig.~\ref{figure:kekulegap0v}b. We also calculate the (relative) fermion parities of the two-vortex sectors to determine the anyon species of each elementary plaquette within the region of mapping. Our computations are performed with (i) $L_1 \times L_2 = 16 \times 12$, $L_1' \times L_2' = 2 \times 8$, (ii) $L_1 \times L_2 = 20 \times 12$, $L_1' \times L_2' = 4 \times 8$, (iii) $L_1 \times L_2 = 20 \times 16$, $L_1' \times L_2' = 4 \times 12$, and (iv) $L_1 \times L_2 = 24 \times 16$, $L_1' \times L_2' = 6 \times 12$. \\

The results are shown in Figs.~\ref{figure:kekulecut}a-\ref{figure:kekulecut}l, which we summarize as follows. Let $(\lambda , \mu , \nu)$ be a cyclic permutation of $(x,y,z)$ and consider the path that linearly interpolates between the strong $J_\mu$ and $J_\nu$ limits. Starting from $t=0$ with a value of $\sim 1$, $\Delta_\psi^{\lambda \lambda} (t)$, $\Delta_\psi^{\lambda \mu} (t)$, and $\Delta_\psi^{\nu \lambda} (t)$ decrease monotonically as $t$ increases to $1/2$, at which they become zero. Numerically, we find that these fermion gaps are of the order of $10^{-9}$ or lower at $t=1/2$. As $t$ further increases from $1/2$ to $1$, they increase monotonically to $\sim 1$. On the other hand, $\Delta_\psi^{\mu \mu} (t)$, $\Delta_\psi^{\nu \nu} (t)$, and $\Delta_\psi^{\mu \nu} (t)$ behave similarly as $\Delta_\psi^{(0\mathrm{v})} (t)$ and remain greater than $0.5$ throughout $t \in [0,1]$. For each $\alpha \beta$, the data $\Delta_\psi^{\alpha \beta}$ of various system sizes overlap nearly perfectly, and $\delta^{\alpha \beta}$ is extremely small such that the error bars are not visible. We also confirm that, for $t \in [ 0 , 1/2 )$, the assignment of anyon species is constant and given by that in the strong $J_\mu$ limit, which is consistent with Corollaries 1 and 2. On the other hand, for $t \in (1/2 , 1]$, the assignment of anyon species is constant and given by that in the strong $J_\nu$ limit, which is again consistent with Corollaries 1 and 2.

\subsubsection{\label{section:auto}Anyon Automorphism}

Let $(\lambda , \mu , \nu)$ be a cyclic permutation of $(x,y,z)$. The strong $J_\mu$ and $J_\nu$ limits can be continuously connected by a path (e.g., one of $UL$, $LR$, and $RU$) along which $\Delta_\psi^{\alpha \beta} \neq 0$ and $\delta^{\alpha \beta} \ll \Delta_\psi^{\alpha \beta}$ is essentially zero for $\alpha \beta = \mu \mu , \nu \nu , \mu \nu$, indicating that any two-vortex sector that does not involve a $\lambda$ hexagon has a finite fermion gap. In either limit, the $\mu$ and $\nu$ hexagons necessarily belong to distinct anyon species. For each unit hexagon $p \in \mu$, one can always finds unit hexagons $p' \in \nu$ and $p_0 \in \mu$ such that the trio $p_0$, $p$, and $p'$ satisfy the conditions (i) and (ii) of Proposition 2. As the model parameters are varied along the path, $\Delta_\psi (p_0 , p)$ and $\Delta_\psi (p_0 , p')$ remain finite, so the anyon species of each of the trio cannot change, as per Proposition 2. In other words, the anyon species of every $\mu$ hexagon is invariant. Similarly, the anyon species of every $\nu$ hexagon is invariant. On the other hand, the anyon species of each $\lambda$ hexagon is allowed to change. Indeed, it must change upon crossing the line $l_{\mu \nu}$, so that the assignment of anyon species is in accordance with the appropriate dimer limit. \\

\begin{figure}
\subfloat[]{\label{figure:autot0}
\includegraphics[scale=0.2]{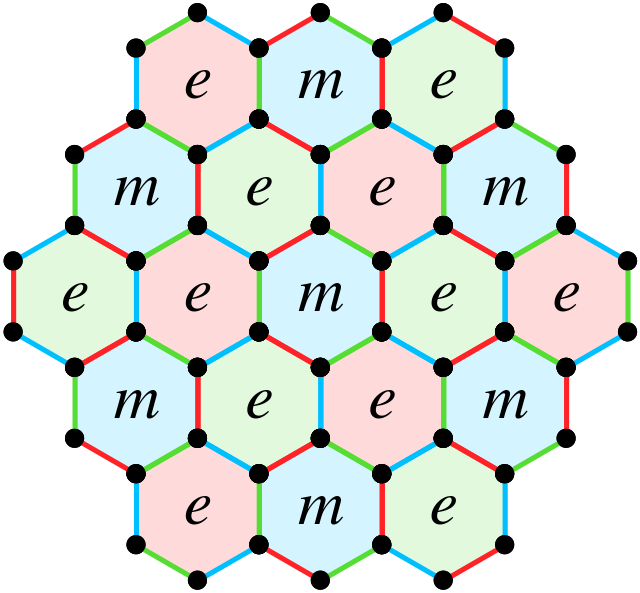}} \hspace{20pt}
\subfloat[]{\label{figure:autot1}
\includegraphics[scale=0.2]{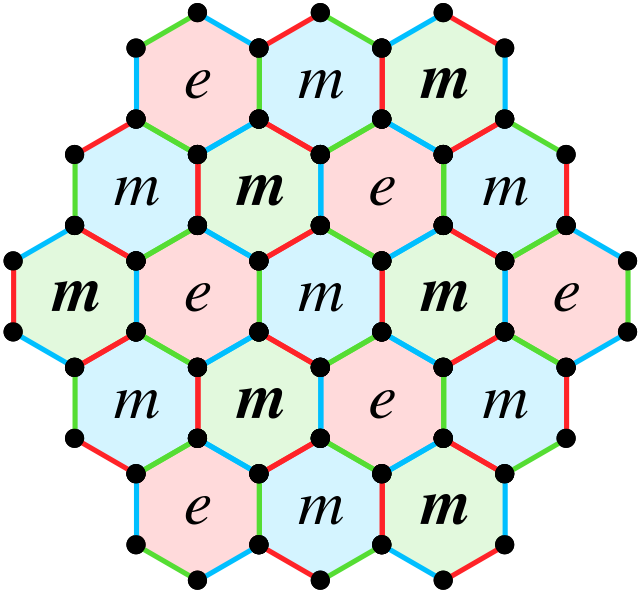}} \\
\subfloat[]{\label{figure:autot2}
\includegraphics[scale=0.2]{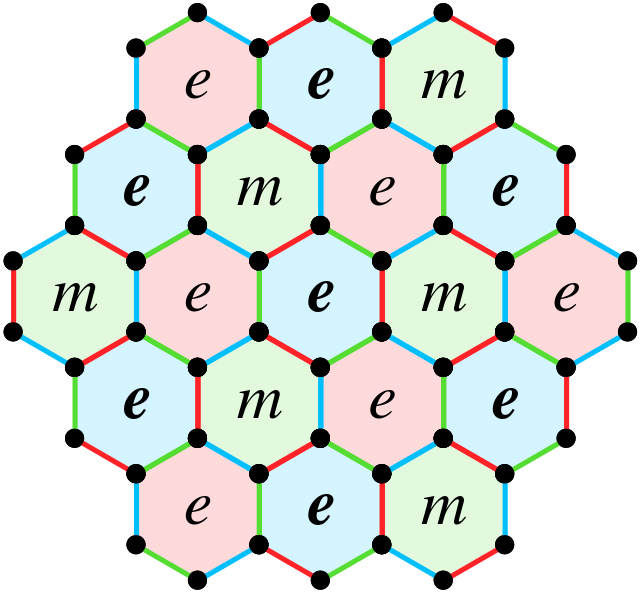}} \hspace{20pt}
\subfloat[]{\label{figure:autot3}
\includegraphics[scale=0.2]{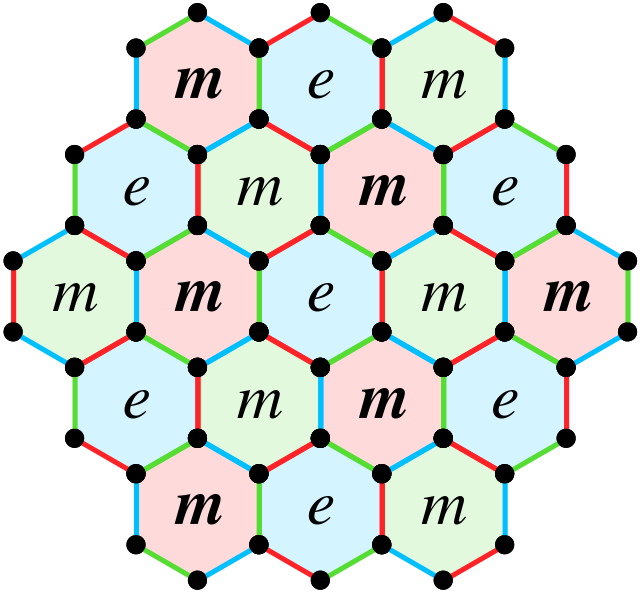}}
\caption{The assignments of anyon species in the Kitaev Kekul\'{e} model at (a) $0 \leq t < 1/2$, i.e., starting from $U$ and before crossing $l_{zx}$, (b) $1/2 < t < 3/2$, i.e., after crossing $l_{zx}$ and before crossing $l_{xy}$, (c) $3/2 < t < 5/2$, i.e., after crossing $l_{xy}$ and before crossing $l_{yz}$, and (d) $5/2 < t \leq 3$, i.e., after crossing $l_{yz}$ and returning to $U$, as the model parameters are varied along the path $ULRU$, which is linearly parametrized by $t \in [0 , 3]$ such that $t=0$, $1$, $2$, and $3$ correspond to $U$, $L$, $R$, and $U$, respectively. Red, green, and blue edges (faces) represent the $x$, $y$, and $z$ bonds (hexagons), respectively. In (b)-(d), the anyon labels that are changed with respect to the previous assignment are indicated with bold fonts.}
\end{figure}

With this understanding, we investigate how the assignment of anyon species changes as the model parameters are varied along the closed path obtained by concatenating $UL$, $LR$, and $RU$, see Fig.~\ref{figure:kekulegap0v}b. We parametrize the path $ULRU$ by $t \in [0,3]$, such that the intervals $[0,1]$, $[1,2]$, and $[2,3]$ correspond to the linear interpolations from $U$ to $L$, from $L$ to $R$, and from $R$ to $U$, respectively. Suppose that the initial assignment of anyon species is such that the $z$ hexagons correspond to $m$, while the $x$ and $y$ hexagons correspond to $e$, see Fig.~\ref{figure:autot0}. Upon crossing $l_{zx}$ at $t=1/2$, the anyon species of the $y$ hexagons changes from $e$ to $m$, see Fig.~\ref{figure:autot1}. Upon crossing $l_{xy}$ at $t=3/2$, the anyon species of the $z$ hexagons changes from $m$ to $e$, see Fig.~\ref{figure:autot2}. Upon crossing $l_{yz}$ at $t=5/2$, the anyon species of the $x$ hexagons changes from $e$ to $m$, see Fig.~\ref{figure:autot3}, which also represents the final assignment of anyon species. Comparing Figs.~\ref{figure:autot0} and \ref{figure:autot3}, one finds that every unit hexagon that was labeled by $e$ ($m$) at $t=0$ is labeled by $m$ ($e$) at $t=3$. In other words, an initially present $e$ ($m$) particle would become an $m$ ($e$) particle upon varying the model parameters along $ULRU$. We say that such a parametric transformation implements an $e \longleftrightarrow m$ automorphism. In general, an automorphism of a topological order is a permutation of anyons that preserve all the fusion and braiding rules \cite{q-2024-08-27-1448}. For the $\mathbb{Z}_2$ topological order, the only nontrivial automorphism is the one that permutes $e$ and $m$ \footnote{The concept of anyon automorphism is closely related to that of anyon permutation symmetry. In this work, the former represents the result of an action, e.g., $e$ and $m$ are permuted under an adiabatic transformation, while the latter reflects the freedom in specifying an initial condition, e.g., whether the vortex at a reference plaquette is assigned to $e$ or $m$ at $t=0$, which will then fix the anyon species of all other vortices.}. \\

More generally, let $\mathscr{T}$ be an adiabatic transformation where the model parameters are varied along a closed path $C$ that winds around the isotropic point $w$ times. If $w$ is odd, then $C$ intersects each of $l_{yz}$, $l_{zx}$, and $l_{xy}$ an odd number of times, so an $e \longleftrightarrow m$ automorphism is implemented by $\mathscr{T}$. If $w$ is even (including $w=0$), then $C$ intersects each of $l_{yz}$, $l_{zx}$, and $l_{xy}$ an even number of times, so one has exactly the same assignment of anyon species before and after $\mathscr{T}$. Let us focus on $w=1$ for concreteness. There exist two ways in the published literature of understanding how $\mathscr{T}$ implements the $e \longleftrightarrow m$ automorphism. The first is to establish an equivalence between the Kitaev Kekul\'{e} model under $\mathscr{T}$ and the honeycomb Floquet code \cite{q-2021-10-19-564}, the latter of which generates logical qubits dynamically through a specific sequence of weight-two measurements of Pauli operators, or a closely related unitary loop model \cite{PhysRevB.108.195134}. The second is to unroll the set of model parameters traced by (a suitably deformed) $C$ in real space \cite{PhysRevB.106.085122}, which creates a localized non-Abelian defect \cite{PhysRevLett.105.030403}. We refer interested readers to Refs.~\cite{PhysRevB.106.085122,PhysRevB.108.195134} for detailed descriptions of these ways. In this work, we offer a new and simple picture---the third way---where the $e \longleftrightarrow m$ automorphism comprises three discrete steps, such that the anyon species of a distinct class ($x$, $y$, or $z$) of unit hexagons is changed at each step. After these three steps, the anyon species of all unit hexagons are changed. \\

We close this subsection by remarking that Corollary 1 (or, equivalently, Proposition 1) is able to map out the anyon species in each dimer limit only up to a global permutation of $e$ and $m$. Therefore, it alone is insufficient to determine which vortices would be converted from one anyon species to the other and which would remain invariant under a continuous change of model parameters that connects different dimer limits. Proposition 2 is indispensable for keeping track of the changes of anyon species of individual vortices under such a parametric transformation that does not involve a fermion-gap-closing transition in the vortex-free sector.

\subsection{\label{section:enrich}Symmetry-Enriched Topological Orders}

\begin{figure}
\subfloat[]{\label{figure:honeycombphase}
\includegraphics[scale=0.22]{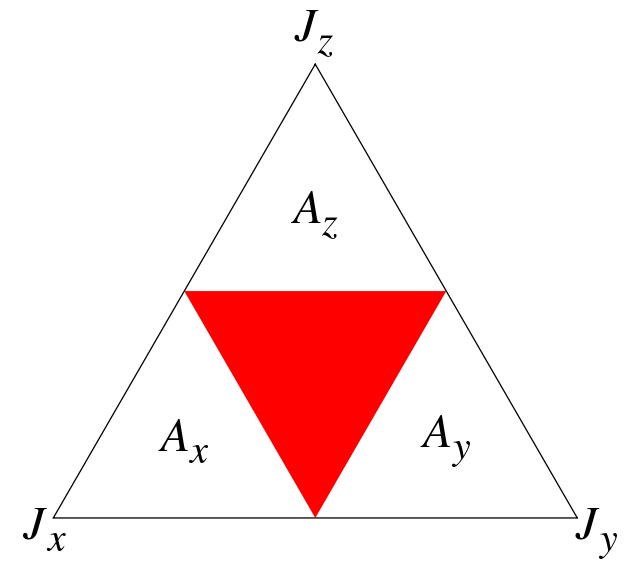}} \hspace{20pt}
\subfloat[]{\label{figure:starphase}
\includegraphics[scale=0.22]{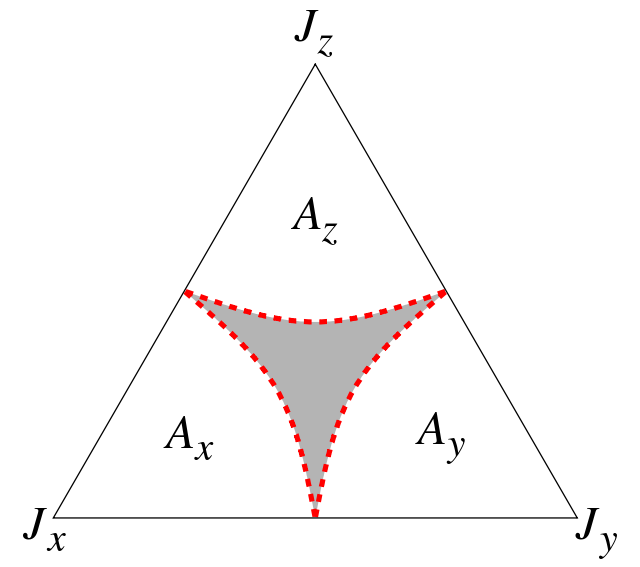}}
\caption{Toric code phases $A_x$, $A_y$, and $A_z$ that are characterized by different assignments of anyon species in the Kitaev (a) honeycomb and (b) star models. In (a), the red filled area indicates the vanishing of the fermion gap in the vortex-free sector and all two-vortex sectors. In (b), the red dashed curves indicate the vanishing of the fermion gap in the vortex-free sector and all two-vortex sectors, while the gray filled area indicates the vanishing of the fermion gap in all two-vortex sectors but not in the vortex-free sector.}
\end{figure}

For the Kitaev square-octagon and Kekul\'{e} models, we see that toric code phases with different assignments of anyon species, which are associated with different dimer limits, are not necessarily separated by a fermion-gap-closing transition in the vortex-free sector. Instead, they are separated by fermion-gap-closing transitions in (some of) the two-vortex sectors, see Figs.~\ref{figure:octagonphase} and \ref{figure:kekulephase}. In contrast, for the Kitaev honeycomb and star models, toric code phases with different assignments of anyon species, which are also associated with different dimer limits, are always separated by a fermion-gap-closing transition in the vortex-free sector, see Figs.~\ref{figure:honeycombphase} and \ref{figure:starphase}. Computational results for the Kitaev honeycomb and star models can be found in Sec.~\ref{section:addresult} in the Supplemental Material \cite{supply}. In this section, we explain the necessity of a $\Delta_\psi^{(0\mathrm{v})}=0$ transition between toric code phases from the perspective of symmetry. \\

We first review several concepts and results pertaining to symmetry-enriched topological (SET) orders that are essential to our discussions. Our presentation in this paragraph mostly follows Sec.~III B of Ref.~\cite{PhysRevX.14.021053} (see also Ref.~\cite{PhysRevB.100.115147}). Distinct SET orders refer to topological orders that have the same set of anyonic excitations but that cannot be smoothly connected to each other in the presence of symmetry \cite{SciPostPhys.15.1.004}. Let $G$ be the symmetry group of the microscopic model, which hosts the topological order in question. The latter may exhibit weak symmetry breaking \cite{KITAEV20062,PhysRevResearch.3.023120}, which means that the ground state preserves $G$ but some of the anyonic excitations are permuted by some elements of $G$. As $\mathbf{g} \in G$ generically changes an anyon $a$ to an anyon $^\mathbf{g}a$ (including the case ${^\mathbf{g}a}=a$), we define the map
\begin{equation}
\rho_\mathbf{g}: a \longrightarrow {^\mathbf{g}a} .
\end{equation}
A given set of $\rho_\mathbf{g}$ thus corresponds to a specific pattern of weak symmetry breaking. Let $\lvert \Psi_{a_1 , a_2 , a_3} \rangle$ represent a state of three anyons $a_1$, $a_2$, and $a_3$ that are far away from each other. $\lvert \Psi_{a_1 , a_2 , a_3} \rangle$ lives in a degenerate state space of dimension $N_{a_1 a_2}^{\overline{a}_3}$, which is the number of independent ways by which $a_1$ and $a_2$ can fuse to $\overline{a}_3$, where the overbar denotes the antiparticle \footnote{If $a_1$, $a_2$, and $a_3$ are Abelian anyons, then $N_{a_1 a_2}^{\overline{a}_3} = 1$.}. The action of $\mathbf{g} \in G$ on $\lvert \Psi_{a_1 , a_2 , a_3} \rangle$ is given by
\begin{equation}
\begin{aligned}[b]
&R_\mathbf{g} \lvert \Psi_{a_1, a_2 , a_3} \rangle \\
&= V_\mathbf{g}^{(1)} V_\mathbf{g}^{(2)} V_\mathbf{g}^{(3)} U_\mathbf{g} ({^\mathbf{g}a_1},{^\mathbf{g}a_2};{^\mathbf{g}\overline{a}_3}) \lvert \Psi_{{^\mathbf{g}a_1} , {^\mathbf{g}a_2} , {^\mathbf{g}a_3}} \rangle ,
\end{aligned}
\end{equation}
where $V_\mathbf{g}^{(i)}$ is a local unitary operator supported only around $a_i$ and $U_\mathbf{g} ({^\mathbf{g}a_1} , {^\mathbf{g}a_2} ; {^\mathbf{g}\overline{a}_3})$ is a unitary matrix of rank $N_{a_1 a_2}^{\overline{a}_3}$ that acts on the degenerate state space. The $V$ operators satisfy
\begin{equation} \label{etasymbol}
\eta_{a_i} (\mathbf{g} , \mathbf{h}) V_{\mathbf{g} \mathbf{h}}^{(i)} \lvert \Psi_{a_1 , a_2 , a_3} \rangle = R_\mathbf{g} V_\mathbf{h}^{(i)} R_\mathbf{g}^{-1} V_\mathbf{g}^{(i)} \lvert \Psi_{a_1 , a_2 , a_3} \rangle
\end{equation}
for $\mathbf{g} , \mathbf{h} \in G$, where $\eta_{a_i} (\mathbf{g} , \mathbf{h})$ are generically nontrivial phase factors that determine the class of symmetry fractionalization \cite{PhysRevB.87.104406}. Given a topological order and a symmetry group $G$, the data $\lbrace \rho_\mathbf{g} ; U_\mathbf{g} (a , b ; c) , \eta_{a} (\mathbf{g} , \mathbf{h}) \rbrace$ completely characterize how this topological order is enriched by $G$. For brevity, we will often refer to $U_\mathbf{g} (a , b ; c)$ and $\eta_{a} (\mathbf{g} , \mathbf{h})$ as the $U$ and $\eta$ symbols, repectively. \\

A generalized Kitaev model of the form \eqref{kitaevmodel} defined on a lattice possesses both time-reversal and spatial symmetries. Regardless of whether time-reversal symmetry is spontaneously broken in the ground state (due to the presence of odd-length elementary plaquettes), we consider only the space group because time reversal should not behave differently in the toric code phases associated with different dimer limits of a given geometry. Moreover, it suffices to consider only the generators of the space group. For generic values of $J_x$, $J_y$, and $J_z$, only spatial transfomations that map $\lambda$ bonds to $\lambda$ bonds for every $\lambda \in \lbrace x,y,z \rbrace$ qualify to be elements of the space group. For the $\mathbb{Z}_2$ topological order, the only nontrivial anyon permutation that can be intertwined with a symmetry of the microscopic model is the one that interchanges $e$ with $m$ \cite{PhysRevX.14.021053}. Given an assignment of anyon species, a space group element permutes (preserves) the $e$ and $m$ particles if it maps each elementary plaquette to one with the opposite (same) anyon label. \\

\begin{table}
\caption{\label{table:honeycomb}The action of a generator $\mathbf{g} \in \lbrace T_1 , T_2 , C_2 \rbrace$ of the plane group $p2$ on the anyon species in the strong $\lambda \in \lbrace x , y , z \rbrace$ bond limit of the Kitaev honeycomb or star model. The entry ``yes'' or ``no'' is the answer to the question ``Does $\mathbf{g}$ permute the $e$ and $m$ anyons in the strong $J_\lambda$ limit?''}
\begin{ruledtabular}
\begin{tabular}{c|ccc}
\hspace{28pt}$\mathbf{g}$\hspace{4pt}{\textbackslash}\hspace{4pt}$\lambda$\hspace{28pt} & $x$ & $y$ & $z$\hspace{28pt} \\ \hline
$T_1$ & yes & no & yes\hspace{28pt} \\
$T_2$ & no & yes & yes\hspace{28pt} \\
$C_2$ & no & no & no\hspace{28pt}
\end{tabular}
\end{ruledtabular}
\end{table}

For each of the Kitaev honeycomb and star models, the space group is the plane group $p2$, which is generated by two translations $T_1$ and $T_2$ and a two-fold rotation $C_2$, see Figs.~\ref{figure:honeycombmodel} and \ref{figure:starmodel}. In the strong $J_x$ limit, $T_1$ permutes the $e$ and $m$ anyons, while $T_2$ and $C_2$ do not. In the strong $J_y$ limit, $T_2$ permutes the $e$ and $m$ anyons, while $T_1$ and $C_2$ do not. In the strong $J_z$ limit, $T_1$ and $T_2$ permute the $e$ and $m$ anyons, while $C_2$ does not. These observations are summarized in Table \ref{table:honeycomb}. We see that these dimer limits exhibit different patterns of weak symmetry breaking, i.e., they have different sets of $\rho_\mathbf{g}$. Therefore, the associated toric code phases $A_x$, $A_y$, and $A_z$ belong to distinct SET orders, which must be separated from each other by $\Delta_\psi^{(0\mathrm{v})}=0$ transitions. \\

For the Kitaev square-octagon model, the space group is the plane group $c2mm$, which is generated by two translations $T_x$ and $T_y$ and two reflections $M_+$ and $M_-$, see Fig.~\ref{figure:octagonmodel}. Table \ref{table:octagon} summarizes the action of each of these operators on the anyon species in each of the strong $J_x$, $J_y$, and $J_z$ limits. The strong $J_z$ limit does not exhibit a weak symmetry breaking, while the strong $J_x$ and $J_y$ limits do. Therefore, the former must be separated from the latter two by a $\Delta_\psi^{(0\mathrm{v})} = 0$ transition. The strong $J_x$ and $J_y$ limits have the same set of $\rho_\mathbf{g}$. Moreover, they have the same set of $U$ symbols, as $U_\mathbf{g} (a , b ; c)$ can be chosen to depend only on $a$, $b$, and $\rho_\mathbf{g}$ \cite{PhysRevX.14.021053}. While solving explicitly for the $\eta$ symbols defined via \eqref{etasymbol} would go beyond the scope of this work, we argue on physical grounds that they should be the same in the strong $J_x$ and $J_y$ limits. First, the symmetry group $G$ is the same in these dimer limits. Second, whether the $e$ and $m$ anyons are permuted by each $\mathbf{g} \in G$ is the same in these dimer limits. Third, the ground-state flux sector (where every elementary plaquette has $\pi$-flux \cite{supply}) is the same throughout the triangular parameter space and no background anyon \cite{zbbd-dqzb} is introduced. Taken together, these similarities should yield the same set of $\eta$ symbols, which indicates the same class of symmetry fractionalization, in the strong $J_x$ and $J_y$ limits. The associated toric code phases $A_x$ and $A_y$ are thus characterized by the same data $\lbrace \rho_\mathbf{g} ; U_\mathbf{g} (a , b ; c) , \eta_{a} (\mathbf{g} , \mathbf{h}) \rbrace$. In other words, they belong to the same SET order and they need not be separated by a $\Delta_\psi^{(0\mathrm{v})}=0$ transition, which is indeed the case. \\

For the Kitaev Kekul\'{e} model, the space group is the plane group $p3m1$, which are generated by two translations $T_1$ and $T_2$, a threefold rotation $C_3$, and a reflection $M_x$, see Fig.~\ref{figure:kekulemodel}. Each of these operators does not permute the $e$ and $m$ anyons in the strong $J_x$, $J_y$, and $J_z$ limits. In other words, there is no weak symmetry breaking in any of these dimer limits. By the same line of reasoning as in the previous paragraph, we infer that the associated toric code phases $A_x$, $A_y$, and $A_z$ belong to the same SET order and they need not be separated from each other by a $\Delta_\psi^{(0\mathrm{v})}=0$ transition, which is indeed the case. \\

\begin{table}
\caption{\label{table:octagon}The action of a generator $\mathbf{g} \in \lbrace T_x , T_y , M_+ , M_- \rbrace$ of the plane group $c2mm$ on the anyon species in the strong $\lambda \in \lbrace x , y , z \rbrace$ bond limit of the Kitaev square-octagon model. The entry ``yes'' or ``no'' is the answer to the question ``Does $\mathbf{g}$ permute the $e$ and $m$ anyons in the strong $J_\lambda$ limit?''}
\begin{ruledtabular}
\begin{tabular}{c|ccc}
\hspace{28pt}$\mathbf{g}$\hspace{4pt}{\textbackslash}\hspace{4pt}$\lambda$\hspace{28pt} & $x$ & $y$ & $z$\hspace{28pt} \\ \hline
$T_x$ & yes & yes & no\hspace{28pt} \\
$T_y$ & yes & yes & no\hspace{28pt} \\
$M_+$ & no & no & no\hspace{28pt} \\
$M_-$ & no & no & no\hspace{28pt}
\end{tabular}
\end{ruledtabular}
\end{table}

We close this subsection with two remarks. First, a point-group generator (rotation or reflection) in each of the four models analyzed above maps at least one elementary plaquette to itself. When the fermion spectrum of every two-vortex sector is gapped, which is the case in any dimer limit, the vortex at each elementary plaquette belongs to a definite anyon species. As a result, the point-group generator cannot interchange $e$ with $m$. The nontrivial anyon permutation can only be implemented by translations, which have no fixed point. Ref.~\cite{zbbd-dqzb} has presented a classification of $\mathbb{Z}_2$ topological orders enriched by only translational symmetries and suggested the possibility of topological vortex bands in some of these phases. Second, the Kitaev square-octagon and Kekul\'{e} models demonstrate that different assignments of anyon species, which cannot be made identical even by a global permutation of $e$ and $m$, can be found within the same SET phase. A necessary condition is the existence of elementary plaquettes that are not related by symmetry, e.g., the unit squares and octagons in the former model. Elementary plaquettes that are related by symmetry are identified as belonging to the same class. If the $e$ and $m$ labels are interchanged throughout a particular class of elementary plaquettes, while the remaining ones are unaffected, then we obtain a different assignment of anyon species but the same pattern of weak symmetry breaking, which makes it possible for the toric code phases before and after the relabeling to be in the same SET order.

\section{\label{section:discuss}Discussion}

In summary, we have investigated the mapping of vortices (i.e., flux excitations) in generalized Kitaev models \eqref{kitaevmodel} on planar graphs to the electric ($e$) and magnetic ($m$) particles of the toric code model \cite{KITAEV20032}, using a combination of the Majorana fermion representation of $\Gamma$ matrices \cite{KITAEV20062,PhysRevB.79.134427}, the fusion rules of Abelian anyons \cite{simontextbook,PhysRevResearch.2.023334}, the fermion parity of physical states \cite{PhysRevB.84.165414,PhysRevB.92.014403}, and the adiabatic theorem of quantum mechanics \cite{BF01343193,JPSJ.5.435,griffithstextbook}. In contrast to previous works that mostly focus on the perturbative limit, where the couplings on a subset of bonds are much larger than the rest, we develop a comprehensive theory for generic parameters that are characterized by a trivial Chern number $\nu=0$. We are able to reproduce the known mapping scheme in the dimer limit \cite{Wootton_2015}, where vortices at two neighboring elementary plaquettes that are separated by a strong (weak) bond are identified as belonging to the same (different) species [\textbf{Corollary 1}]. More importantly, we prove the invariance of a given map under a continuous change of model parameters that does not close the fermion gap in both the vortex-free and two-vortex sectors [\textbf{Corollary 2}]. This property is useful for further dividing a $\nu=0$ phase into multiple regimes associated with different mappings of anyons, as in the Kitaev square-octagon and Kekul\'{e} models \cite{PhysRevB.76.180404,Kamfor_2010,PhysRevB.91.134419}. Our theory also allows one to track the changes of anyon species of individual vortices when the fermion gap closes in some two-vortex sectors but not in the vortex-free sector [\textbf{Proposition 2}]. As a highlight, it depicts the implementation of an $e \longleftrightarrow m$ automorphism \cite{q-2021-10-19-564,PhysRevB.106.085122,PhysRevB.108.195134,q-2024-08-27-1448} in the Kitaev Kekul\'{e} model as the net effect of three successive transitions, each of which involves a change of anyon species on a unique class of unit hexagons. Besides, we show that two different mappings of anyons can nevertheless exhibit the same pattern of weak symmetry breaking \cite{KITAEV20062,PhysRevResearch.3.023120}, and further argue that they belong to the same symmetry-enriched topological order \cite{PhysRevB.100.115147,PhysRevX.14.021053}. A necessary condition is the presence of elementary plaquettes that are not related by symmetry, such as the unit squares and octagons in the Kitaev square-octagon model. \\

We discuss the implications of our results and point out several open questions that may be interesting for future studies. We emphasize that our theory has established the criteria under which the anyon species of a given vortex must remain invariant, which further allows one to determine the extent of a given mapping of anyons, in terms of a set of quantities that can be straightforwardly calculated for a given model. We have thus successfully addressed the questions of how distinct mappings of anyons are fitted into, and what happens to individual vortices when the model parameters are tuned within, a single phase characterized by $\nu=0$. Apart from offering a finer classification of topologically ordered phases based on the mapping of anyons, our results are also potentially useful for encoding quantum information, as elaborated in Sec.~\ref{section:introduce}. The application of our theory to specific models typically requires computing the fermion gaps of multiple two-vortex sectors. For perfect lattices, one may classify the two-vortex sectors according to the types of two elementary plaquettes that host the vortices. For each of the models studied extensively in this work, we find that, to a good approximation, the fermion gap is constant across the two-vortex sectors within each class, as long as the vortices are well-separated. This suggests that it is sufficient to compute the fermion gaps only for a small number (which should be greater than $1$ to exclude the possibility of accidental gapless modes, but perhaps less than $10$) of two-vortex sectors within each class, for an efficient determination of the boundaries between different mappings of anyons. \\

Our theory relies on the knowledge that each $\nu = 0$ phase, which includes the dimer limit, is equivalent to a toric code model at low energies, as per Kitaev's sixteenfold way. In cases where it works, degenerate perturbation theory has the advantage of explicitly demonstrating such an equivalence through an effective Hamiltonian, without any input from Kitaev's sixteenfold way. As noted earlier, for a graph that contains odd-length elementary plaquettes, the absence of the corresponding elementary-plaquette operators in the effective Hamiltonian hinders a complete mapping of anyons. Whether this obstacle can be overcome within the framework of degenerate perturbation theory remains an interesting open question. One may, for instance, analyze operators for extended structures that involve multiple elementary plaquettes, see related discussions in Sec.~\ref{section:perturb} of the Supplemental Material \cite{supply}. Our theory also does not address the interactions between anyons, which typically appear at higher perturbative orders \cite{PhysRevLett.100.057208,PhysRevB.78.125102,PhysRevB.78.245121}. However, we point out that these two approaches can complement each other, especially for graphs with odd-length elementary plaquettes. Our theory identifies the anyon species of the vortices and determines the range of parameters for which a given mapping of anyons is valid, while degenerate perturbation theory reveals how the vortices interact. \\

Our computations for selected models reveal that when the fermion gap of the vortex-free sector closes, the fermion gaps of all two-vortex sectors approach zero as well. In contrast, strictly within a $\nu=0$ phase throughout which the fermion gap of the vortex-free sector is finite, it appears that the fermion gaps of only some, but not all, of the two-vortex sectors can be zero for a given set of couplings. This suggests different nature of the transitions between mappings of anyons in the former and latter cases. In Sec.~\ref{section:addresult} of the Supplemental Material \cite{supply}, we analyze the entire fermion spectra \cite{LAHTINEN20082286}, instead of only the fermion gaps, of the Kitaev square-octagon and Kekul\'{e} models along selected paths in the triangular parameter space. We find that when a change in the mapping of anyons involves a fermion-gap closing in the vortex-free sector, there is always a continuum of fermion modes above the fermion gap in each of the vortex-free and two-vortex sectors. In contrast, when a change in the mapping of anyons occurs without a fermion-gap closing in the vortex-free sector, we observe $0$, $1$, or $2$ gapless fermion modes that are clearly decoupled from a continuous spectrum, in a two-vortex sector with $0$, $1$, or $2$ species-changing vortices, respectively. Excluding all the decoupled fermion modes that are possibly present, we find that the fermion spectrum of the two-vortex sector highly resembles that of the vortex-free sector. The agreement of the number of gapless fermion modes and the number of species-changing vortices is indeed consistent with the fusion rules, but its necessity has neither been proved nor disproved for any flux sector in any generalized Kitaev model. For instance, in a four-vortex sector that involves three species-changing vortices, do we observe exactly three gapless fermion modes? Addressing this issue would be theoretically appealing.

\begin{acknowledgements}
This work was supported in part by the Deutsche Forschungsgemeinschaft via the cluster of excellence ctd.qmat (EXC 2147, project-id 390858490) and SFB 1143 (Project-ID No.~247310070), and by the Engineering and Physical Sciences Research Council (EPSRC) grant No.~EP/V062654/1. 
\end{acknowledgements}

\bibliography{reference260710}

\clearpage

\onecolumngrid

\begin{center}
\textbf{\large Supplemental Material: \\ Mapping vortices to anyons in toric code phases of generalized Kitaev models}
\end{center}
\begin{center}
Li Ern Chern$^{1}$, Roderich Moessner$^{1}$, and Claudio Castelnovo$^{2}$
\end{center}
\begin{center}
{\small
\textit{$^1$Max Planck Institute for the Physics of Complex Systems, 01187 Dresden, Germany} \\
\textit{$^2$T.C.M.~Group, Cavendish Laboratory, University of Cambridge, Cambridge CB3 0HE, United Kingdom}}
\end{center}

\setcounter{equation}{0}
\setcounter{figure}{0}
\setcounter{table}{0}
\setcounter{page}{1}
\setcounter{section}{0}

\renewcommand{\thesection}{S\arabic{section}}
\renewcommand{\theequation}{S\arabic{equation}}
\renewcommand{\thefigure}{S\arabic{figure}}
\renewcommand{\thetable}{S\arabic{table}}

\section{\label{section:constraint}Physical Constraint}

The Majorana fermion representation of either the Pauli or $\Gamma$ matrices introduces unphysical states and enlarges the Hilbert space. The projection to the physical subspace has been analyzed in details in Ref.~\cite{PhysRevB.84.165414} specifically for the Kitaev honeycomb model (see also Ref.~\cite{PhysRevB.92.014403}). In this section, we review the analysis in a way that is applicable to the generalized Kitaev model defined on a graph with the coordination number $z=2n-1$. \\

To see how problems may arise from the Majorana fermion representation, let us begin by discussing the case of $n=2$, e.g., the $S=1/2$ Kitaev honeycomb model. The dimension of the physical Hilbert space at each site, which is spanned by the up and down states of a chosen component of the local spin, is 2. Representing the spin by four Majorana fermions, the dimension of the local Hilbert space is doubled to 4, which can be easily understood as follows. One can combine two Majorana fermions into a complex fermion mode, which can be either occupied or unoccupied, giving a dimension of 2 \footnote{A Majorana fermion is also known as a real fermion, and it has a nominal dimension of $\sqrt{2}$.}. One can make two complex fermions out of four Majorana fermions, so the dimension of the local Hilbert space becomes $2 \times 2 = 4$. For generic $n$, the representation of the $2n-1$ local $\Gamma$ matrices by $2n$ Majorana fermions doubles the dimension of the local Hilbert space from $2^{n-1}$ to $2^n$. The enlargement of the Hilbert space is an indication that some states obtained with the Majorana fermion representation are unphysical and should be discarded. \\

Without loss of generality, we choose the $\Gamma$ matrices such that their product is $\prod_\lambda \Gamma^\lambda = i^{n-1}$, e.g., $\sigma^x \sigma^y \sigma^z = i$ for $n=2$. Expressing the local $\Gamma$ matrices in terms of Majorana fermions, any physical state $\lvert \psi \rangle$ must thus satisfy \cite{KITAEV20062,PhysRevB.79.134427,PhysRevLett.102.217202}
\begin{equation}
D_i \lvert \psi \rangle = \lvert \psi \rangle, \quad D_i \equiv - (-i)^n \left( \prod_\lambda b_i^{\lambda} \right) c_i ,
\end{equation}
for every site $i$. The projector to the physical subspace is given by
\begin{equation} \label{physicalprojector}
\mathcal{P} = \prod_{i \in \Lambda} \left( \frac{1 + D_i}{2} \right) = \frac{1}{2^N} \sum_{\tilde{\Lambda} \subseteq \Lambda} \prod_{i \in \tilde{\Lambda}} D_i ,
\end{equation}
where $\Lambda$ is the set of all sites, $N = \lvert \Lambda \rvert$ is the total number of sites (which is necessarily even), and the summation runs over all possible subsets $\tilde{\Lambda}$ of $\Lambda$. Since $D_i^2=(-1)^{2n^2}=1$, $\prod_{i \in \tilde{\Lambda}} D_i$ and $\prod_{i \in \Lambda \setminus \tilde{\Lambda}} D_i$ differ by a factor of $\prod_{i \in \Lambda} D_i$. The projector \eqref{physicalprojector} thus admits a factorization
\begin{equation} \label{physicalprojectorfactorize}
\mathcal{P} = \left( \frac{1}{2^{N-1}} \sideset{}{'}\sum_{\tilde{\Lambda} \subseteq \Lambda} \prod_{i \in \tilde{\Lambda}} D_i \right) \left( \frac{1+\prod_{i \in \Lambda} D_i}{2} \right) \equiv \mathcal{S} \, \mathcal{P}_0 ,
\end{equation}
where the prime on the summation indicates that it only runs over half of the all possible subsets of $\Lambda$, such that if a subset $\tilde{\Lambda}$ is included, then its complement $\Lambda \setminus \tilde{\Lambda}$ is not. Grouping $c_i$ together, and pairing up $b_i^\lambda$ and $b_j^\lambda$ into $- i u_{ij}^\lambda$ with $\langle ij \rangle$ being a $\lambda$ bond, we can express
\begin{equation} \label{gaugetransformproduct}
\prod_{i \in \Lambda} D_i = (-1)^{\xi'} (-i)^{N_b} \left( \prod_{\langle ij \rangle_\lambda} u_{ij}^\lambda \right) \prod_{i=1}^N c_i ,
\end{equation}
where $\xi'$ is an integer that depends on the geometric details of the graph, e.g., how the sites are connected to each other and how many sites are there in total, which influences the number of $-1$ factors arising from anticommuting the Majorana fermions past each other to obtain the RHS of \eqref{gaugetransformproduct}, and $N_b = (2n - 1) N / 2$ is the total number of bonds. Let $Q \in O (N)$ be the canonical transformation \cite{KITAEV20062} of the quadratic Hamiltonian \eqref{kitaevmodelquadratic} such that
\begin{equation}
Q^\mathrm{T} A Q = \begin{pmatrix} 0 & \varepsilon_1 & & & \\ - \varepsilon_1 & 0 & & & \\ & & \ddots  & & \\ & & & 0 & \varepsilon_{N/2} \\ & & & - \varepsilon_{N/2} & 0 \end{pmatrix} = \bigoplus_{k=1}^{N/2} \begin{pmatrix} 0 & \varepsilon_k \\ - \varepsilon_k & 0 \end{pmatrix} , \quad \varepsilon_k \geq 0 .
\end{equation}
Defining the Majorana fermions $\gamma_i$ such that $c_i=\sum_j Q_{ij} \gamma_j$, we have
\begin{equation} \label{cproduct}
\begin{aligned}[b]
\prod_{i=1}^N c_i &= c_1 \ldots c_N = \sum_{j_1, \ldots, j_N} Q_{1 j_1} \gamma_{j_1} \ldots Q_{N j_N} \gamma_{j_N} = \left[ \sum_P \mathrm{sgn}(P) Q_{1 P(1)} \ldots Q_{N P(N)} \right] ( \gamma_1 \ldots \gamma_N ) \\
&= \det Q \, (-i)^{N/2} \prod_{k=1}^{N/2} i \gamma_{2k-1} \gamma_{2k} = \det Q \, (-i)^{N/2} \prod_{k=1}^{N/2} \left( 2 \psi_k^\dagger \psi_k - 1 \right) ,
\end{aligned}
\end{equation}
where $P$ is a permutation of $N$ elements and $\psi_k \equiv (\gamma_{2k-1} + i \gamma_{2k})/2$ is a complex fermion. To obtain the third equality, note that when $j_m=j_n$ for some $1 \leq m<n \leq N$, we have $\gamma_{j_m} \gamma_{j_n} = 1$ and $\sum_{j_m} Q_{m j_m} Q_{n j_m} = (Q Q^\mathrm{T})_{mn} = \delta_{mn}=0$. Therefore, the LHS of the third equality is nonzero only when $(j_1,\ldots,j_N)$ is a permutation of $(1,\ldots,N)$. Rearranging $\gamma_{j_1} \ldots \gamma_{j_N}$ to $\gamma_1 \ldots \gamma_N$ incurs a factor of $\mathrm{sgn}(P)$, which is $-1$ ($+1$) for an odd (even) permutation $P$. Substituting \eqref{cproduct} in \eqref{gaugetransformproduct}, we obtain
\begin{equation} \label{tripod}
\prod_{i \in \Lambda} D_i = (-1)^\xi \left( \prod_{\langle ij \rangle_\lambda} u_{ij}^\lambda \right) \, \det Q \prod_{k=1}^{N/2} \left( 2 \psi_k^\dagger \psi_k - 1 \right) \equiv (-1)^\xi \, \mathcal{P}_b \det Q \, \mathcal{P}_\psi , \quad \xi \equiv \xi' + \frac{nN}{2} ,
\end{equation}
which is \eqref{projectproduct} in the main text \footnote{Recall that $N$ is necessarily an even number. Also, we will not bother to derive an exact expression of $\xi$ as it is irrelevant for the mapping of anyons.}. \\

The significance of the above analysis is as follows. $\prod_{i \in \Lambda} D_i$ has the eigenvalues $+1$ and $-1$, so $\mathcal{P}_0$ has the corresponding eigenvalues $1$ and $0$. Let $\lvert \phi \rangle$ be an eigenstate of the quadratic Hamiltonian \eqref{kitaevmodelquadratic}, which is a Fock state of the complex fermions $\psi_k$. From \eqref{tripod}, one finds that $\lvert \phi \rangle$ is also an eigenstate of $\prod_{i \in \Lambda} D_i$ and thus $\mathcal{P}_0$. On the other hand, $\mathcal{S}$ generates different gauge-field configurations within the same flux sector when acting on $\lvert \phi \rangle$. By \eqref{physicalprojectorfactorize}, we deduce that $\mathcal{P} \lvert \phi \rangle$ is nonzero iff $\mathcal{P}_0 \lvert \phi \rangle = \lvert \phi \rangle$, i.e., $\prod_{i \in \Lambda} D_i \lvert \phi \rangle = \lvert \phi \rangle$, which is \eqref{physicalconstraint} in the main text. \\

We further show that, for a fixed set of couplings, two sets of gauge fields that differ by a gauge transformation lead to the same physical requirement on the fermion parity. It suffices to consider the gauge transformation $D_i$ at an arbitrary site $i$, as the product of multiple $D_j$ for different $j$ is equivalent to successive applications of these $D_j$ one at a time. Let $\mathcal{P}_b$ and $Q$ be the bond parity and the canonical transformation for a particular set of gauge fields that produces the flux sector in question. These two quantities specify the physical fermion parity $\mathcal{P}_\psi$ via \eqref{physicalconstraint} and \eqref{projectproduct}. We inspect how they change under $D_i$, which flips the $2n-1$ bonds that are connected to $i$. First, $\mathcal{P}_b$ changes by a minus sign. Since the nonzero elements of $A$ are given by $2 J_\lambda u_{jk}^\lambda$, the $i$th row and the $i$th column of $A$ are multiplied by a minus sign. It follows that the $i$th row of $Q$ is multiplied by a minus sign, so $\det Q$ changes by a minus sign. The two minus signs that arise from the changes of $\mathcal{P}_b$ and $\det Q$ cancel out, which implies the invariance of $\mathcal{P}_\psi$ under $D_i$.

\section{\label{section:canonical}Canonical Transformation}

With the Majorana fermion representation, the generalized Kitaev model \eqref{kitaevmodel} is reduced to a quadratic Hamiltonian \eqref{kitaevmodelquadratic}. Let $N$ be the total number of sites. The fermion excitation spectrum is given by the non-negative eigenvalues $\varepsilon_k \geq 0$ of the hermitian matrix $i A$, which is diagonalized by a unitary matrix $U$,
\begin{equation} \label{eigenvaluematrix}
U^\dagger (i A) U = \begin{pmatrix}
- \varepsilon_1 & 0 & & & \\
0 & \varepsilon_1 & & & \\
& & \ddots & & \\
& & & - \varepsilon_{N/2} & 0 \\
& & & 0 & \varepsilon_{N/2}
\end{pmatrix} = \bigoplus_{k=1}^{N/2} \begin{pmatrix} - \varepsilon_k & 0 \\ 0 & \varepsilon_k \end{pmatrix} , \quad \varepsilon_k \geq 0 .
\end{equation}
The eigenvalues of $iA$ come in positive-negative pairs due to the antisymmetry of $A$. We discuss how to construct the canonical transformation $Q$ from $U$ for a gapped fermion excitation spectrum, i.e., $\varepsilon_k > 0$ for all $k$. Let $\mathbf{v}_k$ be the eigenvector of $i A$ corresponding to the eigenvalue $- \varepsilon_k$,
\begin{equation} \label{eigenvalueequation}
i A \mathbf{v}_k = - \varepsilon_k \mathbf{v}_k .
\end{equation}
Moreover, let $\lbrace \mathbf{v}_1 , \ldots , \mathbf{v}_{N/2} \rbrace$ be an orthonormal set, i.e., $\langle \mathbf{v}_k , \mathbf{v}_{k'} \rangle = \delta_{kk'}$. Complex conjugating both sides of \eqref{eigenvalueequation}, we obtain $- i A \mathbf{v}_k^* = - \varepsilon_k \mathbf{v}_k^*$, or
\begin{equation} \label{eigenvalueequationcomplexconjugate}
i A \mathbf{v}_k^* = \varepsilon_k \mathbf{v}_k^* ,
\end{equation}
which says that $\mathbf{v}_k^*$ is the eigenvector of $i A$ corresponding to the eigenvalue $\varepsilon_k$. Since $\varepsilon_k \neq - \varepsilon_{k'}$ and eigenvectors corresponding to different eigenvalues are orthogonal, we deduce that $\langle \mathbf{v}_k^* , \mathbf{v}_{k'} \rangle = 0$ for any pair of $k$ and $k'$. The columns of the unitary transformation are given by the orthonormalized eigenvectors, i.e., $U=(\mathbf{v}_1 , \mathbf{v}_1^* , \ldots , \mathbf{v}_{N/2} , \mathbf{v}_{N/2}^*)$. We observe that the eigenvalue matrix \eqref{eigenvaluematrix} can be brought into a block diagonal form, with each block being a $2 \times 2$ off-diagonal matrix, by another unitary transformation that is also a block diagonal matrix,
\begin{equation}
\left[ \frac{1}{\sqrt{2}} \bigoplus_{k=1}^{N/2} \begin{pmatrix} 1 & - i \\ 1 & i \end{pmatrix} \right]^\dagger \left[ \bigoplus_{k=1}^{N/2} \begin{pmatrix} - \varepsilon_k & 0 \\ 0 & \varepsilon_k \end{pmatrix} \right] \left[ \frac{1}{\sqrt{2}} \bigoplus_{k=1}^{N/2} \begin{pmatrix} 1 & - i \\ 1 & i \end{pmatrix} \right] = \frac{1}{2} \bigoplus_{k=1}^{N/2} \left[ \begin{pmatrix} 1 & 1 \\ i & - i \end{pmatrix} \begin{pmatrix} - \varepsilon_k & 0 \\ 0 & \varepsilon_k \end{pmatrix} \begin{pmatrix} 1 & - i \\ 1 & i \end{pmatrix} \right] = i \bigoplus_{k=1}^{N/2} \begin{pmatrix} 0 & \varepsilon_k \\ - \varepsilon_k & 0 \end{pmatrix} .
\end{equation}
Therefore, the canonical transformation is given by
\begin{equation} \label{reimcolumn}
\begin{aligned}[b]
Q = U \cdot \frac{1}{\sqrt{2}} \bigoplus_{k=1}^{N/2} \begin{pmatrix} 1 & -i \\ 1 & i \end{pmatrix} &= \left( \frac{\mathbf{v}_1 + \mathbf{v}_1^*}{\sqrt{2}}, \frac{- i \mathbf{v}_1 + i \mathbf{v}_1^*}{\sqrt{2}} , \ldots , \frac{\mathbf{v}_{N/2} + \mathbf{v}_{N/2}^*}{\sqrt{2}}, \frac{- i \mathbf{v}_{N/2} + i \mathbf{v}_{N/2}^*}{\sqrt{2}} \right) \\
&= \sqrt{2} (\mathrm{Re} \, \mathbf{v}_1 , \mathrm{Im} \, \mathbf{v}_1 , \ldots , \mathrm{Re} \, \mathbf{v}_{N/2} , \mathrm{Im} \, \mathbf{v}_{N/2}) .
\end{aligned}
\end{equation}
The columns of $Q$ are the real and imaginary parts of $\mathbf{v}_k$ \cite{KITAEV20062}, with the appropriate normalization factor. $Q$ is apparently a real matrix. To show that $Q$ is an orthogonal matrix, we evaluate
\begin{equation}
Q^\mathrm{T} Q = Q^\dagger Q = \left[ \frac{1}{\sqrt{2}} \bigoplus_{k=1}^{N/2} \begin{pmatrix} 1 & -i \\ 1 & i \end{pmatrix} \right]^\dagger U^\dagger U \left[ \frac{1}{\sqrt{2}} \bigoplus_{k=1}^{N/2} \begin{pmatrix} 1 & -i \\ 1 & i \end{pmatrix} \right] = \frac{1}{2} \bigoplus_{k=1}^{N/2} \left[ \begin{pmatrix} 1 & 1 \\ i & - i \end{pmatrix} \begin{pmatrix} 1 & - i \\ 1 & i \end{pmatrix} \right] = \vmathbb{1}.
\end{equation}
Therefore, $Q \in O (N)$ and $\det Q = \pm 1$. \\

We further investigate the change of $\det Q$ under a unitary transformation $W$ that mixes the negative energy states among themselves but not with the positive energy states, and vice versa. We require the mixings to be consistent with the relation between the positive and negative energy states. We have
\begin{equation} \label{canonicalinvariance}
\begin{aligned}[b]
\det Q &= \det \left[ U \cdot \frac{1}{\sqrt{2}} \bigoplus_{k=1}^{N/2} \begin{pmatrix} 1 & -i \\ 1 & i \end{pmatrix} \right] = \det ( \mathbf{v}_1 , \mathbf{v}_1^* , \ldots , \mathbf{v}_{N/2} , \mathbf{v}_{N/2}^* ) \prod_{k=1}^{N/2} \det \left[ \frac{1}{\sqrt{2}} \begin{pmatrix} 1 & -i \\ 1 & i \end{pmatrix} \right] = i^{N/2} (-1)^{N(N-2)/8} \det \tilde{U} , \\
\tilde{U} &\equiv ( \mathbf{v}_1 , \ldots , \mathbf{v}_{N/2} , \mathbf{v}_1^* , \ldots , \mathbf{v}_{N/2}^* ) ,
\end{aligned}
\end{equation}
where we have rearranged columns of $U$ in the third equality, such that the left (right) half of the resulting matrix $\tilde{U}$ consists of $\mathbf{v}_k$ ($\mathbf{v}_k^*$) in the order of increasing $k$, which yields $N(N-2)/8$ factors of $-1$ for the determinant. The unitary transformation $W$ acts on $\tilde{U}$ as a right multiplication and has a block diagonal form $W = w \oplus w'$, where each block is a $U (N/2)$ matrix. Since the positive energy states \eqref{eigenvalueequationcomplexconjugate} are constructed by complex conjugating the negative energy states \eqref{eigenvalueequation}, we require $w' = w^*$ for consistency. Therefore, $\det \tilde{U}$ transform under $W$ as
\begin{equation} \label{unitaryinvariance}
\det \tilde{U} \longrightarrow \det ( \tilde{U} W ) = \det \tilde{U} \det w \det w^* = \det \tilde {U} \det (w w^\dagger) = \det \tilde{U} .
\end{equation}
This shows the invariance of $\det \tilde{U}$, which implies the invariance of $\det Q$ by \eqref{canonicalinvariance}, under a unitary transformation $W$ that mixes the negative (positive) energy states among themselves but not with the positive (negative) energy states \footnote{A similar argument has been made in Ref.~\cite{PhysRevB.90.134404} to show that, in the strong $J_z$ limit of the Kitaev honeycomb model and in an appropriate basis, the determinant of the unitary transformation that mixes the particle and hole operators upon switching on $J_x$ and $J_y$ perturbatively is unity.}. Since an adiabatic transformation of the system that preserves the fermion gap must result in such a unitary transformation of the energy eigenstates, we are led to the following lemma. \\

\noindent \textbf{Lemma 2.} The determinant of the canonical transformation is invariant under an adiabatic transformation that does not close the fermion gap. \\

Finally, we remark that when the fermion excitation spectrum is gapless, i.e., $\varepsilon_k = 0$ for some $k \in \lbrace 1 , \ldots , N/2 \rbrace$, then one cannot construct $Q \in O (N)$ according to \eqref{reimcolumn}, because $\mathbf{v}_k$ and $\mathbf{v}_k^*$ are no longer guaranteed to be orthogonal.

\section{\label{section:adiabatic}Adiabatic Theorem}

In this section, we prove the following adiabatic theorem. Let $\mathcal{P}$ be the projector to the physical subspace and $\lvert \phi \rangle$ be the instantaneous eigenstate of the quadratic Hamiltonian $H (\lbrace u_{ij} \rbrace)$ with all fermion modes unoccupied. Suppose that $\mathcal{P} \lvert \phi \rangle$ is nonzero and the fermion gap of the corresponding flux sector does not close throughout an adiabatic transformation. If the system is initially in $\mathcal{P} \lvert \phi \rangle$, then it remains in $\mathcal{P} \lvert \phi \rangle$ throughout the adiabatic transformation. Our proof is adapted from the presentation in Ref.~\cite{griffithstextbook} that treats the case of non-degenerate energy levels. \\

Let us parametrize the adiabatic transformation by a time variable $t \geq 0$. With the Majorana fermion representation $\Gamma_i^\lambda = i b_i^\lambda c_i$, the generalized Kitaev model \eqref{kitaevmodel} reads
\begin{equation} \label{timedependhamiltonian}
H ( t ) = i \sum_{\lambda} \sum_{\langle ij \rangle \in \lambda} J_\lambda ( t ) \hat{u}_{ij}^\lambda c_i c_j ,
\end{equation}
where the couplings $J_\lambda ( t )$ are now time-dependent and the hat on each gauge-field operator $\hat{u}_{ij}^\lambda \equiv i b_i^\lambda b_j^\lambda$ is introduced to distinguish it from its eigenvalue $u_{ij}^\lambda = \pm 1$. The instantaneous eigenstates $\lvert \lbrace u_{ij}^\lambda \rbrace , \lbrace n_k \rbrace ; t \rangle$ of \eqref{timedependhamiltonian} are uniquely labeled by the gauge fields $u_{ij}^\lambda$, and, for each gauge-field configuration, the occupancies of the complex fermion modes $n_k$. They form a basis of the enlarged Hilbert space that includes unphysical states. This can be seen by counting the total number of distinct eigenstates, which are orthonormal to each other. Let $N$ be the total number of sites in the system. $(2n - 1) N / 2$ bonds lead to $2^{(2n - 1) N / 2}$ different gauge-field configurations. Each gauge-field configuration further leads to $N/2$ complex fermion modes, each of which can be either occupied or unoccupied, giving $2^{N/2}$ Fock states. $2^{(2n - 1) N / 2} \times 2^{N/2} = 2^{nN}$ is exactly twice as large as the dimension $2^{(n-1)N}$ of the physical Hilbert space. We will work with the enlarged Hilbert space because any physical state can be obtained by an appropriate linear combination of $\lvert \lbrace u_{ij}^\lambda \rbrace , \lbrace n_k \rbrace ; t \rangle$. In particular, the coefficients of the instantaneous eigenstates with unphysical fermion parities should be set to zero. \\

Observe that every $\hat{u}_{ij}^\lambda$ commutes with $H ( t )$ at all times, i.e., $u_{ij}^\lambda$ is a constant of motion or a good quantum number. On the other hand, $n_k$ is in general time-dependent because the energy levels depend on $J_\lambda ( t )$, which changes with time. In the rest of this section, we will suppress the superscript on the gauge field that indicates the bond type, i.e., we will write $u_{ij}^\lambda$ simply as $u_{ij}$. \\

A generic wavefunction $\Psi ( t )$ can be expressed as a linear combination of the instantaneous eigenstates of $H ( t )$,
\begin{equation} \label{genericwavefunction}
\Psi ( t ) = \sum_I a_I ( t ) e^{i \theta_I (t)} \lvert I ; t \rangle , \quad \theta_I (t) \equiv - \frac{1}{\hbar} \int_0^t E_I ( t' ) \, \mathrm{d} t' ,
\end{equation}
where we have introduced the compound index $I \equiv ( \lbrace u_{ij} \rbrace , \lbrace n_k \rbrace )$ and $E_I ( t )$ are the instantaneous eigenvalues of $H (t)$. We have also separated a factor containing the dynamic phase $\theta_I ( t ) \in \mathbb{R}$, which would be present even if the Hamiltonian were time-independent, from the coefficient $a_I ( t ) \in \mathbb{C}$ for convenience. The time evolution of $\Psi ( t )$ is governed by the Schr\"{o}dinger equation
\begin{equation} \label{schrodinger}
i \hbar \frac{\mathrm{d} \Psi ( t )}{\mathrm{d} t} = H ( t ) \Psi ( t ) .
\end{equation}
Substituting \eqref{genericwavefunction} in \eqref{schrodinger}, one finds
\begin{equation} \label{schrodingersubstitute}
\sum_I \frac{\mathrm{d} a_I ( t )}{\mathrm{d} t} e^{i \theta_I (t)} \lvert I ; t \rangle = - \sum_I a_I ( t ) e^{i \theta_I (t)} \frac{\mathrm{d}}{\mathrm{d}t} \lvert I ; t \rangle .
\end{equation}
Multiplying $\langle J ; t \rvert$ to both sides of \eqref{schrodingersubstitute} from the left and collecting the exponential factors on the RHS, one obtains
\begin{equation} \label{timederivativegeneric}
\frac{\mathrm{d} a_J ( t )}{\mathrm{d} t} = - \sum_I a_I ( t ) e^{i [\theta_I (t) - \theta_J (t)]} \Big\langle J ; t \Big\lvert \frac{\mathrm{d}}{\mathrm{d}t} \Big\rvert I ; t \Big\rangle .
\end{equation}
The state $\mathrm{d} \lvert I , t \rangle / \mathrm{d} t$ (with a proper normalization factor) can be expanded in terms of $\lvert I' , t \rangle$ since the latter form a complete set of states at each $t$. Moreover, since $\hat{u}_{ij}$ is time-independent, we have
\begin{equation}
\hat{u}_{ij} \frac{\mathrm{d}}{\mathrm{d}t} \lvert I ; t \rangle = \frac{\mathrm{d}}{\mathrm{d}t} ( \hat{u}_{ij} \lvert I ; t \rangle ) = \frac{\mathrm{d}}{\mathrm{d}t} ( u_{ij} \lvert I ; t \rangle ) = u_{ij} \frac{\mathrm{d}}{\mathrm{d}t} \lvert I ; t \rangle
\end{equation}
for every bond $\langle ij \rangle$. This shows that $\mathrm{d} \lvert I ; t \rangle / \mathrm{d} t$ has a definite gauge-field configuration, which is same as that of $\lvert I ; t \rangle$, as expected from the conservation of $u_{ij}$. In other words, $\mathrm{d} \lvert I ; t \rangle / \mathrm{d} t$ has a finite $\lvert I' ; t \rangle$ component if and only if $I'$ labels the same gauge-field configuration as $I$. It follows that $\langle J ; t \lvert (\mathrm{d}/\mathrm{d}t) \rvert I ; t \rangle$ is nonzero if and only if $I$ labels the same gauge-field configuration as $J$. Specializing to $J = ( \lbrace u_{ij} \rbrace , \lbrace \mathbf{0} \rbrace )$, where $\lbrace u_{ij} \rbrace$ is a set of gauge fields that produces the flux sector in question and $\lbrace \mathbf{0} \rbrace$ denote the set in which $n_k = 0$ for all $k$, \eqref{timederivativegeneric} then reads
\begin{equation} \label{timederivativespecific}
\frac{\mathrm{d} a_{\lbrace u_{ij} \rbrace , \lbrace \mathbf{0} \rbrace} (t)}{\mathrm{d}t} = - \sum_{\lbrace n_k \rbrace} a_{\lbrace u_{ij} \rbrace , \lbrace n_k \rbrace} ( t ) \exp \left[ i \theta_{\lbrace u_{ij} \rbrace , \lbrace n_k \rbrace} (t) - i \theta_{\lbrace u_{ij} \rbrace , \lbrace \mathbf{0} \rbrace} (t) \right] \Big\langle \lbrace u_{ij} \rbrace , \lbrace \mathbf{0} \rbrace ; t \Big\lvert \frac{\mathrm{d}}{\mathrm{d}t} \Big\rvert \lbrace u_{ij} \rbrace , \lbrace n_k \rbrace ; t \Big\rangle .
\end{equation}
Note the absence of a sum over different gauge-field configurations on the RHS of \eqref{timederivativespecific}. Let $J' = ( \lbrace u_{ij} \rbrace , \lbrace n_k \rbrace \neq \lbrace \mathbf{0} \rbrace )$. Differentiating both sides of $H (t) \lvert J' ; t \rangle = E_{J'} (t) \lvert J' ; t \rangle$ with respect to $t$, one finds
\begin{equation} \label{timederivativeeigenstate}
\frac{\mathrm{d} H ( t )}{\mathrm{d}t} \lvert J' ; t \rangle + H ( t ) \frac{\mathrm{d}}{\mathrm{d}t} \lvert J' ; t \rangle = \frac{\mathrm{d} E_{J'} ( t )}{\mathrm{d}t} \lvert J' ; t \rangle + E_{J'} ( t ) \frac{\mathrm{d}}{\mathrm{d}t} \lvert J' ; t \rangle .
\end{equation}
Multiplying $\langle J ; t \rvert$ to both sides of \eqref{timederivativeeigenstate} from the left, one obtains
\begin{equation} \label{adiabaticapproximation}
\Big\langle J ; t \Big\lvert \frac{\mathrm{d}}{\mathrm{d}t} \Big\rvert J' ; t \Big\rangle = \Big\langle J ; t \Big\lvert \frac{\mathrm{d}H ( t ) / \mathrm{d}t}{E_{J'} (t) - E_J (t)} \Big\rvert J' ; t \Big\rangle \approx 0 ,
\end{equation}
where $E_{J'} (t) - E_{J} (t)$ is always finite by assumption and the last step follows from the adiabatic approximation, by which the change of $H ( t )$ is made slow enough such that the matrix elements of its time derivative are much smaller than the fermion gap at all times. Substituting \eqref{adiabaticapproximation} in \eqref{timederivativespecific}, one finds
\begin{equation}
\frac{\mathrm{d} a_{\lbrace u_{ij} \rbrace , \lbrace \mathbf{0} \rbrace} (t)}{\mathrm{d}t} = - a_{\lbrace u_{ij} \rbrace , \lbrace \mathbf{0} \rbrace} (t) \, \Big\langle \lbrace u_{ij} \rbrace , \lbrace \mathbf{0} \rbrace ; t \Big\lvert \frac{\mathrm{d}}{\mathrm{d}t} \Big\rvert \lbrace u_{ij} \rbrace , \lbrace \mathbf{0} \rbrace ; t \Big\rangle .
\end{equation}
Solving the above differential equation yields
\begin{equation} \label{timedependcoefficient}
a_{\lbrace u_{ij} \rbrace , \lbrace \mathbf{0} \rbrace} (t) = a_{\lbrace u_{ij} \rbrace , \lbrace \mathbf{0} \rbrace} (0) \exp \left[ i \gamma_{\lbrace u_{ij} \rbrace , \lbrace \mathbf{0} \rbrace} (t) \right] , \quad \gamma_{\lbrace u_{ij} \rbrace , \lbrace \mathbf{0} \rbrace} (t) \equiv i \int_0^t \Big\langle \lbrace u_{ij} \rbrace , \lbrace \mathbf{0} \rbrace ; t' \Big\lvert \frac{\mathrm{d}}{\mathrm{d} t'} \Big\rvert \lbrace u_{ij} \rbrace , \lbrace \mathbf{0} \rbrace ; t' \Big\rangle \, \mathrm{d} t' .
\end{equation}
The moral is that the degenerate states labeled by distinct good quantum numbers evolve independently of each other, as well as other states outside the degenerate subspace, under an adiabatic transformation, as long as they are separated from the rest of the spectrum by a finite gap. $\gamma_{\lbrace u_{ij} \rbrace , \lbrace \mathbf{0} \rbrace} (t)$ is known as the geometric phase. \\

Let $\lbrace u_{ij} \rbrace_1$ be a particular set of gauge fields that produces the flux sector of interest, the physical subspace of which requires an even number of matter fermions. Applying the projector $\mathcal{P}$ to the state $\lvert \lbrace u_{ij} \rbrace_1 , \lbrace \mathbf{0} \rbrace ; 0 \rangle$ generates $2^{N-1}$ states $\lvert \lbrace u_{ij} \rbrace_k , \lbrace \mathbf{0} \rbrace ; 0 \rangle$ with different gauge-field configurations that produce the same flux sector, see \eqref{physicalprojectorfactorize} and related discussions in Sec.~\ref{section:constraint}. Suppose that the system is initially prepared in the properly normalized wavefunction
\begin{equation} \label{initialwavefunction}
\Psi ( 0 ) = \sqrt{2^{N-1}} \, \mathcal{P} \lvert \lbrace u_{ij} \rbrace_1 , \lbrace \mathbf{0} \rbrace ; 0 \rangle =  \frac{1}{\sqrt{2^{N-1}}} \sum_{k=1}^{2^{N-1}} \lvert \lbrace u_{ij} \rbrace_k , \lbrace \mathbf{0} \rbrace ; 0 \rangle .
\end{equation}
According to \eqref{genericwavefunction} and \eqref{timedependcoefficient}, the wavefunction at any time $t>0$ is then
\begin{equation} \label{finalwavefunction}
\begin{aligned}[b]
\Psi ( t ) &= \frac{1}{\sqrt{2^{N-1}}} \sum_{k=1}^{2^{N-1}} \exp\left[ i \theta_{\lbrace u_{ij} \rbrace_k , \lbrace \mathbf{0} \rbrace} (t) \right] \exp \left[ i \gamma_{\lbrace u_{ij} \rbrace_k , \lbrace \mathbf{0} \rbrace} (t) \right] \lvert \lbrace u_{ij} \rbrace_k , \lbrace \mathbf{0} \rbrace ; t \rangle \\
&= \exp\left[ i \theta_{\lbrace u_{ij} \rbrace_1 , \lbrace \mathbf{0} \rbrace} (t) \right] \exp \left[ i \gamma_{\lbrace u_{ij} \rbrace_1 , \lbrace \mathbf{0} \rbrace} (t) \right] \left( \frac{1}{\sqrt{2^{N-1}}} \sum_{k=1}^{2^{N-1}} \lvert \lbrace u_{ij} \rbrace_k , \lbrace \mathbf{0} \rbrace ; t \rangle \right) \\
&= \exp\left[ i \theta_{\lbrace u_{ij} \rbrace_1 , \lbrace \mathbf{0} \rbrace} (t) \right] \exp \left[ i \gamma_{\lbrace u_{ij} \rbrace_1 , \lbrace \mathbf{0} \rbrace} (t) \right] \left( \sqrt{2^{N-1}} \, \mathcal{P} \lvert \lbrace u_{ij} \rbrace_1 , \lbrace \mathbf{0} \rbrace ; t \rangle \right),
\end{aligned}
\end{equation}
where we have used the fact that the dynamic and geometric phases are uniform across the $2^{N-1}$ different gauge-field configurations in the second equality, which are explained as follows. At each $t$, the fermion spectrum depends only on the gauge fluxes. For each $k$, $\lbrace u_{ij} \rbrace_k$ produces the same flux sector as $\lbrace u_{ij} \rbrace_1$, so the energies $E_{\lbrace u_{ij} \rbrace_k , \lbrace \mathbf{0} \rbrace} (t)$ and $E_{\lbrace u_{ij} \rbrace_1 , \lbrace \mathbf{0} \rbrace} (t)$ are equal. Since $\theta_{\lbrace u_{ij} \rbrace_k , \lbrace \mathbf{0} \rbrace} (t)$ is defined as the time integral of $E_{\lbrace u_{ij} \rbrace_k , \lbrace \mathbf{0} \rbrace} (t')$ up to $t'=t$ [see \eqref{genericwavefunction}], we have $\theta_{\lbrace u_{ij} \rbrace_k , \lbrace \mathbf{0} \rbrace} (t) = \theta_{\lbrace u_{ij} \rbrace_1 , \lbrace \mathbf{0} \rbrace} (t)$. For each $k$, we further observe that $\lvert \lbrace u_{ij} \rbrace_k , \lbrace 0 \rbrace ; t \rangle$ is obtained from $\lvert \lbrace u_{ij} \rbrace_1 , \lbrace 0 \rbrace ; t \rangle$ by applying a gauge transformation to the latter. It suffices to consider the gauge transformation $D_r$ at an arbitrary site $r$. Without loss of generality, let $\lvert \lbrace u_{ij} \rbrace_2 , \lbrace \mathbf{0} \rbrace ; t \rangle = D_r \lvert \lbrace u_{ij} \rbrace_1 , \lbrace \mathbf{0} \rbrace ; t \rangle$. Using $D_r^\dagger = D_r$, $D_r^2 = 1$, and the fact that $D_r$ is time-independent, we evaluate
\begin{equation}
\begin{aligned}[b]
\Big\langle \lbrace u_{ij} \rbrace_2 , \lbrace \mathbf{0} \rbrace ; t \Big\lvert \frac{\mathrm{d}}{\mathrm{d} t} \Big\rvert \lbrace u_{ij} \rbrace_2 , \lbrace \mathbf{0} \rbrace ; t \Big\rangle &= \Big\langle \lbrace u_{ij} \rbrace_1 , \lbrace \mathbf{0} \rbrace ; t \Big\lvert D_r \frac{\mathrm{d}}{\mathrm{d} t} D_r \Big\rvert \lbrace u_{ij} \rbrace_1 , \lbrace \mathbf{0} \rbrace ; t \Big\rangle \\
&=\Big\langle \lbrace u_{ij} \rbrace_1 , \lbrace \mathbf{0} \rbrace ; t \Big\lvert D_r^2 \frac{\mathrm{d}}{\mathrm{d} t} \Big\rvert \lbrace u_{ij} \rbrace_1 , \lbrace \mathbf{0} \rbrace ; t \Big\rangle = \Big\langle \lbrace u_{ij} \rbrace_1 , \lbrace \mathbf{0} \rbrace ; t \Big\lvert \frac{\mathrm{d}}{\mathrm{d} t} \Big\rvert \lbrace u_{ij} \rbrace_1 , \lbrace \mathbf{0} \rbrace ; t \Big\rangle ,
\end{aligned}
\end{equation}
which holds for all $t$. From the definition of the geometric phase in \eqref{timedependcoefficient}, we have $\gamma_{\lbrace u_{ij} \rbrace_2 , \lbrace \mathbf{0} \rbrace} (t) = \gamma_{\lbrace u_{ij} \rbrace_1 , \lbrace \mathbf{0} \rbrace} (t)$, which further implies $\gamma_{\lbrace u_{ij} \rbrace_k , \lbrace \mathbf{0} \rbrace} (t) = \gamma_{\lbrace u_{ij} \rbrace_1 , \lbrace \mathbf{0} \rbrace} (t)$ for all $k$. From \eqref{initialwavefunction} and \eqref{finalwavefunction}, we conclude that if the system is initially prepared in $\mathcal{P} \lvert \lbrace u_{ij} \rbrace_1 , \lbrace \mathbf{0} \rbrace ; 0 \rangle$ (with a proper normalization constant), then it remains in $\mathcal{P} \lvert \lbrace u_{ij} \rbrace_1 , \lbrace \mathbf{0} \rbrace ; t \rangle$ at $t>0$ up to an overall phase factor. This completes the proof.

\section{\label{section:perturb}Degenerate Perturbation Theory}

In this section, we review the application of degenerate perturbation theory \cite{PhysRevB.69.064404,KITAEV20062,strongcoupleexpand,PhysRevB.99.104408} in any dimer limit of any generalized Kitaev model and discuss its limitations. \\

Consider the generalized Kitaev model on a graph of coordination number $z=2n-1$, $n \geq 2$ in the strong $J_\lambda$ limit, where the $\lambda$ bonds form a dimer covering of the graph and $J_{\lambda'} \ll J_\lambda$ for every $\lambda' \neq \lambda$. We rewrite the Hamiltonian as the sum of strong and weak interactions,
\begin{equation} \label{modelsplit}
H = H_0 + V , \quad H_0 = - \sum_{\langle ij \rangle \in \lambda} J_\lambda \Gamma_i^\lambda \Gamma_j^\lambda , \quad V = - \sum_{\lambda' \neq \lambda} \sum_{\langle ij \rangle \in \lambda'} J_{\lambda'} \Gamma_i^{\lambda'} \Gamma_j^{\lambda'} .
\end{equation}
In the ground-state subspace of $H_0$, each dimer has a degeneracy of $(2^{n-1})^2/2 = 2^{2n-3}$, so the total degeneracy of all dimers is $2^{(2n-3)N/2}$, which is exponential in the system size. $V$ is treated as a perturbation to $H_0$. Using degenerate perturbation theory, one can derive an effective Hamiltonian $H_\mathrm{eff}$ that encodes the tunneling processes between orthogonal ground states of $H_0$ induced by $V$. The $k$th order contribution to $H_\mathrm{eff}$ is given by \cite{KITAEV20062,PhysRevB.99.104408}
\begin{equation} \label{contributeorder}
H_\mathrm{eff}^{(k)} = \mathcal{P} V \left( \mathcal{Q} \frac{1}{E_0 - H_0} \mathcal{Q} V \right)^{k-1} \mathcal{P} ,
\end{equation}
where $\mathcal{P}$ is the projector to the ground-state subspace of $H_0$, $\mathcal{Q} \equiv \vmathbb{1} - \mathcal{P}$ is the complement of $\mathcal{P}$, and $E_0$ is the ground-state energy. In the language of Green function formalism, where the Green function is given by $G ( E ) = [ E - E_0 - \Sigma ( E ) ]^{-1}$, the approximation of the self-energy $\Sigma ( E )$ by $\Sigma ( E_0 )$ has been made to obtain the expression \eqref{contributeorder}, see Ref.~\cite{KITAEV20062} for details. We refer to such a treatment as the \textit{standard} degenerate perturbation theory when necessary, to distinguish it from other variants. Degenerate perturbation theory is usually carried out to some finite order, and the sum of $H_\mathrm{eff}^{(k)}$ over $k$ in $H_\mathrm{eff}$ is cut off at a sufficiently large $k$. \\

Several limitations of degenerate perturbation theory are listed as follows. First, it is sensitive to the geometric details of the graph and its dimerization, i.e., the distribution of the strong bonds. Whenever one encounters a new graph or a new dimerization on the same graph, one has to apply degenerate perturbation theory order by order and ensure that all the relevant plaquette operators are accounted for, which is often a laborious undertaking. For graphs with no simple repeating structure such as an amorphous solid \cite{s41467-023-42105-9}, its application is simply infeasible. Second, the effective Hamiltonian derived from the standard degenerate perturbation theory in the dimer limit can never contain an operator that acts nontrivially only on an odd-length elementary plaquette. Other variants of degenerate perturbation theory similarly find that the unit triangles on the star lattice \cite{PhysRevB.78.125102,PhysRevB.81.104429} and the unit polygons on the regular hyperbolic tiling $\lbrace p , 3 \rbrace$ with odd $p$ \cite{PhysRevLett.134.256604,mx1t-74dm} only appear in pairs, rather than individually, at the lowest nontrivial orders. This leads to difficulties in interpreting the effective Hamiltonian as a toric code model and determining the anyon species of some vortices, when the graph contains one or more odd-length elementary plaquettes. \\

In the rest of this section, we prove that the effective Hamiltonian $H_\mathrm{eff}$ defined in terms of \eqref{contributeorder} cannot contain an operator that acts nontrivially only on an odd-length elementary plaquette in the dimer limit. To begin with, we observe that $H_\mathrm{eff}$ ought to be hermitian because it is an operator corresponding to energy. We verify this explicitly by showing that $H_\mathrm{eff}^{(k)}$ is hermitian for every $k$,
\begin{equation}
H_\mathrm{eff}^{(k) \dagger} = \mathcal{P}^\dagger \left[ V^\dagger \mathcal{Q}^\dagger \left( \frac{1}{E_0 - H_0} \right )^\dagger \mathcal{Q}^\dagger \right]^{k-1} V^\dagger \mathcal{P}^\dagger = \mathcal{P} \left( V \mathcal{Q} \frac{1}{E_0 - H_0} \mathcal{Q} \right)^{k-1} V \mathcal{P} = \mathcal{P} V \left( \mathcal{Q} \frac{1}{E_0 - H_0} \mathcal{Q} V \right)^{k-1} \mathcal{P} = H_\mathrm{eff}^{(k)} ,
\end{equation}
where we have used the hermicities of $H_0$, $V$, $\mathcal{P}$, and $\mathcal{Q}$. Expanding the bracket in \eqref{contributeorder}, one finds that a potentially finite contribution to $H_\mathrm{eff}^{(k)}$ has the form
\begin{equation} \label{weakproduct}
c \, \mathcal{P} \hat{V}_{i_k j_k} \hat{V}_{i_{k-1} j_{k-1}} \ldots \hat{V}_{i_2 j_2} \hat{V}_{i_1 j_1} \mathcal{P} , \quad c \in \mathbb{R} ,
\end{equation}
where $\hat{V}_{ij} \equiv \Gamma_i^{\lambda'} \Gamma_j^{\lambda'}$ for a weak bond $\langle ij \rangle_{\lambda'}$, and the coefficient $c \sim J_{\lambda'}^k / J_\lambda^{k-1}$ depends on the couplings. Moreover, the bond operators in \eqref{weakproduct} are restricted such that for every ket $\lvert \psi \rangle$ in the ground-state subspace, $\hat{V}_{i_l j_l} \ldots \hat{V}_{i_1 j_1} \lvert \psi \rangle$ is an excited state for every $l<k$, while $\hat{V}_{i_k j_k} \ldots \hat{V}_{i_1 j_1} \lvert \psi \rangle$ is one of the degenerate ground states. We also allow $\langle i_l j_l \rangle = \langle i_{l'} j_{l'} \rangle$ for $l \neq l'$. \\

\begin{figure}
\subfloat[]{\label{figure:perturbhigh}
\includegraphics[scale=0.2]{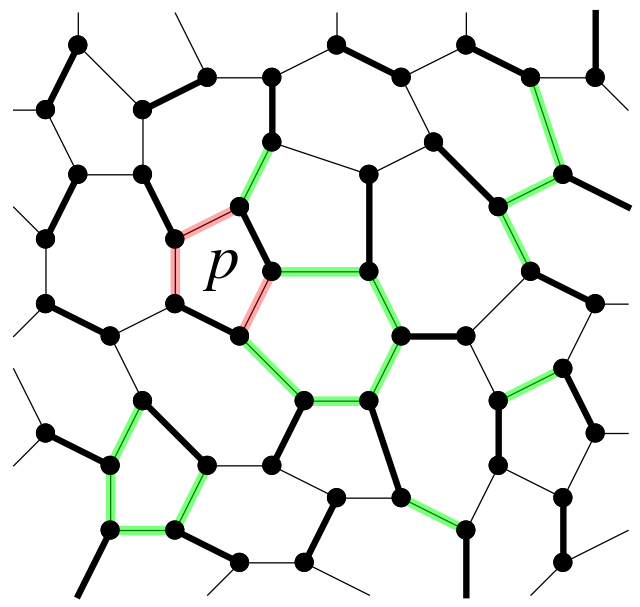}} \hspace{25pt}
\subfloat[]{\label{figure:perturbsingle}
\includegraphics[scale=0.2]{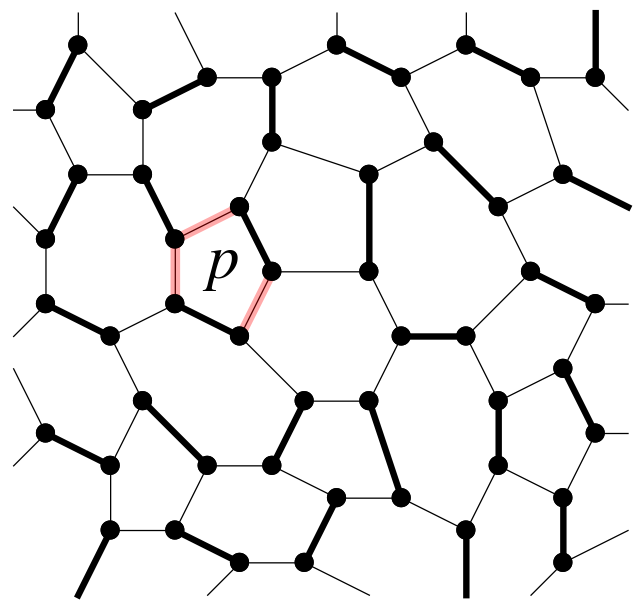}} \hspace{25pt}
\subfloat[]{\label{figure:perturbdouble}
\includegraphics[scale=0.2]{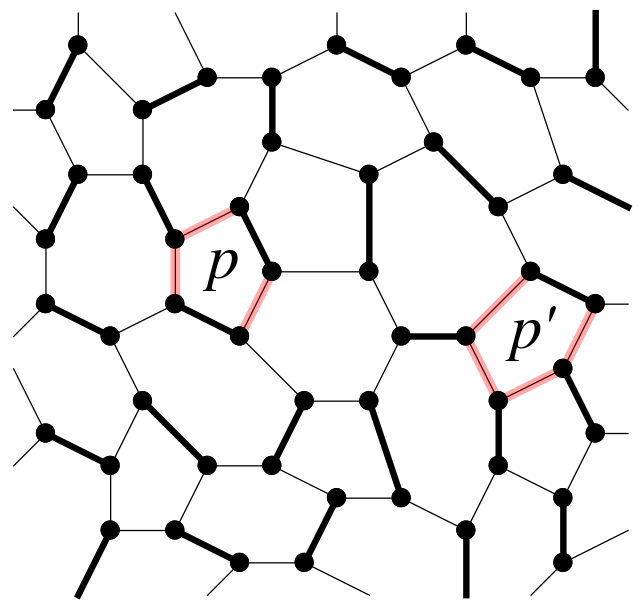}}
\caption{(a) Example of an operator that results from the standard degenerate perturbation theory in a dimer limit and acts nontrivially only on the odd-length elementary plaquette $p$ with $\lvert \partial p \rvert = 5$, see \eqref{weakproduct}. The strong (weak) bonds are indicated by thick (thin) black lines. Each weak bond highlighted by red (green) indicates that the corresponding interaction is applied an odd number of times (an even number of times greater than $0$). (b) For operators such as (a) that act nontrivially only on the elementary plaquette $p$, it sufficies to consider the operator in which the interaction on every weak bond in $p$ is applied exactly once and any other interaction is not applied at all, exploiting the fact that each local $\Gamma$ matrix squares to the identity, see \eqref{weakproductreduce}. (c) Example of an operator that acts nontrivially on two disconnected odd-length elementary plaquettes $p$ and $p'$.}
\end{figure}

For an operator of the form \eqref{weakproduct} that acts nontrivially only on the local states at the sites of an odd-length elementary plaquette (see Fig.~\ref{figure:perturbhigh} for example), we show that its coefficient must be zero. To simplify the discussions, we call $i$ the site index and $\lambda'$ the bond index of $\Gamma_i^{\lambda'}$. We also denote the number of strong or weak bonds in an elementary plaquette $p$ by $n_s (\partial p)$ or $n_w (\partial p)$, respectively. First, note that the bond operator satisfies $\hat{V}_{ij}^2 = 1$. If the same bond operator appears more than once in \eqref{weakproduct}, every two copies can be brought together and removed from the product \footnote{A minus sign may arise from moving bond operators past others, but it will preserve the realness of the overall coefficient.}. Furthermore, \eqref{weakproduct} acts trivially on the local state at a site $i$ iff each distinct $\hat{V}_{ij}$ with one end at $i$ is applied an even number (including $0$) of times. These observations imply that, for the operator in question, it is sufficient to consider
\begin{equation} \label{weakproductreduce}
\hat{O}_p = \mathcal{P} \left( \prod_{\substack{\langle ij \rangle_{\lambda'} \in \partial p \\ \lambda' \neq \lambda}} \hat{V}_{ij} \right) \mathcal{P} ,
\end{equation}
where $p$ labels the elementary plaquette and each weak-bond operator in $p$ appears once and only once in the product (see Fig.~\ref{figure:perturbsingle} for example). The order of the perturbation is equal to the number of weak bonds in the elementary plaquette plus an even number, i.e., $k = n_w (\partial p) + 2 q , q \geq 0$. Next, we group the $\Gamma$ matrices in \eqref{weakproductreduce} according to their site indices. A site $i$ in $p$ is the point of contact between (i) two weak bonds in $p$ or (ii) a weak bond and a strong bond both in $p$. In the former case, there are two $\Gamma$ matrices with the site index $i$ (but different bond indices $\mu$ and $\nu$). In the latter case, there is only one $\Gamma$ matrix with the site index $i$ (and, say, the bond index $\kappa$). Since each dimer in $p$ contributes two sites to $p$, we have $2 n_s (\partial p)$ of site indices that appear once in \eqref{weakproductreduce}. The number of (unique) site indices that appear twice in \eqref{weakproductreduce} is then $\lvert \partial p \rvert - 2 n_s (\partial p)$, which is odd when $\lvert \partial p \rvert$ is odd. Equation \eqref{weakproductreduce} can thus be arranged into the form
\begin{equation} \label{weakproductsite}
\hat{O}_p = \mathcal{P} \left( \prod_{l=1}^{2 n_s (\partial p)} \Gamma_{i_l}^{\kappa_l} \right) \left( \prod_{l=1}^{\lvert \partial p \rvert - 2 n_s (\partial p)} \Gamma_{j_l}^{\mu_l} \Gamma_{j_l}^{\nu_l} \right) \mathcal{P} ,
\end{equation}
where the sites $i_1, \ldots, i_{2 n_s (\partial p)} , j_1 , \ldots , j_{\lvert \partial p \rvert - 2 n_s (\partial p)}$ are distinct from each other. We are now ready to show that $\hat{O}_p$ cannot be multiplied by a finite coefficient. Suppose that $H_\mathrm{eff}$ contains the term $c \, \hat{O}_p$, where the coefficient $c$ is necessarily real, see \eqref{weakproduct}. We have argued that the effective Hamiltonian is hermitian, i.e., $H_\mathrm{eff}^\dagger = H_\mathrm{eff}$. Taking the hermitian conjugation of \eqref{weakproductsite}, we find
\begin{equation}
\hat{O}_p^\dagger = \mathcal{P} \left( \prod_{l=1}^{2 n_s (\partial p)} \Gamma_{i_l}^{\kappa_l} \right) \left( \prod_{l=1}^{\lvert \partial p \rvert - 2 n_s (\partial p)} \Gamma_{j_l}^{\nu_l} \Gamma_{j_l}^{\mu_l} \right) \mathcal{P} = (-1)^{\lvert \partial p \rvert - 2 n_s (\partial p)} \mathcal{P} \left( \prod_{l=1}^{2 n_s (\partial p)} \Gamma_{i_l}^{\kappa_l} \right) \left( \prod_{l=1}^{\lvert \partial p \rvert - 2 n_s (\partial p)} \Gamma_{j_l}^{\mu_l} \Gamma_{j_l}^{\nu_l} \right) \mathcal{P} = - \hat{O}_p ,
\end{equation}
which follows from $P^\dagger = P$, $\Gamma_i^{\kappa \dagger} = \Gamma_i^{\kappa}$, $(\Gamma_j^\mu \Gamma_j^\nu)^\dagger = \Gamma_j^\nu \Gamma_j^\mu = - \Gamma_j^\mu \Gamma_j^\nu$, and the fact that $\Gamma$ matrices with different site indices commute with each other. Therefore, $c$ must be $0$. We are led to the following lemma. \\

\noindent \textbf{Lemma 0.} The effective Hamiltonian, which is derived from the standard degenerate perturbation theory in any dimer limit of any generalized Kitaev model, cannot contain an operator that acts nontrivially only on the local states at the sites of an odd-length elementary plaquette. \\

To put it more concisely, one cannot have an odd-length elementary-plaquette operator in $H_\mathrm{eff}$. As pointed out earlier, this creates obstructions in interpreting $H_\mathrm{eff}$ as a toric code model and determining the anyon species of some vortices, when the graph contains odd-length elementary plaquettes. Whether these challenges can be overcome within the framework of degenerate perturbation theory is unclear. Possible solutions include studying operators that act nontrivially on an even number of disconnected odd-length elementary plaquettes (see Fig.~\ref{figure:perturbdouble} for example), the coefficients of which are not forced to zero by Lemma 0, and combining two or more operators that appear in $H_\mathrm{eff}$ to obtain the desired odd-length elementary-plaquette operator. In any case, the order of the perturbation will necessarily be larger than the number of weak bonds in the elementary plaquette $p$, i.e.,~$k > n_w ( \partial p )$, for $\lvert \partial p \rvert \, \mathrm{mod} \, 2 = 1$. This is in contrast to the case of $\lvert \partial p \rvert \, \mathrm{mod} \, 2 = 0$, e.g., the Kitaev honeycomb and square-octagon models, where one obtains the elementary-plaquette operator at $k = n_w ( \partial p )$ with a finite coefficient. Alternatively, one may directly analyze elementary-plaquette operators made up of the interactions on the weak bonds (see Fig.~\ref{figure:perturbsingle} for example), without attempting to interpret $H_\mathrm{eff}$. In spite of this, determining the anyon species for elementary-plaquette operators expressed in terms of various $\Gamma$ matrices can be challenging, especially for $n \geq 3$, cf.~the ``star'' and ``plaquette'' operators in the original toric code model \cite{KITAEV20032}, which consist purely of $x$ and $z$ Pauli matrices, respectively, and correspond apparently to two distinct classes. \\

Finally, we remark that the diagrammatic perturbation theory of Ref.~\cite{PhysRevB.90.134404}, which is formulated with Majorana fermions instead of the original spins, also leads to a conclusion similar to Lemma 0. Although results from the diagrammatic and standard perturbation theories do agree in several instances \cite{PhysRevB.90.134404}, it is unclear whether these methods are formally equivalent.

\section{\label{section:lieb}Ground-State Flux Sector}

Consider a generalized Kitaev model \eqref{kitaevmodel} defined on some lattice. Lieb's theorem \cite{PhysRevLett.73.2158,Macris1996} states that if the lattice is bipartite and, for each elementary plaquette $p$, if there exists a reflection symmetry $M$ of the model such that the mirror plane does not contain any site and $p$ is mapped to itself under $M$, then the ground-state energy is minimized by a $\pi$ ($0$) flux through $p$ for $\lvert \partial p \rvert \, \mathrm{mod} \, 4 = 0$ ($2$). While not rigorously proven, numerical evidence also points to an extended flux phase conjecture \cite{s41467-023-42105-9} that covers Lieb's theorem and applies to any graph. According to this conjecture, the ground-state flux sector is such that $W_p = - (\pm i)^{\lvert \partial p \rvert}$ for each elementary plaquette $p$, where the sign of $i$ inside the bracket is a global choice, which reflects a two-fold degenerate ground state when $\lvert \partial p \rvert$ is odd for some $p$. \\

\begin{figure}
\subfloat[]{\label{figure:honeycombgauge}
\includegraphics[scale=0.2]{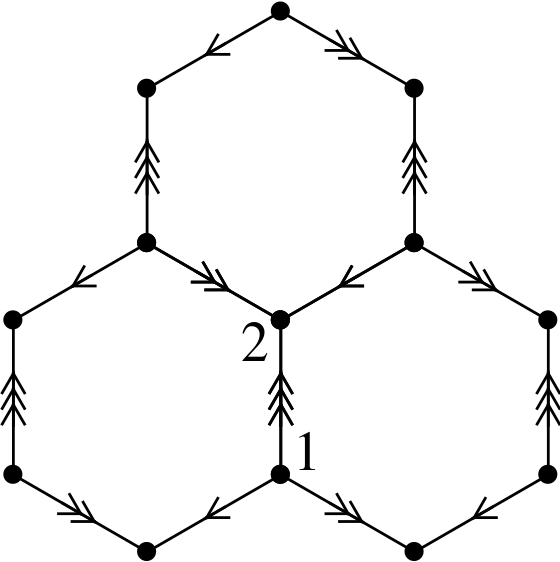}} \hspace{25pt}
\subfloat[]{\label{figure:octagongauge}
\includegraphics[scale=0.2]{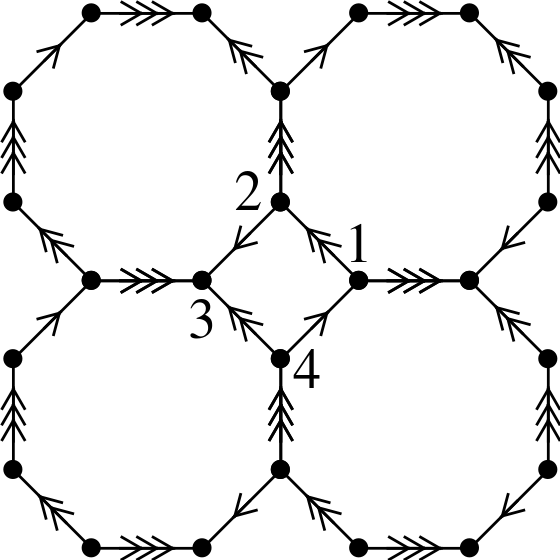}} \hspace{25pt}
\subfloat[]{\label{figure:stargauge}
\includegraphics[scale=0.2]{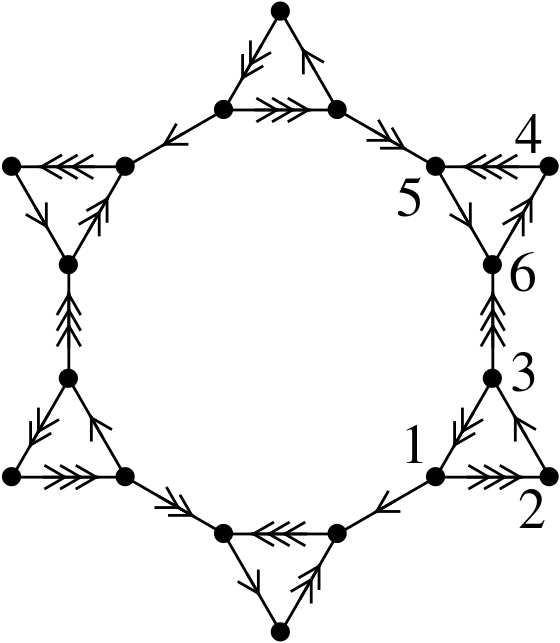}} \hspace{25pt}
\subfloat[]{\label{figure:kekulegauge}
\includegraphics[scale=0.2]{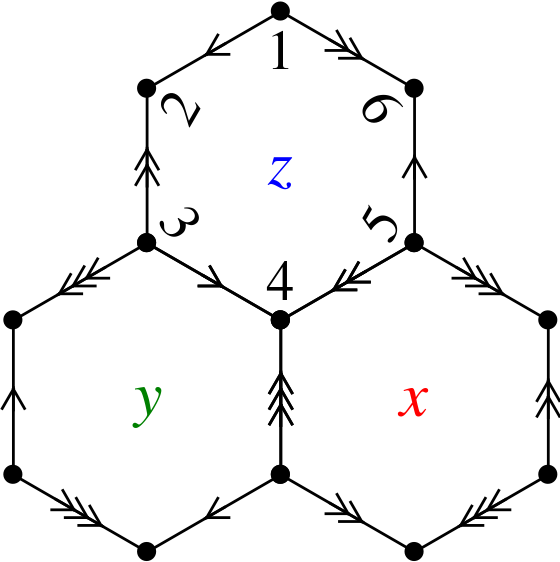}}
\caption{Translationally invariant gauge-field configurations that produce the ground-state flux sectors of the Kitaev (a) honeycomb, (b) square-octagon, (c) star, and (d) Kekul\'{e} models. The single, double, and triple arrows represent the couplings $J_x$, $J_y$, and $J_z$ respectively. On each bond, the gauge field is $u_{ij}^\lambda = +1$ ($-1$) along (against) the direction of the arrow(s). The integers represent distinct sublattices. In (d), $x$, $y$, and $z$ label the three distinct types of unit hexagons in the Kitaev Kekul\'{e} model.}
\end{figure}

For the Kitaev square-octagon model, the lattice is bipartite and one can find the required reflection symmetries for the application of Lieb's theorem by adequate translations of $M_+$ or $M_-$, see Fig.~\ref{figure:octagonmodel}. Since each elementary plaquette $p$ is either a unit square of length $4$ or a unit octagon of length $8$, we always have $\lvert \partial p \rvert \, \mathrm{mod} \, 4 = 0$. Lieb's theorem implies that every elementary plaquette has $\pi$-flux, i.e., $W_p=-1$ for all $p$, in the ground state. \\

The Kitaev honeycomb and Kekul\'{e} models are both defined on the honeycomb lattice, which is bipartite. Different colorings of the bonds lead to different symmetries in these models, see Figs.~\ref{figure:honeycombmodel} and \ref{figure:kekulemodel}. In the Kitaev Kekul\'{e} model, one can find the required reflection symmetries for the application of Lieb's theorem by adequate translations of $M_x$. Since each elementary plaquette $p$ is a unit hexagon of length $6$, we have $\lvert \partial p \rvert \, \mathrm{mod} \, 4 = 2$. Lieb's theorem implies that every unit hexagon has $0$-flux, i.e., $W_p=+1$ for all $p$, in the ground state. In contrast, the Kitaev honeycomb model does not have a reflection symmetry unless two of the three couplings are equal, so Lieb's theorem is inapplicable for generic values of $J_x$, $J_y$, and $J_z$. Nevertheless, a comparison between different flux sectors finds that the energy is lowest when every unit hexagon has $0$-flux, independent of what the couplings are \cite{KITAEV20062}. We assume that such a uniform flux sector indeed yields the ground state. \\

For the Kitaev star model (see Fig.~\ref{figure:starmodel}), the lattice is non-bipartite, so Lieb's theorem is inapplicable. Earlier numerical studies \cite{PhysRevLett.99.247203,PhysRevLett.115.087203} indicated that $W_p=-1$ for every unit dodecagon $p$ and $W_{p'} = \pm i$ for every unit triangle $p'$, where the sign of $i$ is uniform across all unit triangles, in the ground state. Such a pair of degenerate flux sectors, which are related by time reversal, is consistent with the extended flux phase conjecture mentioned earlier in this section. For concreteness, we choose the ground-state flux sector where every unit triangle has $\pi/2$-flux (i.e., $W_{p'} = +i$) for the calculations in this work (see Sec.~\ref{section:addresult}). \\

For each of the aforementioned models, a translationally invariant gauge-field configuration $\lbrace u_{ij}^\lambda \rbrace$ that produces the ground-state flux sector is shown in one of Figs.~\ref{figure:honeycombgauge}-\ref{figure:kekulegauge}. This allows us to compute the fermion gap $\Delta_\psi$, as well as the Chern number $\nu$ where it is well-defined, in the thermodynamic limit via a Fourier transform. We use a uniform grid of $400 \times 400$ $\mathbf{q}$ points in the first Brillouin zone and identify $\min_\mathbf{q} \varepsilon ( \mathbf{q} )$ as $\Delta_\psi$, which is plotted over the triangular parameter space in Fig.~\ref{figure:octagondata}b, \ref{figure:kekulegap0v}a, \ref{figure:honeycombdata}b, or \ref{figure:stardata}b.

\section{\label{section:transmute}Anyon Transmutation}

Following Ref.~\cite{PhysRevB.90.134404}, we provide an example of how anyon transmutation can arise due to the topology of the torus. Consider the Kitaev honeycomb model in the strong $z$ bond limit. The $e$ and $m$ particles live on alternating rows of unit hexagons, as shown in Fig.~\ref{figure:honeycombanyonz}. If the torus has an odd number of rows, then one cannot consistently assign the $e$ and $m$ labels to the rows throughout the torus. An alternative viewpoint is that one must have two adjacent rows with the same label, which defines a defect line between them. When an $e$ ($m$) particle crosses this defect line, it becomes an $m$ ($e$) particle, a phenomenon known as anyon transmutation \cite{PhysRevB.90.134404}. Nevertheless, this issue can be easily resolved by enforcing an even number of rows on the torus. We emphasize that such a potential complication is genuinely due to the model and the geometry on which it is defined, irrespective of the method that is used to map out the anyon species of the elementary plaquettes, whether it is degenerate perturbation theory or the one outlined in Sec.~\ref{section:method}.

\section{\label{section:lemma1}Proof of First Lemma}

In this section, we prove the following lemma. \\

\noindent \textbf{Lemma 1.} Let $l_1$ and $l_2$ be two strings of bonds that connect the same pair of elementary plaquettes. Suppose that $l_1$ and $l_2$ form a contractible loop. Then, $n_w (l_1)$ and $n_w (l_2)$ have the same parity. \\

\begin{figure}
\subfloat[]{\label{figure:honeycombremove}
\includegraphics[scale=0.2]{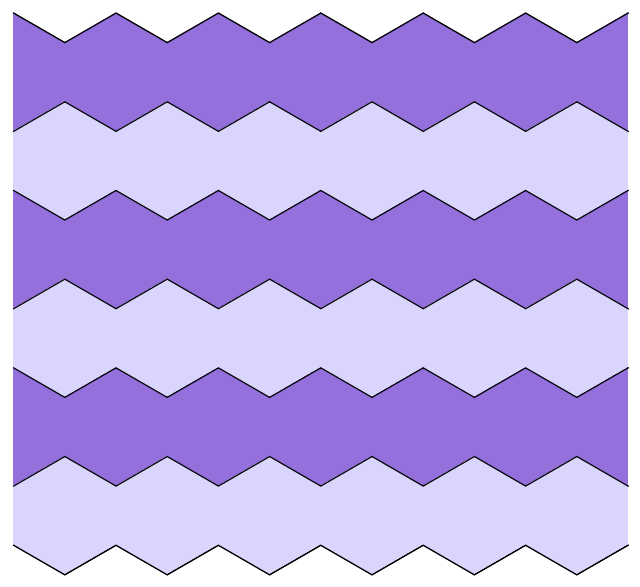}} \hspace{20pt}
\subfloat[]{\label{figure:octagonremove}
\includegraphics[scale=0.2]{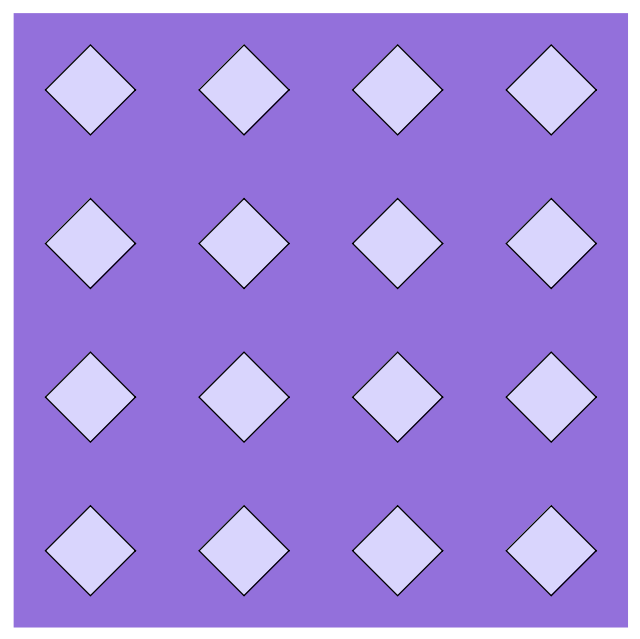}} \hspace{20pt}
\subfloat[]{\label{figure:amorphousremove}
\includegraphics[scale=0.2]{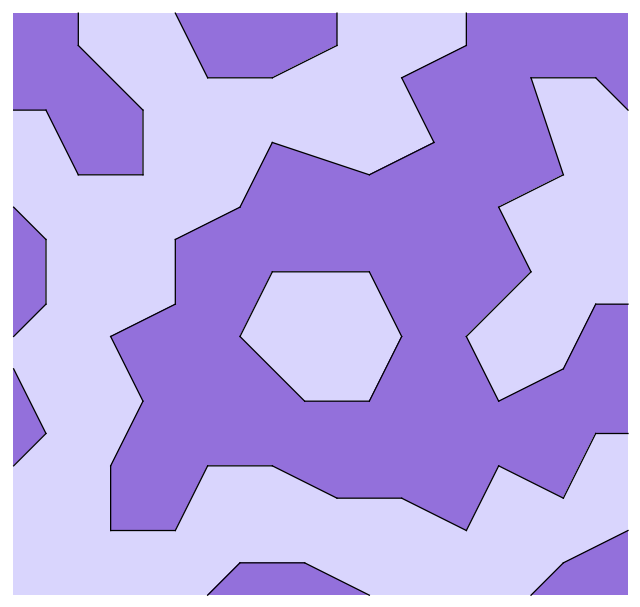}} \hspace{20pt}
\subfloat[]{\label{figure:shastryremove}
\includegraphics[scale=0.2]{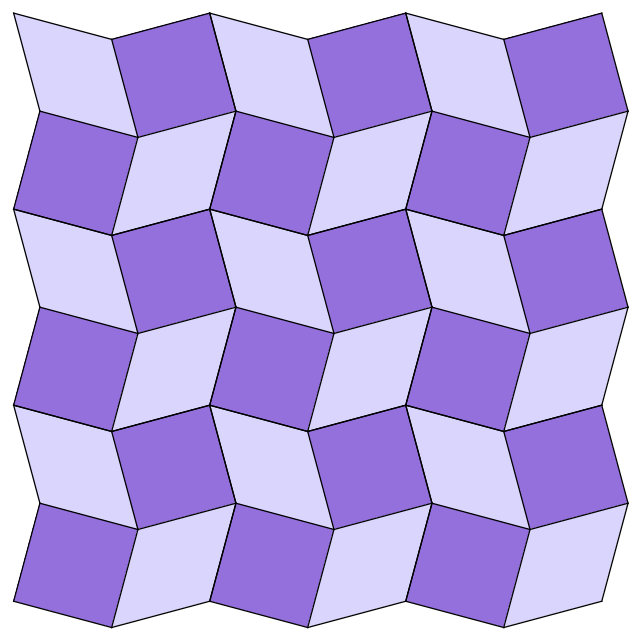}} \\
\subfloat[]{\label{figure:honeycombremiso}
\includegraphics[scale=0.2]{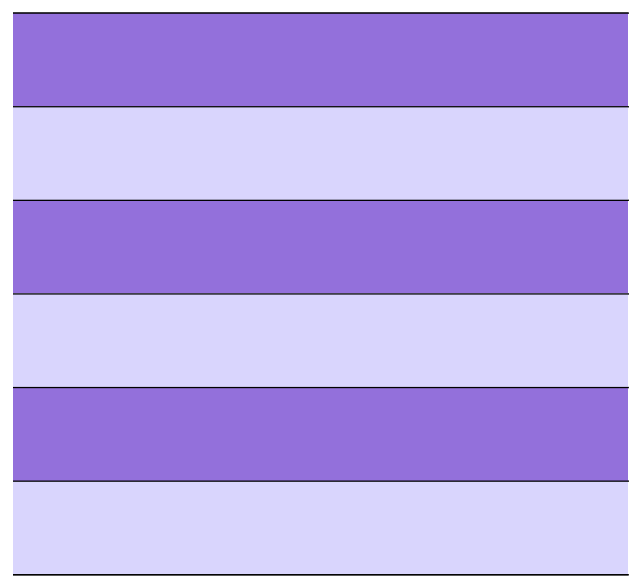}} \hspace{20pt}
\subfloat[]{\label{figure:octagonremiso}
\includegraphics[scale=0.2]{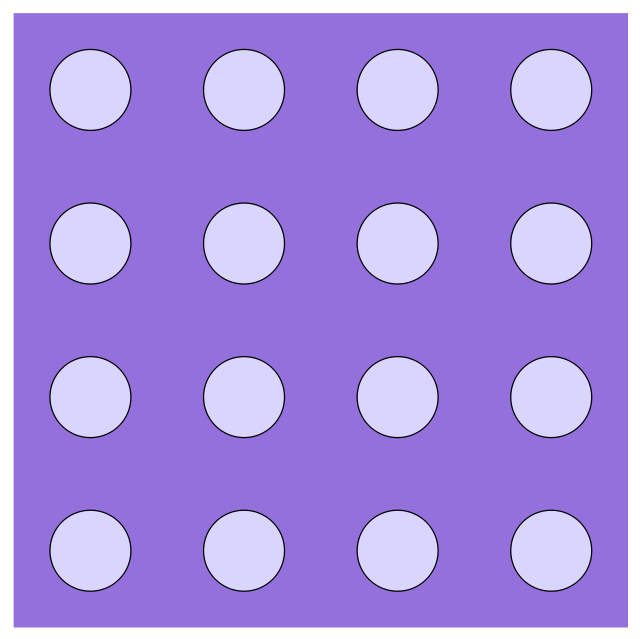}} \hspace{20pt}
\subfloat[]{\label{figure:amorphousremiso}
\includegraphics[scale=0.2]{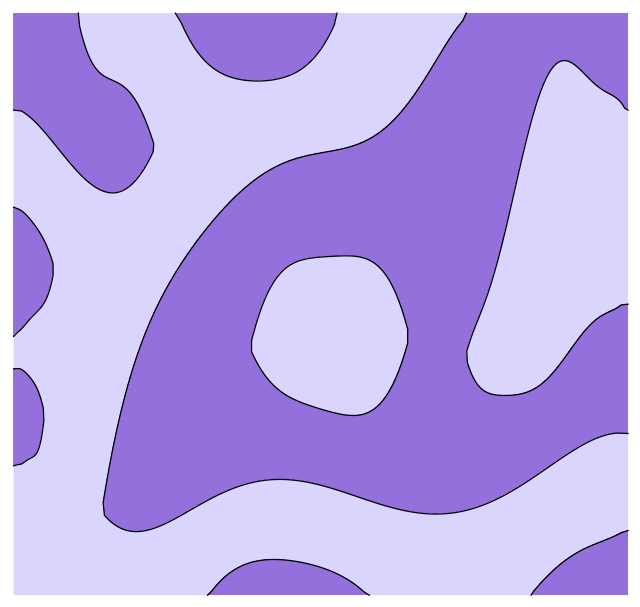}} \hspace{20pt}
\subfloat[]{\label{figure:shastryremiso}
\includegraphics[scale=0.2]{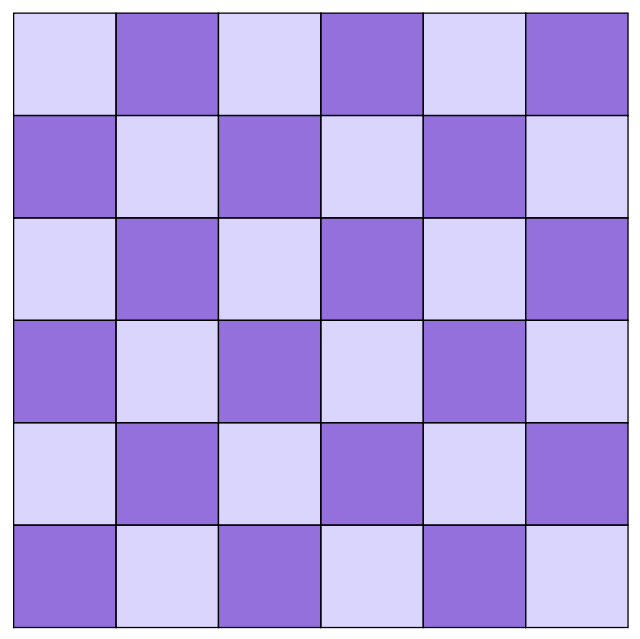}}
\caption{The spatial structures upon removing the strong $z$ or blue bonds of (a) the Kitaev honeycomb model, (b) the Kitaev square-octagon model, (c) the amorphous Kitaev model, and (d) the Kitaev Shastry-Sutherland model. The remaining weak bonds partition each region of interest into two classes of subregions, which are indicated by light and dark purple colors. (e)-(h) are obtained from (a)-(d) by smoothly deforming the boundaries between the subregions.}
\end{figure}

\noindent \textbf{Proof.} The generalized Kitaev model \eqref{kitaevmodel} is defined on a planar graph $\Lambda$ with the coordination number $z=2n-1$, which is an odd number. In the dimer limit, removing all the strong bonds in $\Lambda$ results in a planar graph $\Lambda'$ with an even coordination number. In the following, we consider the cases $n=2$ and $n>2$ separately. \\

\noindent $n=2$.\quad Since $l_1$ and $l_2$ form a contractible loop, we can focus on a region $\mathcal{R}$ containing them that is topologically equivalent to a disk. Each site in $\Lambda'$ has a coordination number of $2$. In other words, the weak bonds form one-dimensional (1D) manifolds, which can either be a loop inside $\mathcal{R}$ or a curve with both endpoints on the boundary of $\mathcal{R}$ \cite{Wootton_2015}. A curve can never have an open end inside $\mathcal{R}$. Different loops and curves do not intersect each other. These 1D manifolds then partition $\mathcal{R}$ into two classes of subregions. Crossing these 1D manifolds an odd (even) number of times is a sufficient and necessary condition for reaching the opposite (same) class of subregion. Each of $l_1$ and $l_2$ defines a path that intersects every weak bond contained in it once and only once (and does not intersect any weak bond not contained in it). The endpoints of these paths, which correspond to the elementary plaquettes $p_i$ and $p_f$, each belongs to a particular class of subregion. Travelling from $p_i$ to $p_f$ by $l_1$ and returning from $p_f$ to $p_i$ by $l_2$, one must cross an even number of weak bonds, for otherwise one ends up in a subregion that does not contain $p_i$. In other words, $n_w ( l_1 ) + n_w ( l_2 )$ is an even number, which implies $n_w ( l_1 ) \, \mathrm{mod} \, 2 = n_w ( l_2 ) \, \mathrm{mod} \, 2$. \\

\noindent $n > 2$.\quad Each site in $\Lambda'$ has a coordination number of $2n-2 \geq 4$. Consider the dual graph $\Lambda_*'$ of $\Lambda'$, whose vertices are located at the plaquette centers of $\Lambda'$. Each elementary loop on $\Lambda'_*$ thus has a length of $2n-2$. Since any contractible loop on $\Lambda'_*$ can be constructed by combining elementary loops, its length must be even. Every string $l$ of bonds connecting two elementary plaquettes on $\Lambda$ can be mapped to a string $l'$ of bonds on $\Lambda'$ simply by removing all the strong bonds contained in $l$. By construction, clearly, the number of weak bonds contained in $l$ is equal to that contained in $l'$, i.e., $n_w (l) = n_w (l')$. Furthermore, each bond (plaquette) of $\Lambda'$ can be uniquely mapped to an edge (vertex) of $\Lambda_*'$. Let us denote the image of $l'$ under this dual map by $l_*'$, which is a path on $\Lambda_*'$, and the number of edges in $l_*'$ by $\lvert l_*' \rvert$. Suppose that $l_1$ and $l_2$ are two different strings that connect the same pair of elementary plaquettes on $\Lambda$. Upon removing all the strong bonds in $l_1$ and $l_2$, we obtain $l_1'$ and $l_2'$, respectively. By assumption, $l_{1*}'$ and $l_{2*}'$ form a contractible loop on $\Lambda_*'$, the length of which must be even, so $\lvert l_{1*}' \rvert \, \mathrm{mod} \, 2 = \lvert l_{2*}' \rvert \, \mathrm{mod} \, 2$. Since $n_w (l_1) = n_w (l_1') = \lvert l_{1*}' \rvert$ and $n_w (l_2) = n_w (l_2') = \lvert l_{2*}' \rvert$, we deduce $n_w (l_1) \, \mathrm{mod} \, 2 = n_w (l_2) \, \mathrm{mod} \, 2$. \qed \\

Examples of the spatial structures resulting from the removal of dimers on various graphs are given in Figs.~\ref{figure:honeycombremove}-\ref{figure:shastryremiso}. Note in particular that for tri-coordinated graphs, each set of connected weak bonds can be viewed as a 1D manifold, see Figs.~\ref{figure:honeycombremiso}-\ref{figure:amorphousremiso}.

\section{\label{section:statistics}Nontrivial Mutual Statistics}

\begin{figure}
\subfloat[]{\label{figure:encircle}
\includegraphics[scale=0.25]{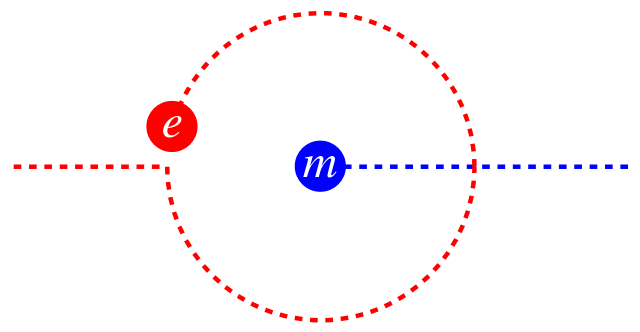}} \hspace{30pt}
\subfloat[]{\label{figure:crosstri}
\includegraphics[scale=0.25]{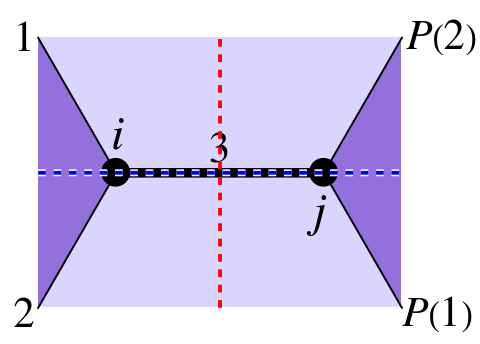}}
\caption{(a) When an $e$ particle encircles an $m$ particle, the associated $e$ and $m$ strings must intersect an odd number of times, which changes the wavefunction by an overall factor of $-1$. (b) The local environment of a strong bond $\langle ij \rangle$, which is indexed by $3$ and indicated by a thick black line, on a tri-coordinated lattice. The weak bonds are indexed by $1$ and $2$ and indicated by thin black lines. $P$ is a permutation of the same two elements. Elementary plaquettes (not drawn in full) filled with light and dark purple colors indicate that the respective vortices are $e$ and $m$ particles. The red and blue dashed lines represent segments of the $e$ and $m$ strings, respectively.}
\end{figure}

In this section, we demonstrate that the $e$ and $m$ particles assigned according to Corollary 1 in any dimer limit indeed have the correct braiding statistics. Let us start with a continuum description of what is expected. Since the $e$ particles can only be created pairwise, if we find a single $e$ particle in some local region in space, then it must be connected to another $e$ particle at some point outside this region by an extended string operator, which we call an $e$ string. An $m$ string is similarly defined. When an $e$ particle goes around an $m$ particle or vice versa, the associated $e$ and $m$ strings must intersect an odd number of times, as illustrated in Fig.~\ref{figure:encircle}. The wavefunction acquires a $-1$ factor for each such intersection. It suffices to consider one particular point of intersection in space. On a lattice, this means that we can focus on the local environment of a strong bond without loss of generality, and study the various possibilities of how a crossing between the $e$ and $m$ strings can take place. \\

For the coordination number $z=3$, there is only one possibility, in which the $e$ particle travels \textit{across} the strong bond while the $m$ particle travels \textit{along} the strong bond (or vice versa), as illustrated in Fig.~\ref{figure:crosstri}. This situation is essentially the same as that analyzed in Ref.~\cite{PACHOS20071254} for the strong $J_z$ limit of the Kitaev honeycomb model. For generality, we have indexed the strong bond by $3$ and the weak bonds by $1$ and $2$, while $P$ is a permutation of two elements. We say that an operator is allowed if it does not take the wavefunction out of the low-energy subspace by exciting a dimer, which would cost an energy of $\sim 2 J_3$. The allowed operators that move an $e$ particle across the strong bond are $\sigma_i^3$ and $\sigma_j^3$. The allowed operators that move an $m$ particle along the strong bond are $\sigma_i^1 \sigma_j^{P (2)}$ and $\sigma_i^2 \sigma_j^{P (1)}$. Let $\hat{O}_e$ be either $\sigma_i^3$ or $\sigma_j^3$, and $\hat{O}_m$ be either $\sigma_i^1 \sigma_j^{P (2)}$ or $\sigma_i^2 \sigma_j^{P (1)}$. $\hat{O}_e$ and $\hat{O}_m$ always anticommute, i.e., $\hat{O}_e \hat{O}_m = - \hat{O}_m \hat{O}_e$, which implies the nontrivial mutual statistics of the $e$ and $m$ particles. \\

\begin{figure}
\subfloat[]{\label{figure:crosspenta}
\includegraphics[scale=0.25]{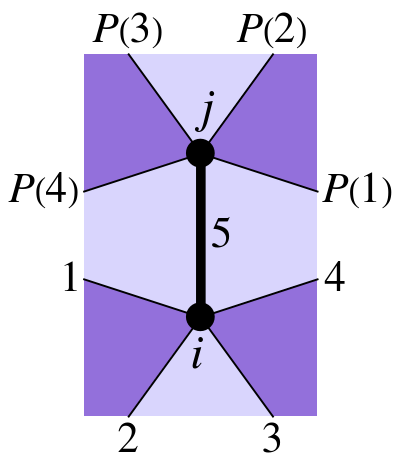}} \;
\subfloat[]{\label{figure:ehophor}
\includegraphics[scale=0.25]{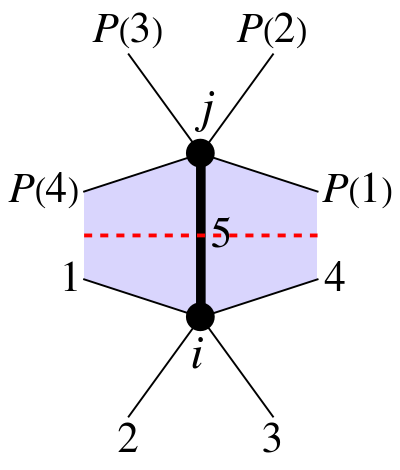}} \;
\subfloat[]{\label{figure:ehopver}
\includegraphics[scale=0.25]{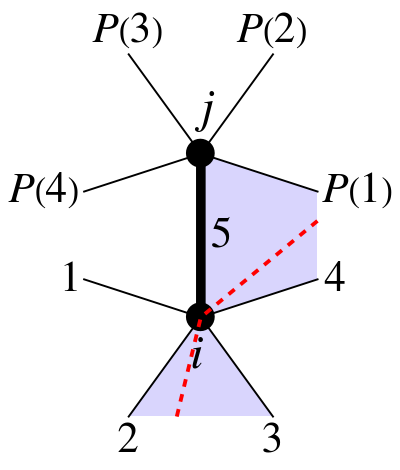}} \;
\subfloat[]{\label{figure:mhopver}
\includegraphics[scale=0.25]{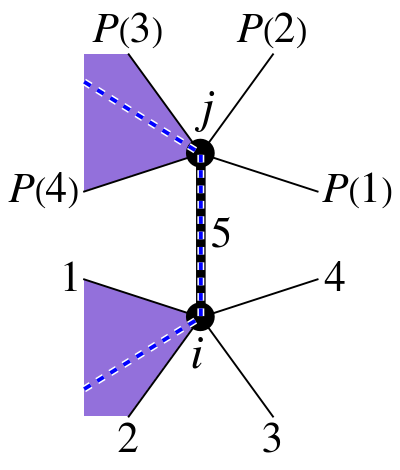}} \;
\subfloat[]{\label{figure:mhophor}
\includegraphics[scale=0.25]{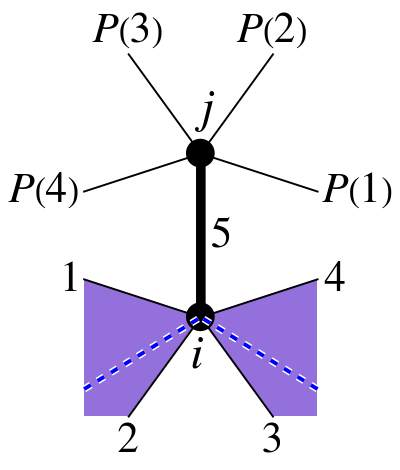}} \;
\subfloat[]{\label{figure:mhopdia}
\includegraphics[scale=0.25]{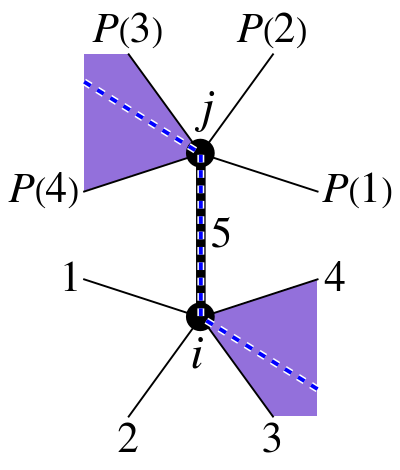}}
\caption{(a) The local environment of a strong bond $\langle ij \rangle$, which is indexed by $5$ and indicated by a thick line, on a penta-coordinated lattice. The weak bonds are indexed by $1$, $2$, $3$, and $4$ and indicated by thin lines. $P$ is a permutation of the same four elements. Elementary plaquettes (not drawn in full) filled with light and dark purple colors indicate that the respective vortices are $e$ and $m$ particles. Each of (b)-(f) represents the hopping of a vortex from one colored plaquette to another, assuming that a vortex is initially present on one of the colored plaquettes. The segments of the $e$ and $m$ strings contributed by these hoppings are indicated by red and blue dashed lines, respectively. (b) The hopping of an $e$ particle across the bond $\langle ij \rangle$. (c) The hopping of an $e$ particle across the site $i$. (d) The hopping of an $m$ particle along the bond $\langle ij \rangle$. (e) The hopping of an $m$ particle across the site $i$. (f) The hopping of an $m$ particle obtained by combining (d) and (e).}
\end{figure}

For the coordination number $z=5$, the analysis is similar to $z=3$ but slightly more tedious, as there exist more than one possibility of how the crossing between the $e$ and $m$ strings can arise. Fig.~\ref{figure:crosspenta} shows the local environment of a strong bond $\langle ij \rangle$, which is indexed by the number $5$ without loss of generality, on a penta-coordinated lattice. The four distinct weak bonds are indexed by $1$, $2$, $3$, and $4$, while $P$ is a permutation of four elements. It suffices to study the two hopppings of $e$ in Figs.~\ref{figure:ehophor} and \ref{figure:ehopver} as others can be obtained by symmetry and combinations \footnote{Each of Figs.~\ref{figure:ehophor}-\ref{figure:mhopdia} can also represents the creation or annihilation of a pair of vortices on the colored plaquettes, if the initial number of vortices on these plaquettes is zero or two, respectively.}. \\

The hopping of $e$ across the bond $\langle ij \rangle$ as depicted in Fig.~\ref{figure:ehophor} is carried out by the operator $\hat{O}_e = \Gamma_i^5$ or $\Gamma_j^5$. The hopping of $m$ along the bond $\langle ij \rangle$ as depicted in Fig.~\ref{figure:mhopver} is carried out by the operator $\hat{O}_m = \Gamma_i^1 \Gamma_j^{P (4)}$. The segments of the $e$ and $m$ string contributed by these hoppings intersect unambiguously, and one indeed finds $\hat{O}_e \hat{O}_m = - \hat{O}_m \hat{O}_e$. On the other hand, the hopping of $e$ across the site $i$ as depicted in Fig.~\ref{figure:ehopver} is carried out by the operator $\hat{O}_e = \Gamma_i^3 \Gamma_i^4$. The hopping of $m$ across the site $i$ as depicted in Fig.~\ref{figure:mhophor} is carried out by the operator $\hat{O}_m = \Gamma_i^2 \Gamma_i^3$. Once again, the segments of the $e$ and $m$ strings contributed by these hoppings intersect unambiguously, and one finds $\hat{O}_e \hat{O}_m = - \hat{O}_m \hat{O}_e$. In either case, the anticommutation of $\hat{O}_e$ and $\hat{O}_m$ implies the nontrivial mutual statistics of the $e$ and $m$ particles. \\

We make two further observations. First, the $e$ and $m$ strings in Figs.~\ref{figure:ehophor} and \ref{figure:mhophor}, as well as those in Figs.~\ref{figure:ehopver} and \ref{figure:mhopver}, do not result in an unambiguous crossing. Indeed, it is easy to verify that $\hat{O}_e$ of Fig.~\ref{figure:ehophor} commutes with $\hat{O}_m$ of Fig.~\ref{figure:mhophor}; similarly for Figs.~\ref{figure:ehopver} and \ref{figure:mhopver}. Second, combining Figs.~\ref{figure:mhopver} and \ref{figure:mhophor} leads to the hopping of $m$ as depicted in Fig.~\ref{figure:mhopdia}. There is an unambiguous crossing between the $m$ string in Fig.~\ref{figure:mhopdia} and the $e$ string in each of Figs.~\ref{figure:ehophor} and \ref{figure:ehopver}. Indeed, $\hat{O}_m$ of Fig.~\ref{figure:mhopdia}, which is the product of those of Figs.~\ref{figure:mhopver} and \ref{figure:mhophor}, anticommutes with $\hat{O}_e$ of each of Figs.~\ref{figure:ehophor} and \ref{figure:ehopver}. \\

We expect that the above analyses can be straightforwardly generalized to demonstrate the correct braiding statistics of the $e$ and $m$ particles for higher coordination numbers $z=2n-1, n \geq 4$. However, we have relied on the assumption of a dimer limit, where the coupling strengths on the strong and weak bonds differ by orders of magnitude, and studied one- and two-body operators that create, annihilate, or move the vortices without exciting a dimer. Advanced analyses using perturbative continuous unitary transformations (PCUTs) \cite{PhysRevLett.100.057208,PhysRevLett.100.177204,PhysRevB.78.245121} reveal that these operators can create fermionic quasiparticles, each of which is a hard core boson attached to a string operator, in addition to a pair of vortices when acting on the ground state. The probability of creating such quasiparticles is given in powers of the ratio of the weak to strong couplings, which are negligible in the dimer limit. For parameters that are less anisotropic yet still in the toric code phase, one could employ PCUTs to systematically include corrections to the aforementioned operators such that they create only the vortices but not the fermions. It would be interesting to look for instances where PCUTs as well as other perturbative approaches possibly break down. For example, in the Kitaev square-octagon model, there exists a parameter regime where $J_z > J_x > J_y$ that belongs to the same toric code phase as the strong $J_x$ limit. We leave this topic for possible future investigations. \\

\section{\label{section:square}Isolated-Square Limit}

Using degenerate perturbation theory, Ref.~\cite{Kells_2011} derives an effective Hamiltonian $H_\mathrm{eff}$, which is equivalent to a toric code model, in the isolated-square limit of the Kitaev square-octagon model where $(J_x^2 + J_y^2)^{1/2} \gg J_z$. We compare their approach and result to ours in the dimer limits $J_x \gg J_y , J_z$ and $J_y \gg J_z , J_x$. First, while $J_z$ is treated as a perturbation, $J_x$ and $J_y$ are treated on an equal footing in Ref.~\cite{Kells_2011}, such that nontrivial contributions to $H_\mathrm{eff}$ only begin to appear at the $4$th order. If there were a further hierarchy among $J_x$ and $J_y$, e.g., $J_x \gg J_y$, then nontrivial contributions to $H_\mathrm{eff}$ would appear at the $2$nd order. In essence, Ref.~\cite{Kells_2011} is investigating the limit $J_x \approx J_y \gg J_z$, which is different from the limits $J_x \gg J_y , J_z$ and $J_y \gg J_z, J_x$ where Corollary 1 is applicable. Second, each unit square, which consists of four sites connected by $x$ and $y$ bonds, is treated as a low-energy doublet in the degenerate perturbation theory of Ref.~\cite{Kells_2011}. The resulting $H_\mathrm{eff}$ is essentially defined on a square lattice, the elementary plaquettes (sites) of which correspond to the unit octagons (squares) of the square-octagon lattice. Anyons appear only on the unit octagons, but not the unit squares, of the square-octagon lattice. In contrast, our method treats the strong bonds as low-energy doublets and reveals the toric code physics on the same planar graph that is used to construct the generalized Kitaev model, such that each elementary plaquette on the square-octagon lattice can host an anyon of a particular species. Despite these differences, we notice one similarity between the results of Ref.~\cite{Kells_2011} and ours. The centers of the unit octagons form a square lattice, which is bipartite. In the $( J_x^2 + J_y^2 )^{1/2} \gg J_z$ limit of Ref.~\cite{Kells_2011}, as well as in both the $J_x \gg J_y , J_z$ and $J_y \gg J_z , J_x$ limits of our work, we find that anyons of one species live on the odd sublattice, while anyons of the other species live on the even sublattice, cf.~Fig.~4 in Ref.~\cite{Kells_2011} and Fig.~4 in our work.

\section{\label{section:addresult}Additional Results}

In this section, we present various results from the studies of the Kitaev honeycomb and star models that lead to Figs.~\ref{figure:honeycombphase} and \ref{figure:starphase} in the main text. We also examine the entire fermion spectra of distinct two-vortex sectors along selected paths for the Kitaev square-octagon and Kekul\'{e} models. As in the main text, the two endpoints of a path in the triangular parameter space are chosen from the sets of couplings $\mathbf{J}^{(U)}=(0.1,0.1,0.8)$, $\mathbf{J}^{(L)}=(0.8,0.1,0.1)$, and $\mathbf{J}^{(R)}=(0.1,0.8,0.1)$. For instance, the path $LR$ is given by the linear interpolation from $\mathbf{J}^{(L)}$ to $\mathbf{J}^{(R)}$ and parametrized by a single variable $t \in [0,1]$.

\subsection{\label{section:honeycomb}Kitaev honeycomb model}

\begin{figure*}
\includegraphics[scale=0.22]{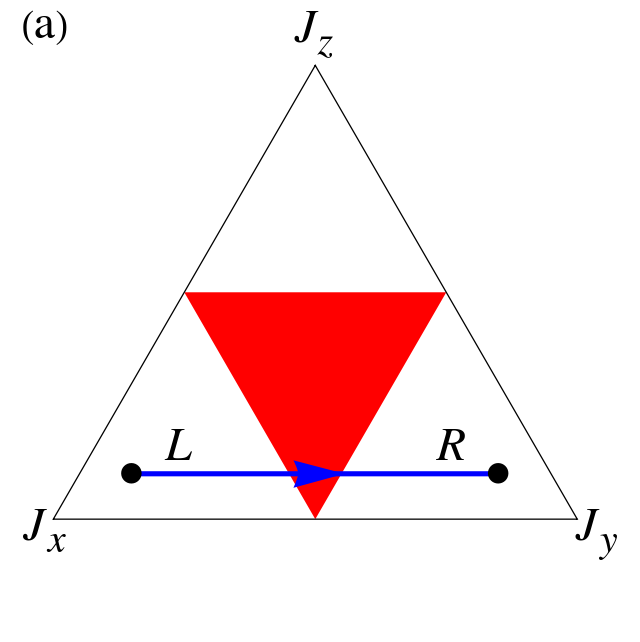} \hspace{20pt}
\includegraphics[scale=0.22]{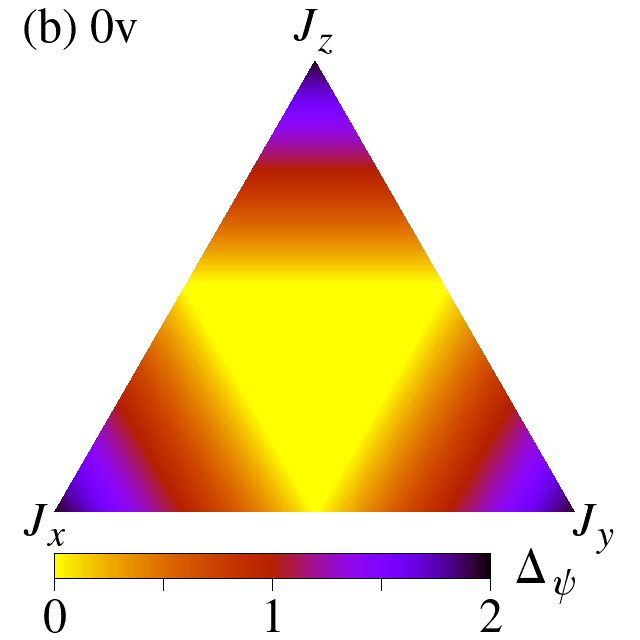} \hspace{20pt}
\includegraphics[scale=0.22]{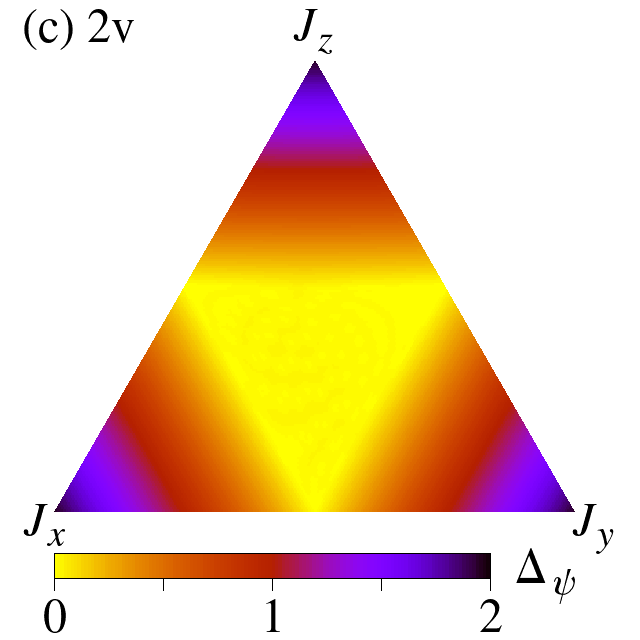} \hspace{20pt}
\includegraphics[scale=0.22]{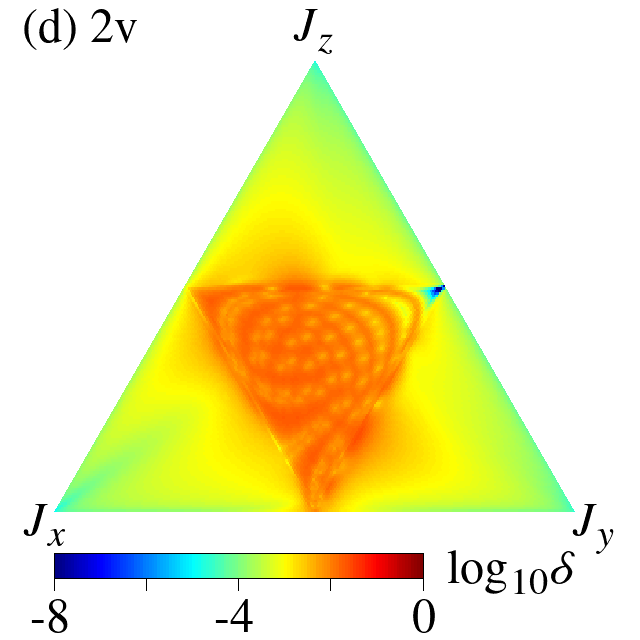} \\ \vspace{5pt}
\includegraphics[scale=0.25]{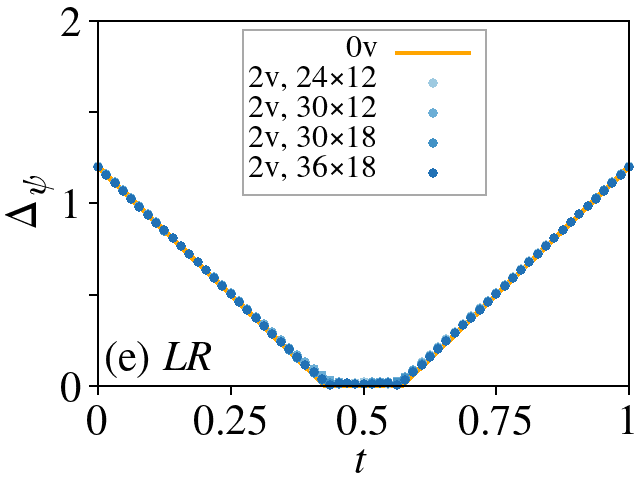} \;
\includegraphics[scale=0.25]{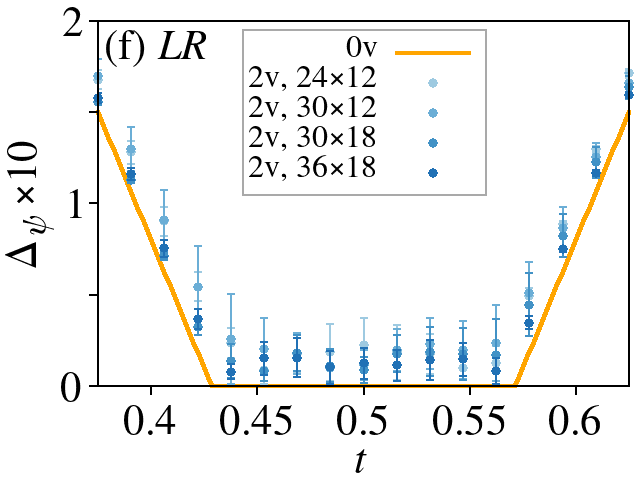}
\caption{\label{figure:honeycombdata}Various results from the study of the Kitaev honeycomb model. The triangular parameter space in (a)-(d) is defined by $J_x + J_y + J_z = 1$, where the vertex labeled by $J_\lambda$ corresponds to the $J_\lambda \longrightarrow 1$ limit. $0\mathrm{v}$ and $2\mathrm{v}$ stand for ``vortex-free'' and ``two-vortex'', respectively. (a) The red filled area indicates the closing of the fermion gap in the vortex-free sector and all two-vortex sectors. The blue solid line with an arrow represents the selected path along which we perform finite-size scaling analyses of the fermion gaps and determine the assignments of anyon species numerically. (b) The fermion gap $\Delta_\psi^{(0\mathrm{v})}$ of the vortex-free sector. (c) The averaged fermion gap $\Delta_\psi^{(2\mathrm{v})}$ of two-vortex sectors. The standard deviation $\delta$ is shown in (d) on a logarithmic scale. (e) The averaged fermion gap of two-vortex sectors (blue data) along the path $LR$, which is linearly parametrized by $t \in [0,1]$. The error bars represent the standard deviations. The fermion gap of the vortex-free sector (orange data) is also plotted for comparison. (f) Zoom-in of (e) near zero energy. In (e) and (f), the product of two integers associated with each set of 2v data indicates the size of the system ($L_1 \times L_2$ unit cells) on which the computations are performed.}
\end{figure*}

The Kitaev honeycomb model (Fig.~\ref{figure:honeycombmodel}) has only one type of elementary plaquettes, namely the unit hexagons. The fermion gap $\Delta_\psi^{(0\mathrm{v})}$ of the vortex-free sector closes in the parameter regime that satisfies $J_\lambda \leq J_\mu + J_\nu$ for all three cyclic permutations $(\lambda , \mu , \nu)$ of $(x , y , z)$ \cite{KITAEV20062}, see Fig.~\ref{figure:honeycombdata}b. This results in three gapped regimes that are separated from each other and smoothly connected to the strong $J_x$, $J_y$, and $J_z$ limits, respectively. Each of these regimes is characterized by $\nu = 0$, i.e., it hosts the $\mathbb{Z}_2$ topological order. With $L_1 \times L_2 = 24 \times 18$ and $L_1' \times L_2' = 4 \times 12$, we compute the averaged fermion gap $\Delta_\psi^{(2 \mathrm{v})}$ of two-vortex sectors and its error $\delta$, which are plotted over the triangular parameter space in Figs.~\ref{figure:honeycombdata}c and \ref{figure:honeycombdata}d. The profile of $\Delta_\psi^{(2 \mathrm{v})}$ highly resembles that of $\Delta_\psi^{(0 \mathrm{v})}$. In particular, at the parameters where the fermion gap of the vortex-free sector is zero (finite), our data suggests that the fermion gaps of two-vortex sectors are zero (finite) as well. We also find that the error is largest near and inside the gapless regime, where it is of the order of $10^{-2}$. \\

We further provide a detailed convergence analysis of $\Delta_\psi^{(2\mathrm{v})}$ with respect to the sizes of the system and the region of mapping, along the path $LR$ as indicated in Fig.~\ref{figure:honeycombdata}a. Our computations are performed with (i) $L_1 \times L_2 = 24 \times 12$, $L_1' \times L_2' = 4 \times 8$, (ii) $L_1 \times L_2 = 30 \times 12$, $L_1' \times L_2' = 8 \times 8$, (iii) $L_1 \times L_2 = 30 \times 18$, $L_1' \times L_2' = 8 \times 12$, and (iv) $L_1 \times L_2 = 36 \times 18$, $L_1' \times L_2' = 12 \times 12$. The results are plotted in Fig.~\ref{figure:honeycombdata}e, where $\Delta_\psi^{(0\mathrm{v})}$ is also plotted for comparison. Starting from $t=0$ with a value of $\sim 1$, $\Delta_\psi^{(2\mathrm{v})} (t)$ decreases monotonically as $t$ increases to $3/7$, at which point $\Delta_\psi^{(0\mathrm{v})} (t)$ becomes zero and remains so until $t=4/7$. We find that $\Delta_\psi^{(2\mathrm{v})} (t)$ is smallest for $t \in [3/7 , 4/7]$, where it is of the order of $10^{-2}$ or lower, which suggests that the fermion gap of every two-vortex sector also vanishes in this interval. As $t$ further increases from $4/7$ to $1$, $\Delta_\psi^{(2\mathrm{v})} (t)$ increases monotonically to $\sim 1$. The discrepancies between the data of different system sizes, as well as the error bars, are largest near and inside the critical regime $3/7 \leq t \leq 4/7$, see Fig.~\ref{figure:honeycombdata}f. Far from the critical regime, the data of different system sizes overlap rather perfectly, and the error bars are small. Overall, $\Delta_\psi^{(2\mathrm{v})} (t)$ behaves similarly to $\Delta_\psi^{(0\mathrm{v})} (t)$, as noted in the last paragraph. By calculating the (relative) fermion parities of the two-vortex sectors, we also confirm that, for $t \in [0 , 3/7)$ and $t \in (4/7 , 1]$, the assignments of anyon species are given by those in the strong $J_x$ and $J_y$ limits, respectively, which is consistent with Corollaries 1 and 2.

\subsection{\label{section:star}Kitaev star model}

\begin{figure*}
\includegraphics[scale=0.22]{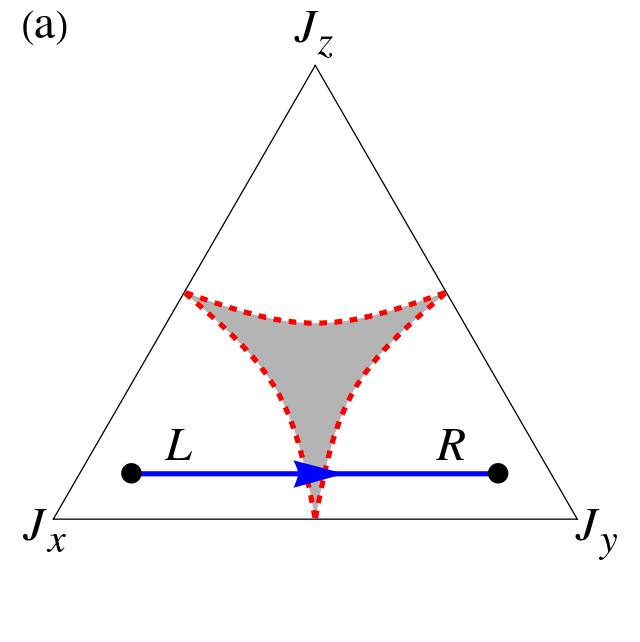} \hspace{20pt}
\includegraphics[scale=0.22]{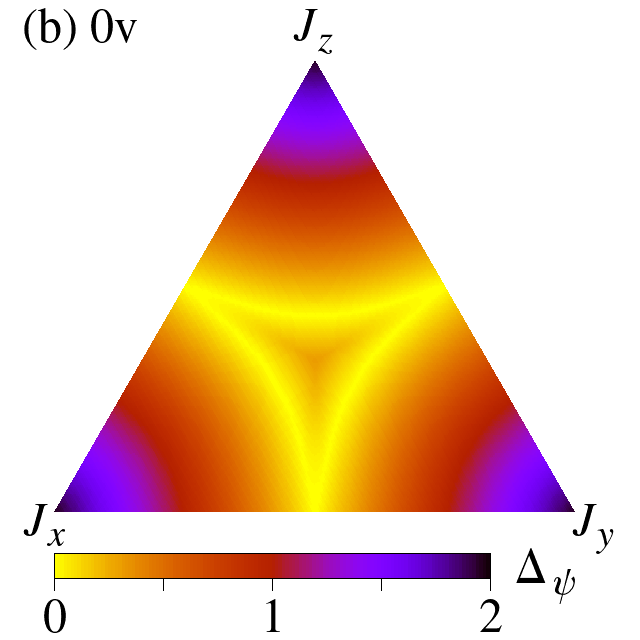} \hspace{20pt}
\includegraphics[scale=0.22]{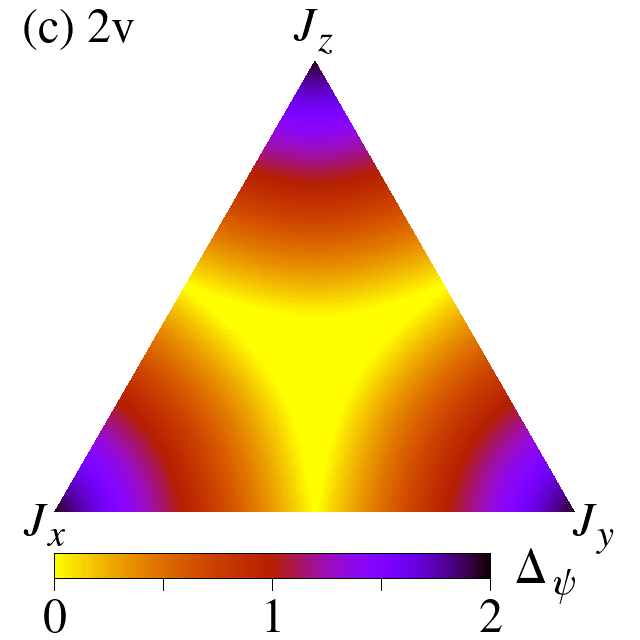} \hspace{20pt}
\includegraphics[scale=0.22]{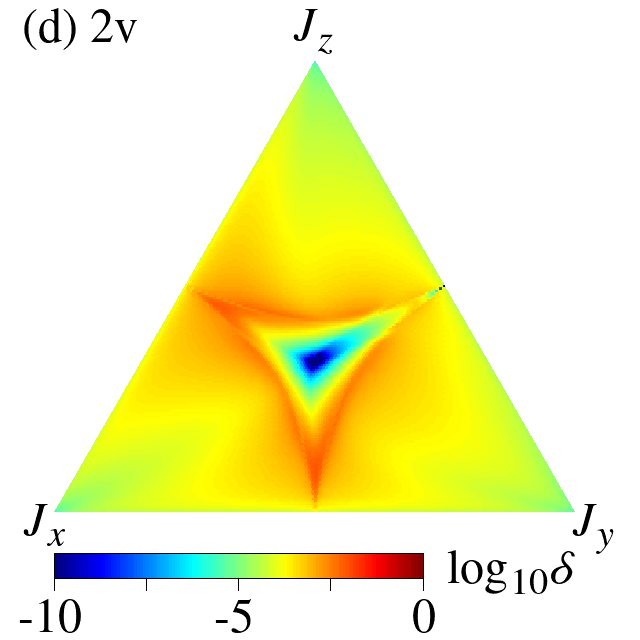} \\ \vspace{5pt}
\includegraphics[scale=0.25]{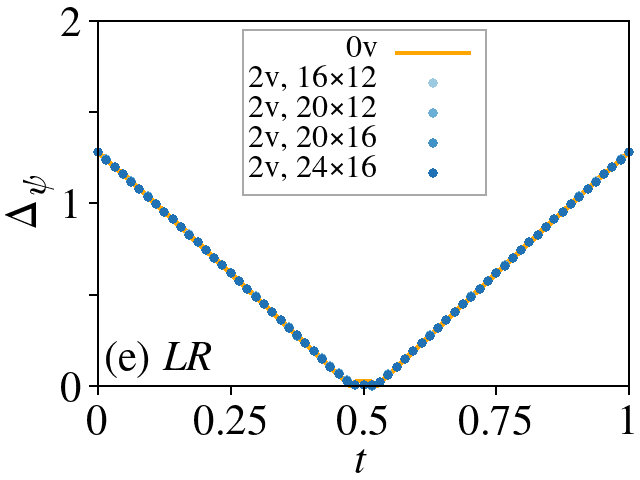} \;
\includegraphics[scale=0.25]{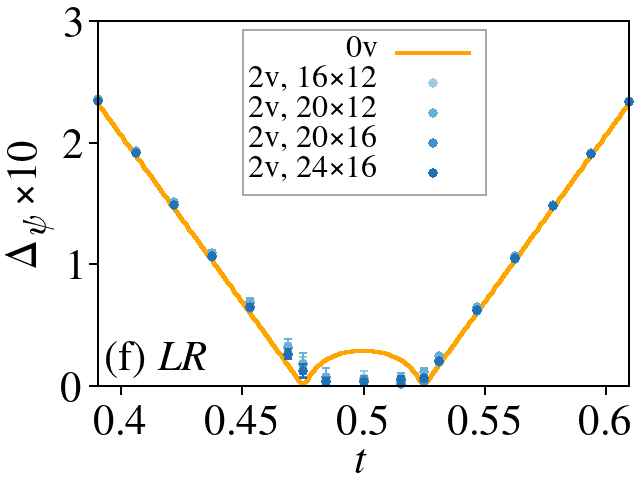}
\caption{\label{figure:stardata}Various results from the study of the Kitaev star model. The triangular parameter space in (a)-(d) is defined by $J_x + J_y + J_z = 1$, where the vertex labeled by $J_\lambda$ corresponds to the $J_\lambda \longrightarrow 1$ limit. $0\mathrm{v}$ and $2\mathrm{v}$ stand for ``vortex-free'' and ``two-vortex'', respectively. (a) The red dashed curves indicate the closing of the fermion gap in the vortex-free sector and all two-vortex sectors. The gray filled area indicates the closing of the fermion gap in all two-vortex sectors, while the vortex-free sector has a finite fermion gap. The blue solid line with an arrow represents the selected path along which we perform finite-size scaling analyses of the fermion gaps and determine the assignments of anyon species numerically. (b) The fermion gap $\Delta_\psi^{(0\mathrm{v})}$ of the vortex-free sector. (c) The averaged fermion gap $\Delta_\psi^{(2\mathrm{v})}$ of two-vortex sectors $(p_0 , p_i)$, where $p_0$ is a unit dodecagon and $p_i$ runs over all three types of elementary plaquettes. The standard deviation $\delta$ is shown in (d) on a logarithmic scale. (e) The averaged fermion gap (blue data) of two-vortex sectors $(p_0 , p_i)$, where $p_0$ is a unit dodecagon and $p_i$ runs over all three types of elementary plaquettes, along the path $LR$, which is linearly parametrized by $t \in [0,1]$. The error bars represent the standard deviations. The fermion gap of the vortex-free sector (orange data) is also plotted for comparison. (f) Zoom-in of (e) near zero energy. In (e) and (f), the product of two integers associated with each set of 2v data indicates the size of the system ($L_1 \times L_2$ unit cells) on which the computations are performed.}
\end{figure*}

The Kitaev star model (Fig.~\ref{figure:starmodel}) has three elementary plaquettes per unit cell, namely a unit dodecagon, an up triangle, and a down triangle. The fermion gap $\Delta_\psi^{(0\mathrm{v})}$ of the vortex-free sector closes at the parameters that satify $J_\lambda^3 - J_\mu^3 - J_\nu^3 - J_\lambda J_\mu J_\nu = 0$ for any of the three cyclic permutations $(\lambda , \mu , \nu)$ of $(x , y , z)$ \cite{PhysRevB.82.174412}, see Fig.~\ref{figure:stardata}b. The gapped regime in the middle, which is bounded by the three curves, is characterized by $\nu = -1$, i.e., it hosts the Ising topological order. The remaining three gapped regimes, which are separated from each other and smoothly connected to the strong $J_x$, $J_y$, and $J_z$ limits, respectively, are characterized by $\nu=0$, i.e., they host the $\mathbb{Z}_2$ topological order. With $L_1 \times L_2 = 16 \times 16$ and $L_1' \times L_2' = 2 \times 12$, we compute the average fermion gap $\Delta_\psi^{(2\mathrm{v})}$ of two-vortex sectors with one vortex fixed at a unit dodecagon and the other going over all three types of elementary plaquettes within the region of mapping, as well as its error $\delta$. The results are plotted in Figs.~\ref{figure:stardata}c and \ref{figure:stardata}d. We also compute $\Delta_\psi^{(2\mathrm{v})}$ and $\delta$ with one vortex fixed at an up triangle and the other going over all three types of elementary plaquettes within the region of mapping. The results are the same as Figs.~\ref{figure:stardata}c and \ref{figure:stardata}d, so they are not shown for brevity. We find that $\Delta_\psi^{(2\mathrm{v})}$ goes to zero on the topological phase boundaries $J_\lambda^3 - J_\mu^3 - J_\nu^3 - J_\lambda J_\mu J_\nu = 0$ that separate the $\nu=-1$ regime from the $\nu=0$ regimes. In addition, $\Delta_\psi^{(2\mathrm{v})}$ vanishes in the $\nu = -1$ region, where all vortices map to a single species $\sigma$. The $\sigma$ anyons are non-Abelian as two of them can fuse into the vacuum $1$ or a fermion $\psi$. The degenerate state space of two $\sigma$ anyons thus has a dimension of $2$, which manifests as a zero-energy (complex) fermion mode in every two-vortex sector \cite{LAHTINEN20082286,PhysRevResearch.2.023334}. Occupying (unoccupying) such a mode corresponds to the fusion outcome $\psi$ ($1$) \footnote{An alternative, but perhaps equivalent, way of understanding the necessity of $\Delta_\psi^{(2\mathrm{v})}=0$ is described as follows. When $\nu$ is odd, each vortex carries an unpaired Majorana mode. Two vortices thus contribute to a complex fermion mode, which can be occupied or unoccupied, at zero energy.}. We also find that the error is largest in the vicinity of the topological phase boundaries, where it is of the order of $10^{-3}$. The smallness of the error, together with the invariance of the results upon moving one of the vortices from a unit dodecagon to a unit triangle, indicates that the fermion gap of each two-vortex sector is independent of the types of the elementary plaquettes that host the vortices. \\

We further provide a detailed convergence analysis of $\Delta_\psi^{(2\mathrm{v})}$ with respect to the sizes of the system and the region of mapping, along the path $LR$ as indicated in Fig.~\ref{figure:stardata}a. Our computations are performed with (i) $L_1 \times L_2 = 16 \times 12$, $L_1' \times L_2' = 2 \times 8$, (ii) $L_1 \times L_2 = 20 \times 12$, $L_1' \times L_2' = 4 \times 8$, (iii) $L_1 \times L_2 = 20 \times 16$, $L_1' \times L_2' = 4 \times 12$, and (iv) $L_1 \times L_2 = 24 \times 16$, $L_1' \times L_2' = 6 \times 12$. The results with one vortex fixed at a unit dodecagon and the other going over all three types of elementary plaquettes within the region of mapping are plotted in Fig.~\ref{figure:stardata}e, where $\Delta_\psi^{(0\mathrm{v})}$ is also plotted for comparison. The results with the fixed vortex at an up triangle instead of a unit dodecagon highly resemble those in Fig.~\ref{figure:stardata}e, so they are not shown for brevity. Starting with a value of $\sim 1$ at $t=0$, $\Delta_\psi^{(2\mathrm{v})} (t)$ decreases monotonically as $t$ increases to $0.475$ (to three significant digits), at which $\Delta_\psi^{(0\mathrm{v})} (t) = 0$. We find that $\Delta_\psi^{(2\mathrm{v})} (t)$ is smallest at $t \in (0.475 , 0.525)$, where it is of the order of $10^{-3}$ and $\nu = -1$. As $t$ increases from $0.525$ (to three significant digits), at which $\Delta_\psi^{(0\mathrm{v})} (t)=0$ again, to $1$, $\Delta_\psi^{(2\mathrm{v})} (t)$ increases monotonically to $\sim 1$. The discrepancies between the data of different system sizes, as well as the error bars, are largest near and inside the $\nu = -1$ regime, see Fig.~\ref{figure:stardata}f. Far from the $\nu = -1$ regime, the data of different system sizes overlap nearly perfectly, and the error bars are small. Overall, $\Delta_\psi^{(2\mathrm{v})} (t)$ behaves similarly as $\Delta_\psi^{(0\mathrm{v})} (t)$ outside the $\nu = -1$ regime. By calculating the (relative) fermion parities of the two-vortex sectors, we also confirm that, for $t \in [0 , 0.475)$ and $t \in (0.525 , 1]$, the assignments of anyon species are given by those in the strong $J_x$ and $J_y$ limits, respectively, which is consistent with Corollaries 1 and 2.

\subsection{\label{section:spectrum}Fermion spectrum}

In the main text, we have analyzed the fermion gaps, each of which is given by the smallest non-negative eigenvalue of a Hamiltonian, for the Kitaev square-octagon and Kekul\'{e} models. In this section, we further inspect the entire fermion spectra, each of which consists of all non-negative eigenvalues $\varepsilon_1 , \ldots , \varepsilon_{N/2}$ of a Hamiltonian, for these models along selected paths in the triangular parameter space. The eigenvalues are indexed such that $\varepsilon_1 \leq \ldots \leq \varepsilon_{N/2}$. For each $1 \leq k \leq N/2$, we compute the averaged $k$th eigenvalue $\varepsilon_k^{\alpha \beta}$ over a number of two-vortex sectors $(p_0 , p_i)$, where $p_0$ ($p_i$) is an elementary plaquette of type $\alpha$ ($\beta$), and take the standard deviation as the error. The procedure is essentially the same as that for the calculation of $\Delta_\psi^{\alpha \beta}$, see Sec.~\ref{section:setup}. For comparison, we also compute the fermion spectrum $\lbrace \varepsilon_k^{(0 \mathrm{v})} \vert 1 \leq k \leq N/2 \rbrace$ of the vortex-free sector. All these computations are performed on a finite system with periodic boundary conditions. \\

\begin{figure*}
\includegraphics[scale=0.25]{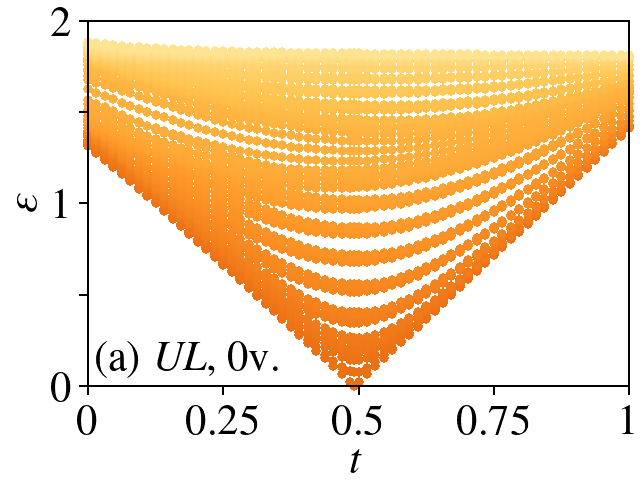} \;
\includegraphics[scale=0.25]{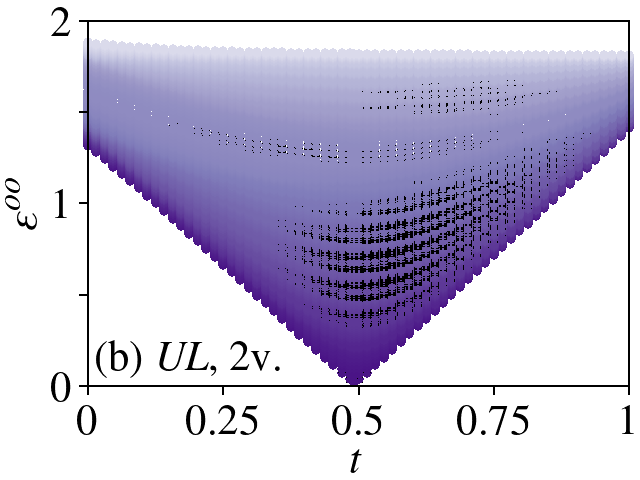} \;
\includegraphics[scale=0.25]{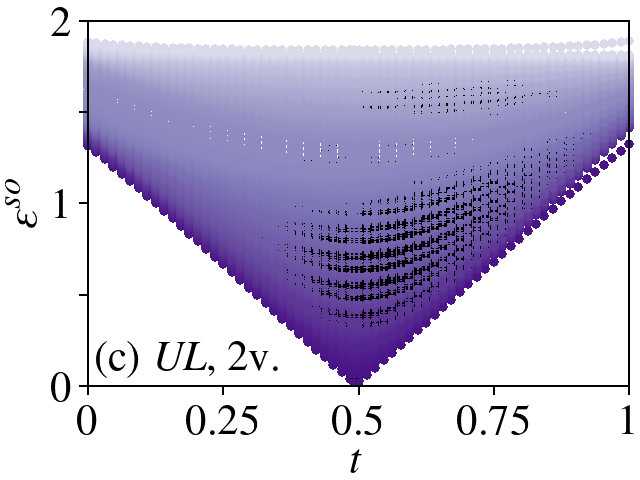} \;
\includegraphics[scale=0.25]{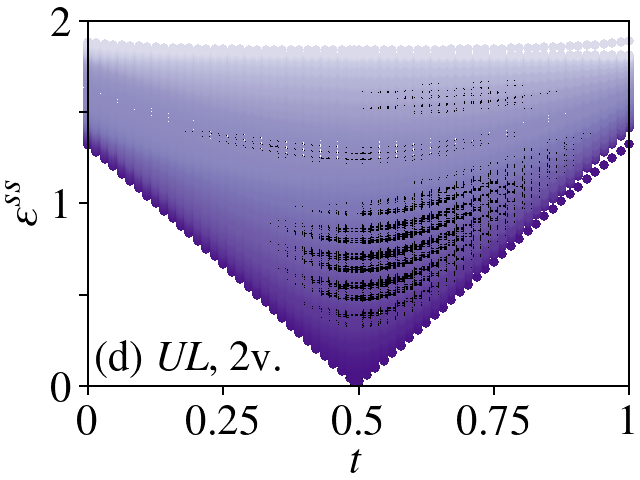} \\
\includegraphics[scale=0.25]{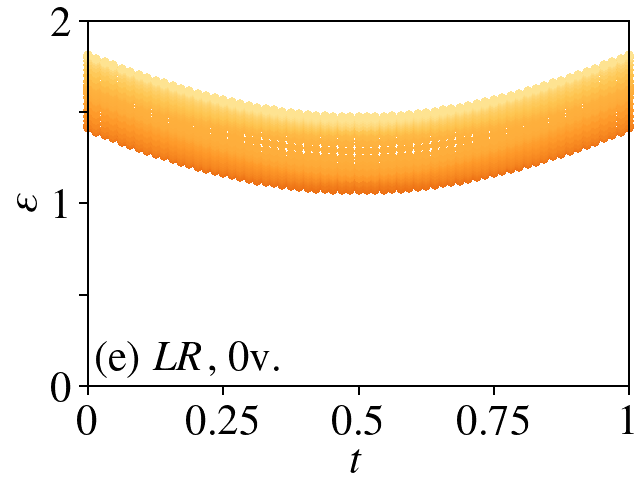} \;
\includegraphics[scale=0.25]{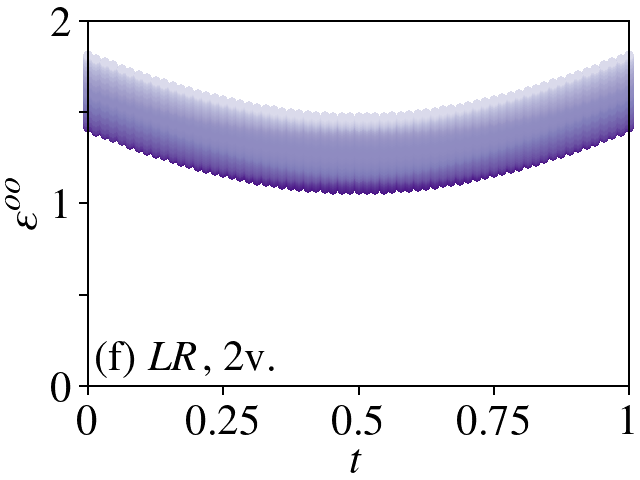} \;
\includegraphics[scale=0.25]{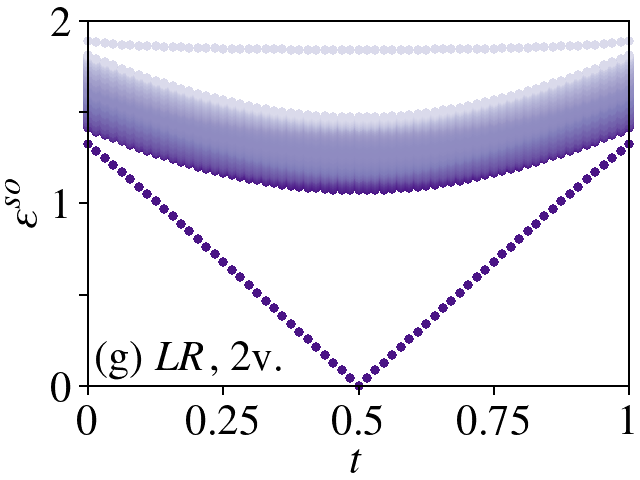} \;
\includegraphics[scale=0.25]{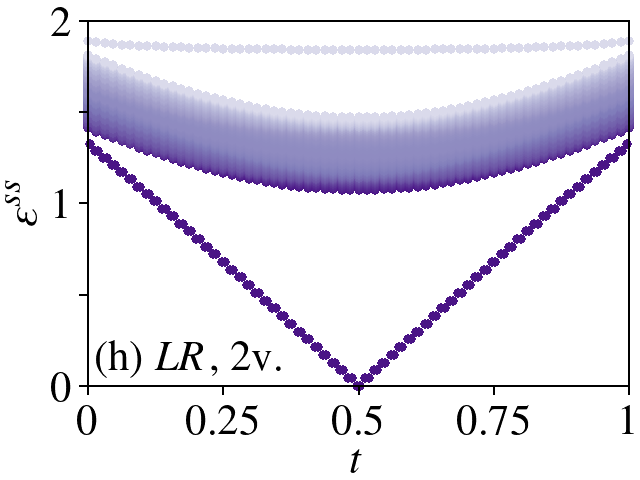}
\caption{\label{figure:octagonspectrum}Further results from the study of the Kitaev square-octagon model. $0\mathrm{v}$ and $2\mathrm{v}$ stand for ``vortex-free'' and ``two-vortex'', respectively. (a) The fermion spectrum $\lbrace \varepsilon_k^{(0\mathrm{v})} \rbrace$ of the vortex-free sector (orange dots of varying shades) along the path $UL$, which is linearly parametrized by $t \in [0,1]$. (b,c,d) The averaged fermion spectra $\lbrace \varepsilon_k^{oo} \rbrace$, $\lbrace \varepsilon_k^{so} \rbrace$, and $\lbrace \varepsilon_k^{ss} \rbrace$ of two-vortex sectors (purple dots of varying shades) with both vortices at unit octagons, one vortex at a unit square and the other at a unit octagon, and both vortices at unit squares, respectively, along the path $UL$. (e) The fermion spectrum $\lbrace \varepsilon_k^{(0\mathrm{v})} \rbrace$ of the vortex-free sector (orange dots of varying shades) along the path $LR$, which is linearly parametrized by $t \in [0,1]$. (f,g,h) The averaged fermion spectra $\lbrace \varepsilon_k^{oo} \rbrace$, $\lbrace \varepsilon_k^{so} \rbrace$, and $\lbrace \varepsilon_k^{ss} \rbrace$ of two-vortex sectors (purple dots of varying shades) with both vortices at unit octagons, one vortex at a unit square and the other at a unit octagon, and both vortices at unit squares, respectively, along the path $LR$. In (a)-(h), lighter colors indicate larger $k$. In (b)-(d) and (f)-(h), the (black) error bars, which represent the standard deviations, are mostly smaller than the dots or hidden in the continuum of data. In (h), the two lowest ($k=1,2$) fermion modes are shifted slightly away from each other for visibility.}
\end{figure*}

\textit{Kitaev square-octagon model.} With $L_x \times L_y = 16 \times 16$ and $L_x' \times L_y' = 4 \times 12$, we compute $\lbrace \varepsilon_k^{(0 \mathrm{v})} \rbrace$ and $\lbrace \varepsilon_k^{\alpha \beta} \rbrace$ for $\alpha \beta = oo , so , ss$. Along the path $UL$, $\lbrace \varepsilon_k^{\alpha \beta} (t) \rbrace$ behave similarly as $\lbrace \varepsilon_k^{(0\mathrm{v})} (t) \rbrace$ for all $\alpha \beta$, see Figs.~\ref{figure:octagonspectrum}a-\ref{figure:octagonspectrum}d. In each of these spectra, we observe a continuum of fermion modes above the fermion gap at all $t$, including the critical point $t=0.492$ (to three significant digits) where $\Delta_\psi^{(0\mathrm{v})} (t) = 0$ and $\Delta_\psi^{\alpha \beta} (t) \approx 0$. In contrast, along the path $LR$, $\lbrace \varepsilon_k^{oo} (t) \rbrace$, $\lbrace \varepsilon_k^{so} (t) \rbrace$, and $\lbrace \varepsilon_k^{ss} (t) \rbrace$ differ from each other qualitatively. Recall that $\Delta_\psi^{\alpha \beta} (t) = 0$ at $t=1/2$ if and only if $\alpha = s$ or $\beta = s$, which further indicates a change (the invariance) of anyon species on every unit square (octagon) as $t$ increases or decreases past $1/2$, while $\Delta_\psi^{(0\mathrm{v})} (t)$ remains finite throughout $LR$. $\lbrace \varepsilon_k^{oo} (t) \rbrace$ highly resembles $\lbrace \varepsilon_k^{(0\mathrm{v})} (t) \rbrace$, where the fermion modes form a continuum above $\varepsilon = 1$, see Figs.~\ref{figure:octagonspectrum}e and \ref{figure:octagonspectrum}f. $\lbrace \varepsilon_k^{so} (t) \rbrace$ exhibits a continuum similar to that of the vortex free sector, but the lowest ($k=1$) fermion mode is decoupled from the continuum, see Fig.~\ref{figure:octagonspectrum}g. The separation of this mode from the continuum is larger when $t$ is closer to $1/2$. $\lbrace \varepsilon_k^{ss} (t) \rbrace$ also exhibits a continuum similar to that of the vortex-free sector, but the two lowest ($k=1,2$) fermion modes are decoupled from the continuum, see Fig.~\ref{figure:octagonspectrum}h. The separation of these modes, which are degenerate throughout $LR$, from the continuum is larger when $t$ is closer to $1/2$. The largest error of $\varepsilon_k^{\alpha \beta} (t)$ is of the order of $10^{-2}$ ($10^{-3}$) along the path $UL$ ($LR$), which indicates that the two-vortex sectors being averaged over have rather similar fermion spectra. \\

\begin{figure*}
\includegraphics[scale=0.25]{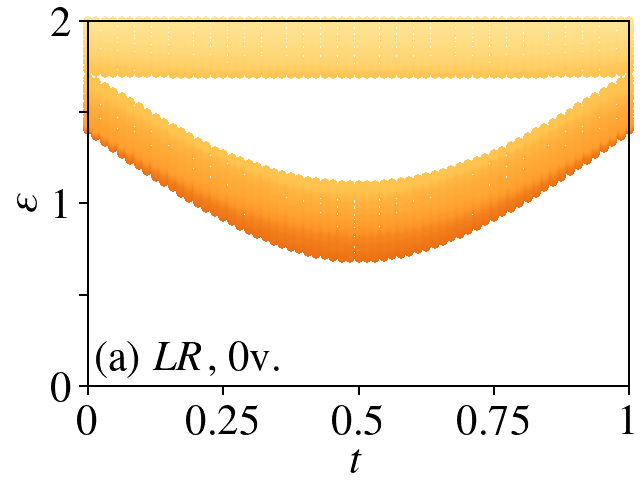} \;
\includegraphics[scale=0.25]{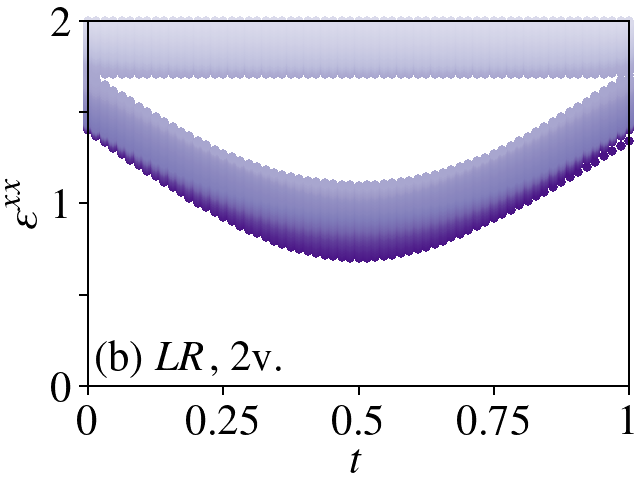} \;
\includegraphics[scale=0.25]{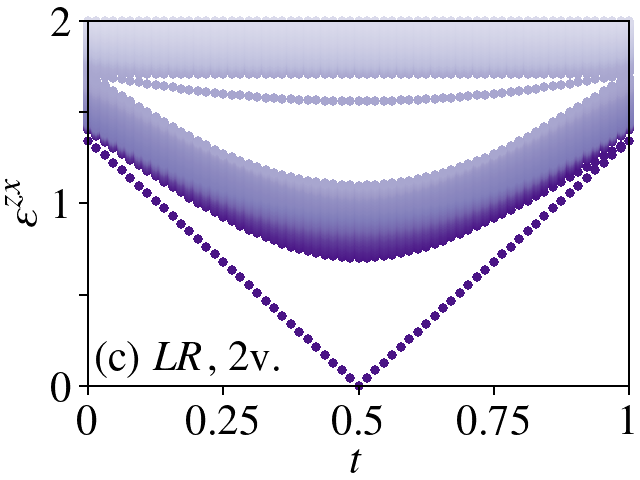} \;
\includegraphics[scale=0.25]{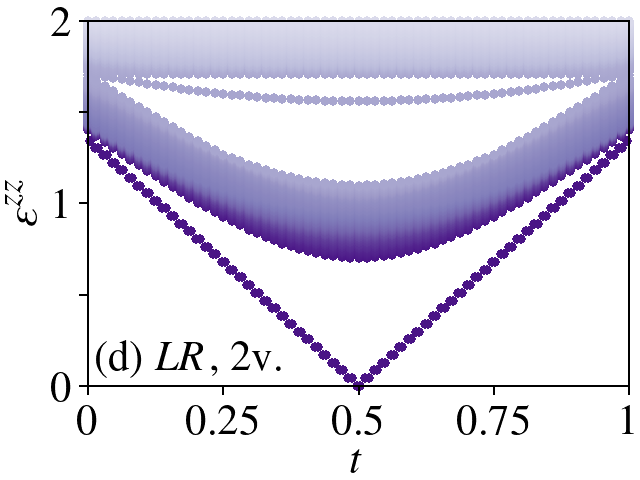}
\caption{\label{figure:kekulespectrum}Further results from the study of the Kitaev Kekul\'{e} model. $0\mathrm{v}$ and $2\mathrm{v}$ stand for ``vortex-free'' and ``two-vortex'', respectively. (a) The fermion spectrum $\lbrace \varepsilon_k^{(0\mathrm{v})} \rbrace$ of the vortex-free sector (orange dots of varying shades) along the path $LR$, which is linearly parametrized by $t \in [0,1]$. (b,c,d) The averaged fermion spectra $\lbrace \varepsilon_k^{xx} \rbrace$, $\lbrace \varepsilon_k^{zx} \rbrace$, and $\lbrace \varepsilon_k^{zz} \rbrace$ of two-vortex sectors (purple dots of varying shades) with both vortices at $x$ hexagons, one vortex at a $z$ hexagon and the other at an $x$ hexagon, and both vortices at $z$ hexagons, respectively, along the path $LR$. In (a)-(d), lighter colors indicate larger $k$. In (b)-(d), the (black) error bars, which represent the standard deviations, are either smaller than the dots or hidden in the continuum of data. In (d), the two lowest ($k=1,2$) fermion modes are shifted slightly away from each other for visibility.}
\end{figure*}

\textit{Kitaev Kekul\'{e} model.} With $L_1 \times L_2 = 16 \times 16$ and $L_1' \times L_2' = 2 \times 12$, we compute $\lbrace \varepsilon_k^{(0\mathrm{v})} \rbrace$ and $\lbrace \varepsilon_k^{\alpha \beta} \rbrace$ for $\alpha \beta = xx , yy , zz , yz , zx , xy$ along the path $LR$. Recall that $\Delta_\psi^{\alpha \beta} = 0$ at $t = 1/2$ if and only if $\alpha = z$ or $\beta = z$, which further indicates a change (the invariance) of anyon species on every $z$ ($x$ and $y$) hexagon, while $\Delta_\psi^{(0\mathrm{v})} (t)$ remains finite throughout $LR$. We find that the fermion spectra of the two-vortex sectors can be qualitatively distinguished according to the number $n \in \lbrace 0 , 1 , 2 \rbrace$ of vortices that are hosted by the $z$ hexagons. We take $\alpha \beta = xx$, $zx$, $zz$ as representatives of the cases $n = 0$, $1$, and $2$, respectively. $\lbrace \varepsilon_k^{xx} (t) \rbrace$ highly resembles $\lbrace \varepsilon_k^{(0\mathrm{v})} (t) \rbrace$, where the fermion modes form two continua above $\varepsilon = 0.5$, see Figs.~\ref{figure:kekulespectrum}a and \ref{figure:kekulespectrum}b. $\lbrace \varepsilon_k^{zx} (t) \rbrace$ exhibits two continua similar to those of the vortex-free sector, but the lowest ($k=1$) fermion mode is unambiguously decoupled from the lower continuum for $t$ not near $0$ or $1$, see Fig.~\ref{figure:kekulespectrum}c. The separation of this mode from the continuum is larger when $t$ is closer to $1/2$. $\lbrace \varepsilon_k^{zz} (t) \rbrace$ also exhibits two continua similar to those of the vortex-free sector, but the two lowest ($k=1,2$) fermion modes are unambiguously decoupled from the lower continuum for $t$ not near $0$ or $1$, see Fig.~\ref{figure:kekulespectrum}d. The separation of these modes, which are degenerate throughout $LR$, from the continuum is larger when $t$ is closer to $1/2$. The largest error of $\varepsilon_k^{\alpha \beta} (t)$ is of the order of $10^{-3}$, which indicates that the two-vortex sectors being averaged over have rather similar fermion spectra. \\

Along the path $LR$ in either model, we also notice the presence of decoupled fermion mode(s) at high energies for an extended range of $t$ in concurrence with the presence of decoupled fermion mode(s) at low energies, see Figs.~\ref{figure:octagonspectrum}g, \ref{figure:octagonspectrum}h, \ref{figure:kekulespectrum}c, and \ref{figure:kekulespectrum}d. Nevertheless, the assignment of anyon species is not affected by the high-energy modes, so we focus on the low-energy modes in the discussions above. \\

Our results suggest the following scenario. Suppose that the fermion gaps of some but not all of the two-vortex sectors become zero, while the fermion gap of the vortex-free sector remains finite, at a critical point. If a two-vortex sector involves $n \in \lbrace 0 , 1 , 2 \rbrace$ vortices whose anyon species are changed upon the transition, then there exist exactly $n$ gapless fermion modes that are decoupled from the rest of the spectrum. For $n>0$, the presence of at least $n$ gapless fermion modes is consistent with the fusion rules, but the discontinuity between the $n$th and $(n+1)$th modes, though allowed, is unexpected. In addition, removing the decoupled fermion modes at both low and high energies results in a spectrum strongly resembling that of the vortex-free sector.

\end{document}